\documentclass[11pt]{article}
\pdfoutput=1

\usepackage{jheppub} 

\usepackage{amsmath,amssymb,epsfig,amsfonts,tikz-cd}
\usepackage{graphicx,subfigure}
\usepackage{epsf}
\usepackage{makeidx}
\usepackage{float}
\restylefloat{table}
\usepackage{pdflscape}
\usepackage{tikz}
\usepackage{array}
\usepackage{youngtab}
\usepackage{multirow}
\usepackage{color}
\usepackage{ulem}
\usepackage{hyperref}
\usepackage{cleveref}
\usepackage{braket}
\usepackage{stmaryrd}
\usepackage{mathtools}
\usepackage{bm}
\usepackage[makeroom]{cancel}
\usepackage{ulem}
\usepackage{amsmath, pstricks-add}

\usepackage{booktabs,multirow}
\usepackage{array}
\usepackage{makecell}
 \usepackage{tabu}
 
\usepackage{environ}
\usepackage{tikz-cd}

\usetikzlibrary{backgrounds}

\usepackage{tensor}

\newcommand{\be}{\begin{equation}}
\newcommand{\ee}{\end{equation}}
\newcommand{\ba}{\begin{aligned}}
\newcommand{\ea}{\end{aligned}}

\newcommand{\bs}{\begin{split}}
\newcommand{\es}{\end{split}}

\def\tM3{\mathcal{M}_3}

\NewEnviron{eq}[1]{\begin{equation}\begin{split}\label{eq: #1} \BODY \end{split}\end{equation}}

\renewcommand{\arraystretch}{1.55}

\newcommand{\lb}{\left(}
\newcommand{\rb}{\right)}
\newcommand{\lbb}{\left[}
\newcommand{\rbb}{\right]}
\newcommand{\lbbb}{\left\{}
\newcommand{\rbbb}{\right\}}
\newcommand{\tn}[1]{\textnormal{#1}}

\newlength{\sswidth}

\newcommand{\C}{\mathbb{C}}
\renewcommand{\P}{\mathbb{P}}

\newcommand{\bea}{\begin{eqnarray}}
\newcommand{\eea}{\end{eqnarray}}

\newcommand{\R}{{\mathbb R}}
\newcommand{\Z}{{\mathbb Z}}

\def\diag{\mathop{\mathrm{diag}}\nolimits}

\def\bra#1{{\langle{#1}|}}

\def\ket#1{{|{#1}\rangle}}

\def\unit{{1\kern-.65ex {\rm l}}}
\def\1{{1\kern-.65ex {\rm l}}}

\newcommand{\del}{\partial}

\def\CB{{\cal B}}
\def\CC{{\cal C}}

\def\CH{{\cal H}}

\def\CL{{\cal L}}
\def\CM{{\cal M}}
\def\CN{{\cal N}}
\def\CO{{\cal O}}

\def\CQ{{\cal Q}}

\def\CS{{\cal S}}

\newcount\hour \newcount\minute
\hour=\time \divide \hour by 60
\minute=\time
\def\now{%
\ifnum \hour<13
  \ifnum \hour=0 \advance \hour by 12 \number\hour:\else \number\hour:\fi%
     \ifnum \minute<10 0\fi%
     \number\minute%
\ A.M.%
\else \advance \hour by -12 \number\hour:%
  \ifnum \minute<10 0\fi%
  \number\minute%
  \ P.M.%
\fi%
}

\makeatother

\begin{document}

\baselineskip=18pt  
\numberwithin{equation}{section}  


%
%


\thispagestyle{empty}



%
%


\thispagestyle{empty}


\vspace*{1cm} 
\begin{center}
{\Huge \noindent{Local $G_2$-Manifolds, Higgs Bundles\smallskip 

\noindent ~and a Colored Quantum Mechanics}
}

\vspace*{1.5cm}
Max H\"ubner\\
\vspace*{0.5cm} 
{\it Mathematical Institute, University of Oxford, \\
Andrew-Wiles Building,  Woodstock Road, Oxford, OX2 6GG, UK}\\
\vspace*{1cm}
\end{center}
\noindent M-theory on local $G_2$-manifolds engineers 4d minimally supersymmetric gauge theories. We consider ALE-fibered $G_2$-manifolds and study the 4d physics from the view point of a partially twisted 7d supersymmetric Yang-Mills theory and its Higgs bundle. Euclidean M2-brane instantons descend to non-perturbative effects of the 7d supersymmetric Yang-Mills theory, which are found to be in one to one correspondence with the instantons of a colored supersymmetric quantum mechanics. We compute the contributions of M2-brane instantons to the 4d superpotential in the effective 7d description via localization in the colored quantum mechanics. Further we consider non-split Higgs bundles and analyze their 4d spectrum.

%

\newpage

\tableofcontents
\vspace{40pt}
\section{Introduction}

M-theory compactified on compact $G_2$-manifolds gives rise to 4d $\CN=1$ gauge theories coupled to gravity \mbox{\cite{Acharya:1996ci, Acharya:1998pm, Acharya:2000gb, Witten:2001uq, Atiyah:2001qf, Acharya:2001gy, Acharya:2004qe, Kennon2018G_2ManifoldsAM}}. Favourably, these constructions involve purely geometric backgrounds which, unlike other known construction of minimally supersymmetric 4d vacua, need not be supplemented with additional data. The challenges in these constructions lie in understanding the complicated geometry of $G_2$-manifolds and their metric moduli spaces. Further the list of smooth compact $G_2$-manifolds \cite{joyce1996I, joyce1996II, MR2024648, MR3109862, Corti:2012kd, Joyce2017ANC} is short and contains no examples of singular compact $G_2$-manifolds with the required codimension 4 and 7 singularities necessary to engineer non-abelian gauge symmetries and chiral matter in 4d respectively. 

The class of $G_2$-manifolds referred to as twisted connected sum $G_2$-manifolds gives a landscape of roughly $10^6$ compact geometries and have recently been studied, together with their singular limits, in the physics literature \cite{Earp:2013jea, Halverson:2014tya, Guio:2017zfn, Braun:2017ryx, Braun:2017uku, Fiset:2018huv, Braun:2018vhk, Xu:2020nlh, Cvetic:2020piw}. The gauge theory sector of these compactifications can be isolated and studied in local models of the geometry \cite{Pantev:2009de, Braun:2018vhk, Barbosa:2019bgh} using techniques involving Higgs bundles and spectral covers previously fruitful in F-theory model building \cite{Donagi1993SpectralC, Friedman:1997yq, Donagi:2008ca, Beasley:2008dc, Hayashi:2008ba, Blumenhagen:2009yv, Marsano:2009gv, Marsano:2009wr, Donagi:2009ra, Hayashi:2009ge, Hayashi:2010zp, Marsano:2011hv}. Independently, local $G_2$-manifolds have given key insights into the physics at conical codimenion 7 singularities  \cite{Acharya:2001gy, Witten:2001uq, Berglund:2002hw}, dualities between 4d theories \cite{Atiyah:2001qf, Aganagic:2001ug, Cachazo:2001jy, Curio:2002ja}, confinement and domain wall theories in 4d \cite{Atiyah:2000zz, Acharya:2001dz, Eckhard:2018raj} and more. 

Semi-realistic field theories are engineered by local $G_2$-manifolds $X_7$ realizing an ADE gauge group in 4d. These necessarily have a description in terms of an ALE fibration over a supersymmetric 3-cycle $M_3$
\be\label{eq:ALE101}
\widetilde{\C^2/\Gamma_{\text{ADE}}} ~ \xhookrightarrow{} ~ X_7 ~ \rightarrow ~ M_3\,,
\ee
and have been studied in \cite{Acharya:2001gy,Atiyah:2001qf, Pantev:2009de, Braun:2018vhk}. The 4d $\CN=1$ gauge theory engineered by M-theory on $X_7$ can be derived in two steps. A reduction along the ALE fibers $\widetilde{\C^2/\Gamma_{\text{ADE}}}$ produces an effective 7d partially twisted supersymmetric Yang Mills theory on $M_3\times\R^{1,3}$ and subsequently compactifying this theory on $M_3$ the 4d $\CN=1$ gauge theory follows. Amongst the 4d data the superpotential proves most challenging to derive. It only receives contributions from Euclidean M2-branes instantons wrapped on supersymmetric 3-cycles in $X_7$. Contributions of single M2-brane instantons to the superpotential are computed in \cite{Harvey:1999as} but it is hard to gain insight into the global structure of these instantons directly in M-theory. The effective 7d SYM remedies this situation by translating the data of the ALE geometry \eqref{eq:ALE101} into a Higgs bundle with 3d base $M_3$ and associated spectral cover. This gives shape to the global structure of the ALE geometry at the cost of obscuring the effects of the M2-brane instantons, which simply descend to non-perturbative physics of the 7d SYM.

In \cite{Pantev:2009de, Braun:2018vhk} these non-perturbative effects in 7d SYM were studied in the context of a Higgs bundle with split spectral cover. Here it was found that the non-perturbative effects, due to M2-brane instantons, which generate the quadratic terms of the superpotential can be understood and computed using Witten's supersymmetric quantum mechanics (SQM) \cite{Witten:1982im}. More precisely, the gradient flow line instantons of the SQM were in correspondence with some of the supersymmetric cycles in the ALE geometry $X_7$. This proved sufficient for computing the 4d spectrum and demonstrate its chirality. However, it remained unclear how to interpret the supersymmetric cycles generating higher terms of the superpotential in this SQM frame work and compute their non-perturbative contributions in the effective 7d SYM. Similarly, the analysis did not apply to more general ALE fibrations with non-split spectral covers. The reason for these limitations lies in the SQM only ever encoding the information of a single sheet of the spectral cover. Both non-split spectral covers, where the sheets are mixed by monodromy effects, and the non-perturbative effects generating Yukawa couplings simultaneously involve multiple sheets of the spectral cover.

In this paper we present a colored $\CN=(1,1)$ SQM whose instantons are in one to one correspondence with all non-perturbative effects of the 7d SYM, which in turn originate from M2-brane instantons in M-theory on $X_7$. We compute the non-perturbative contributions to the superpotential of individual M2-brane instantons in the effective 7d SYM and comment on the global structure of all such contributions. Further, we discuss Higgs bundles over the 3d base $M_3$ with non-split spectral covers, give explicit examples and analyze the problem of zero-mode counting for these configurations.

This paper is structured as follows. In section \ref{sec:Recap} we establish notation and cover background material on local $G_2$-manifolds, partially twisted 7d SYM and the effective 4d field theories these engineer. Extended discussions on the reviewed topics can be found in \cite{Acharya:1998pm, Acharya:2004qe, Pantev:2009de, Braun:2018vhk}. Section \ref{sec:Harmonic} concentrates on 3d Higgs bundles with non-split spectral covers. Here we note their general structure and how a large class of such configurations follow (implicitly) from TCS $G_2$-manifolds. We also give an explicit class of examples and discuss the topology of the sheets of such covers which function as the target space of the colored SQM. In section \ref{sec:GeneralHiggsBundleSQM} we introduce the colored SQM in all generality. In the presented form it is applicable to the study of all BPS vacua of the 7d SYM, in particular vacua with flux, and we discuss the perturbative ground states and flow tree instantons of the SQM. Section \ref{sec:Split} then studies the colored SQM for split Higgs bundles which were the focus of \cite{Pantev:2009de, Braun:2018vhk}. We demonstrate how to understand the colored SQM as multiple interacting copies of Witten's SQM. Further we present the localization computation determining the contributions of the Euclidean M2-brane instantons to the Yukawa couplings in 4d. In section \ref{sec:NonSplit} we rerun the arguments from section \ref{sec:Split} for non-split Higgs bundles focussing on the 4d spectrum. We explain the consequence of the non-split cover in the 7d SYM, consider an explicit examples and determine their spectrum. Finally, in section \ref{sec:4d} we give a brief summary before ending with some concluding remarks in section \ref{sec:CONC}.

\section{M-theory on ALE-fibered $G_2$-Manifolds}
\label{sec:Recap}

At low energies M-theory on the non-compact $G_2$-manifolds of \eqref{eq:ALE101} is well approximated by a partially twisted 7d supersymmetric Yang-Mills theory on $\R^{1,3}\times M_3$. The geometry of each ALE fiber is encoded in the background value of the Higgs field of the SYM theory and the metric equations ensuring the $G_2$-holonomy of the ALE fibration generalize to the BPS equations of the gauge theory. These BPS equations read
\be\label{eq:BPS101}
iF_A+[\phi,\phi]=0\,, \qquad d_{A}\phi=0\,, \qquad d_A*\phi=0\,,
\ee
where the connection $A$ and the Higgs field $\phi$ are Lie algebra valued 1-forms on $M_3$. These equations are a 3d generalisation of Hitchin's equations \cite{Hitchin87theself-duality, Pantev:2009de, Braun:2018vhk} and solved by complex flat connections on $M_3$ satisfying a gauge fixing constraint. Solutions to \eqref{eq:BPS101} are the supersymmetric vacua of the partially twisted 7d SYM 
which fully determine 4d $\CN=1$ gauge theory when compactified on $M_3$.

Here we introduce ALE-fibered $G_2$-manifolds and discuss their geometry. We expand on the partially twisted 7d SYM they engineer in M-theory and the minimally supersymmetric gauge theories these give rise to in 4d. Of particular interest to us are the 3-cycles of the geometry which wrapped by M2-branes generate the 4d superpotential. These are most favourably discussed in a spectral cover picture of the set-up which we introduce for abelian solutions to the BPS system \eqref{eq:BPS101}. 

\subsection{ALE-fibered $G_2$-manifolds}
\label{sec:LocalG2}

We begin with a non-compact $G_2$-manifold with ADE singularities supported along an associative submanifold $M_3$. Partial, minimal resolutions of the singularities lead to ALE-fibered manifolds
\be\label{eq:ALEFiberedSpace}
\widetilde{\C^2/\Gamma_{\text{ADE}}} ~ \xhookrightarrow{} ~ X_7 ~ \rightarrow ~ M_3\,.
\ee
Each ALE fiber $\widetilde{\C^2/\Gamma_{\text{ADE}}}$ is Hyperk\"ahler with a triplet of K\"ahler forms $\omega_i$ which vary across the base $M_3$. Whenever the space $X_7$ admits a metric $g_{ij}$ of special holonomy $\text{Hol}\,(X_7,g_{ij})=G_2$ there exists an induced 3-form $\Phi_3$ satisfying
\be\label{eq:G2Constraints}
d\Phi_3=0\,, \qquad d*_{\Phi_3}\Phi_3=0\,.
\ee
For the ALE-fibered geometries \eqref{eq:ALEFiberedSpace} it can be constructed from the Hyperk\"ahler triplet \cite{ Acharya:1998pm} and given with respect to a locally flat frame on $M_3$ by
\be\label{eq:G23Form}
\Phi_3=dx^1\wedge dx^2\wedge dx^3+ dx^i \wedge \omega_i\,.
\ee 
For further discussion on $G_2$ geometry we refer to \cite{joyce2000compact, 6852, Acharya:2004qe}.

\begin{figure}
\centering
\begin{minipage}[t]{.40\textwidth}
  \centering
  \includegraphics[width=1\linewidth]{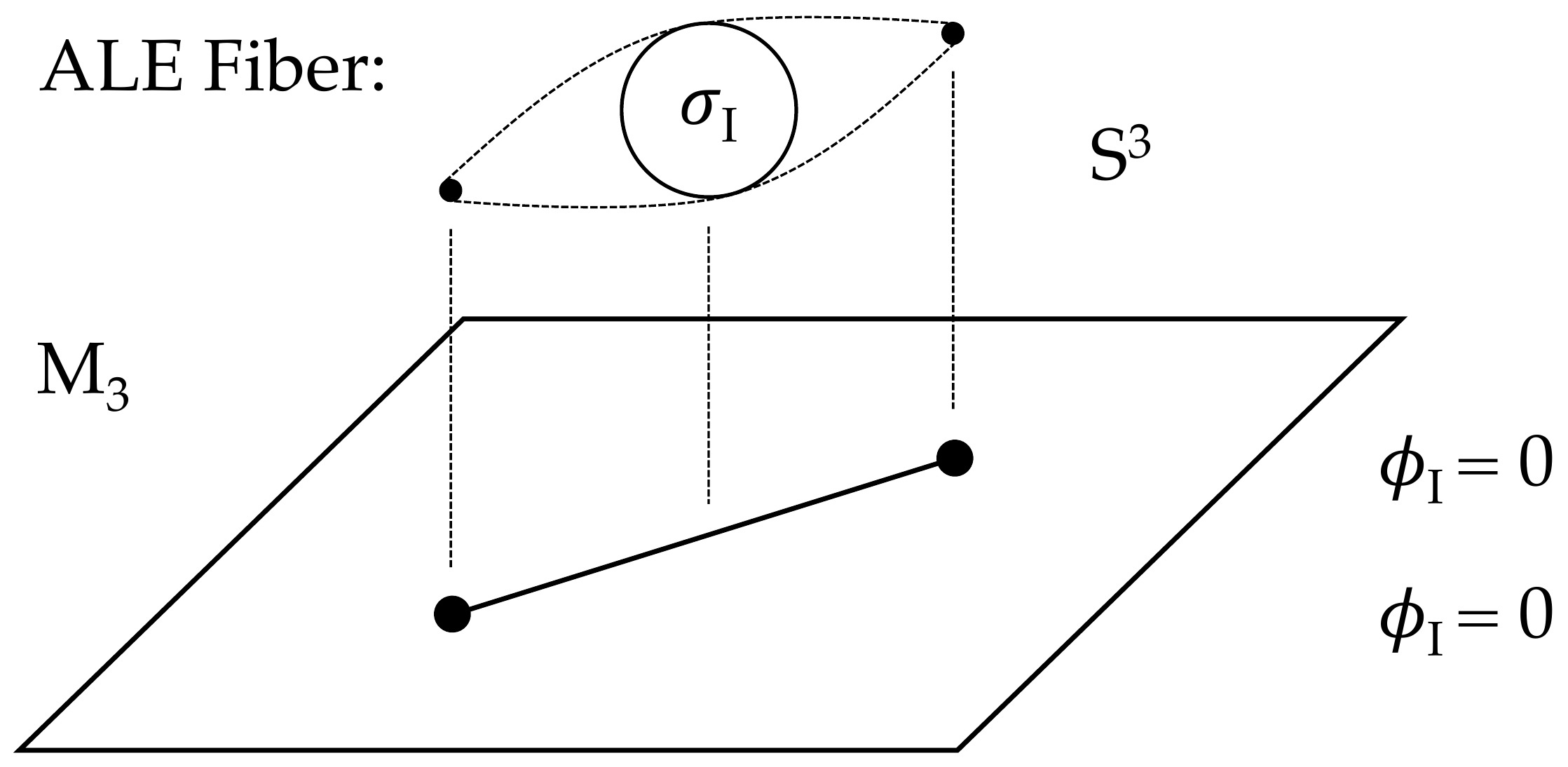}
\end{minipage}\qquad \qquad
\begin{minipage}[t]{.40\textwidth}
  \centering
  \includegraphics[width=1\linewidth]{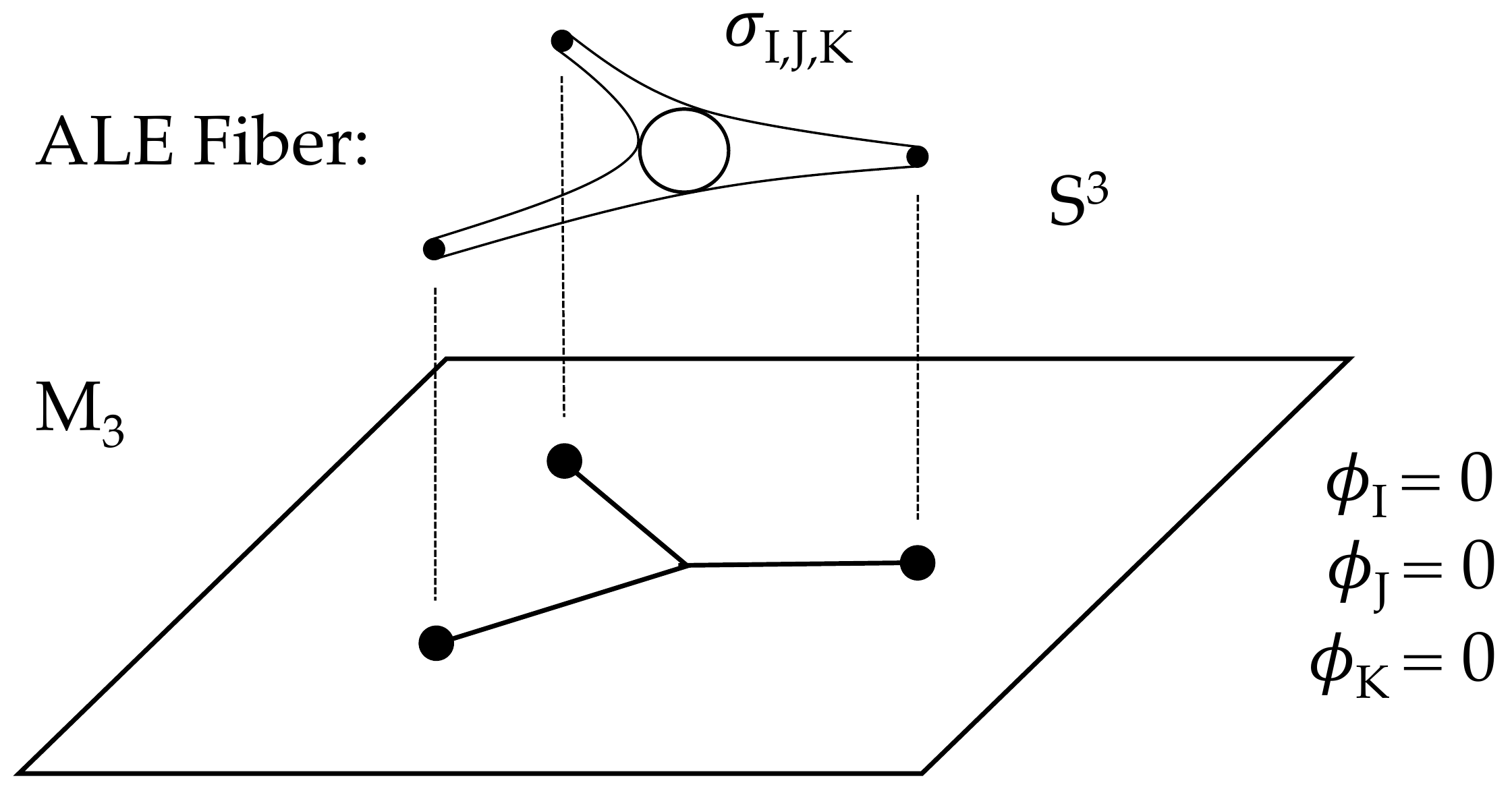}
\end{minipage}
\caption{The pictures show two supersymmetric 3-spheres $S^3$ in the local $G_2$-manifold $X_7$ of \eqref{eq:ALEFiberedSpace}. In M-theory these are wrapped by Euclidean M2-brane instantons which contribute to the 4d superpotential. The picture on the left shows a 3-sphere traced out by the 2-sphere $\sigma_I$ along a path in $M_3$ connecting two fibers in which the cycle collapses. The picture on the right shows a 3-sphere traced out in similar fashion by three linearly dependent 2-spheres $\sigma_{I,J,K}$.}
\label{fig:Relevant3Spheres}
\end{figure}

The second homology group of a fully resolved ALE fiber is generated by a basis of $R$ 2-cycles $\sigma_I\in H_2(\widetilde{\C^2/\Gamma_{\text{ADE}}})$ introduced by resolutions. Here the number $R$ is the rank of the corresponding Lie group $G_{\tn{ADE}}$. The 2-cycles $\sigma_I\cong S^2$ are 2-spheres. Integrating the $G_2$ 3-form $\Phi_3$ against the cycles $\sigma_I$ in each fiber gives rise to $R$ local 1-forms $\phi_I$ which collect the Hyperk\"ahler periods of the 2-cycle as its component functions
\be\label{eq:HiggsComponents}
\phi_I=\lb \int_{\sigma_I} \omega_i\rb dx^i\,, \qquad I=1,\dots, R\,.
\ee
The vanishing locus of $\phi_I$ therefore correspond to fibers in which the cycle $\sigma_I$ collapses. If the initial ADE singularity is only partially resolved then the 1-forms $\phi_I$ associated to the collapsed vanishing cycles vanish globally on $M_3$. The vanishing locus of non-zero local 1-forms $\phi_I$ is cut out by 3 equations on $M_3$ and thus generically consists of points. These correspond to isolated singularity enhancements in the ALE-fibered $G_2$-manifold $X_7$ 
\be\label{eq:SingEnhance}
\tn{Singularity Enhancement in } X_7\,:\qquad \phi_I(x)=0\,.
\ee
Paths in $M_3$ connecting points above which the 2-cycle $\sigma_I$ collapses lift to 3-spheres in the ALE fibration $X_7$. More generally, tree-like graphs connecting points above which one of a linearly dependent collection of 2-cycles collapses also lift to 3-spheres. These 3-spheres constitute supersymmetric cycles whenever their associated graphs are piecewise solutions to flow equations set by the Cartan components \eqref{eq:HiggsComponents} of the Higgs field \cite{Acharya:2001gy, Pantev:2009de, Braun:2018vhk}. In figure \ref{fig:Relevant3Spheres} we have sketched two such 3-spheres and their projections to the base $M_3$.

The equations \eqref{eq:G2Constraints} integrate to constraints on the local 1-forms $\phi_I$ given in \eqref{eq:HiggsComponents}  
\be\ba\label{eq:HiggsHarmonicity}
&\tn{F-term\,:}\qquad 0=d\phi_I\,, \\
&\tn{D-term\,:}\qquad 0=d*\phi_I\,,
\ea\ee
and hold on the associative submanifold $M_3$ with respect to the $G_2$ metric $g_{ij}$ pulled back to $M_3$. We refer to the equations of \eqref{eq:HiggsHarmonicity} as F-term and D-term equations respectively and to the collection of 1-forms $\phi_I$ as the Cartan components of the Higgs field of the ALE-fibered $G_2$-manifold $X_7$. 

\subsection{Gauge Theory Sector}
\label{sec:GaugeTheorySector}

In M-theory the 3-spheres depicted in figure \ref{fig:Relevant3Spheres} are wrapped by M2-branes and give rise to a non-perturbatively generated superpotential for the 4d $\CN=1$ theory. Contributions of single such Euclidean M2-brane instantons to the superpotential were computed in \cite{Harvey:1999as}. Alternatively, these contributions can be derived from non-perturbative effects in an effective 7d SYM description. Here they are associated with the tree-like graphs which lift to the supersymmetric 3-spheres. With this in mind we now briefly discuss this 7d SYM and refer to \cite{Braun:2018vhk} for an extended discussion. 

The effective description of M-theory on the fibration $\C^2/\Gamma_{\text{ADE}} \, \xhookrightarrow{} \, X_7 \, \rightarrow \, M_3$ is a partially twisted 7d supersymmetric Yang-Mills theory on $\R^{1,3}\times M_3$ with gauge group $G_{\text{ADE}}$ \cite{Acharya:2000gb, Braun:2018vhk}. The global symmetries organizing the spectrum are the 4d Lorentz symmetry $SO(1,3)_L$ and an internal $SO(3)_{\text{twist}}=\text{diag}(SO(3)_{M_3},SO(3)_R)$ which follows from topologically twisting the local Lorentz group $SO(3)_{M_3}$ on the supersymmetric submanifold $M_3\subset X_7'$ with the R-symmetry group $SO(3)_R$ of the 7d $\CN=1$ supersymmetry algebra. After the twist the single vector multiplet of a 7d SYM decomposes into the gauge field $V_\mu$ and its associated gaugino $\eta_\alpha$, which transform as
\be\label{eq:4dGauge}
(V_\mu)\equiv 
 (\bm{2},\bm{2};\bm{1})\,, \qquad (\eta_\alpha,\bar{\eta}_{\dot\alpha})\equiv (\bm{2},\bm{1};\bm{1})\oplus(\bm{1},\bm{2};\bm{1})\,,
\ee
under $SO(1,3)_L\times SO(3)_\tn{twist}$, and the connection $A_i$ and the twisted scalars $\phi_i$ along $M_3$ together with their superpartners $\psi_{i\alpha}$ transform as
\be\label{eq:4dScalar}
(A_i)\equiv (\bm{1},\bm{1};\bm{3})\,,\qquad (\phi_i)\equiv (\bm{1}, \bm{1};\bm{3})\,, \qquad (\psi_{i\alpha},\bar{\psi}_{i\dot\alpha})\equiv (\bm{2},\bm{1};\bm{3})\oplus(\bm{1},\bm{2};\bm{3})\,.
\ee
The twisted scalars $\phi_i$ are called the Higgs field. The connection $A_i$ and Higgs field $\phi_i$ naturally complexify to $\varphi_i=\phi_i+iA_i$. Compactifying on $M_3$ to $\R^{1,3}$ the fields \eqref{eq:4dGauge} and \eqref{eq:4dScalar} descend to 4d $\CN=1$ vector and chiral multiplets respectively. The fields $(V_\mu,\eta_\alpha)$ and $(\varphi_i,\psi_{i\alpha})$ transform as scalars and 1-forms under the new local Lorentz symmetry $SO(3)_{\tn{twist}}$ of the submanifold $M_3$ and are therefore identified as
\be
V_\mu,\eta_\alpha,\bar\eta_{\dot\alpha} \in \Omega^0\lb M_3,\tn{ad}\,P_{\text{ADE}}\rb \,, \qquad A,\phi,\psi_{\alpha},\bar{\psi}_{\dot\alpha} \in \Omega^1\lb M_3,\tn{ad}\,P_{\text{ADE}}\rb \,.
\ee
Here $P_{\text{ADE}}$ is the principle bundle associated to the gauge group $G_{\text{ADE}}$ and $\tn{ad}\,P_{\text{ADE}}$ the associated vector bundle via the adjoint representation. Their geometry is determined by the background value of the connection $A$. Equivalently, the fields of \eqref{eq:4dGauge} and \eqref{eq:4dScalar} are Lie-algebra valued functions and 1-forms on $M_3$.

The supersymmetric vacua of the partially twisted 7d SYM are determined by its BPS equations which formulate a Hitchin system 
\be\ba\label{eq:BPS}
&\tn{F-term\,:}\qquad 0=iF_A+[\phi,\phi]\,, \quad 0=d_A\phi\\
&\tn{D-term\,:}\qquad 0=d_A* \phi\,,
\ea\ee
where $F_A$ is the curvature of the connection $A$ and the Hodge star and exterior derivative are taken on compact manifold $M_3$. Expanded in components the individual equations read
\be\ba\label{eq:BPSExpanded}
0&=i\lb\del_iA_j-\del_jA_i+i[A_i,A_j] \rb+[\phi_i,\phi_j]\,, \\
0&=\del_i\phi_j-\del_j\phi_i+i[A_i,\phi_j]-i[A_j,\phi_i]\,, \\
0&=g^{ij}\lb \del_i\phi_j+i[A_i,\phi_j] \rb\,.
\ea\ee
The Higgs field $\phi$ and connection $A$ define a complexified connection $\CQ$ on $M_3$ which by the F-term equations is flat
\be\label{eq:ComplexifiedConnection}
\CQ=d+\lbb\varphi,\cdot\,\rbb\,,\qquad \qquad\varphi=\phi+iA\,,\qquad \qquad\CQ\circ \CQ=0\,.
\ee
The D-term can be understood as complex gauge fixing condition. Given a 7d supersymmetric vacuum in terms a solution to \eqref{eq:BPS} the 4d physics follows from a compactification on the cycle $M_3$. The zero modes of the compactification are determined by both $\CQ$ and its complex conjugate as well as their adjoint operators. It is therefore natural to identify half of the fields with their Hodge dual images
\be
\eta_\alpha~ \rightarrow ~*\eta_\alpha\in \Omega^3\lb M_3,\tn{ad}\,P_{\text{ADE}}\rb\,, \qquad \bar\psi_{\dot\alpha}~ \rightarrow ~*\bar\psi_{\dot\alpha}\in \Omega^2\lb M_3,\tn{ad}\,P_{\text{ADE}}\rb\,.
\ee
After this identification the massless spectrum in 4d is counted by the zero modes of only the operator \eqref{eq:ComplexifiedConnection} and its adjoint on the supersymmetric submanifold $M_3$.

Non-trivial backgrounds for $A,\phi$ break the gauge symmetry $G_{\tn{ADE}}$ and its adjoint representation as
\be\ba\label{eq:GeneralSplittingText}
G_{\tn{ADE}}\quad&\rightarrow\quad G_{\tn{GUT}}\times H\,,\\
\tn{Ad}\, G\quad&\rightarrow\quad \lb \tn{Ad}\,G_{\tn{GUT}} \otimes {\bf 1}\rb \oplus \lb {\bf 1}\otimes \tn{Ad}\,H\rb \oplus \sum_n {\bf R}_n\otimes {\bf S}_n \,,
\ea\ee
where $G_{\tn{GUT}}$ is the commutant of the backgrounds for $A,\phi$. If the flat complexified connection $\CQ$ is not fully reducible the symmetry group $H$ may be further broken by monodromy effects to the stablizer of $\varphi=\phi+iA$ as explained in \cite{Chung:2014qpa}, we return to this case in section \ref{sec:NonSplit}. The decomposition \eqref{eq:GeneralSplittingText} lifts to the level of gauge bundles and we denote the bundle associated to ${\bf S}_n$ by $\CS_n$. Consequentially fermions valued in ${\bf R}_n$ are sections of $\CS_n$. The zero modes valued in ${\bf R}_n$ leading to massless 4d fields are therefore counted by
\be\ba\label{eq:Cohomologies}
\eta_\alpha\in H_{\CQ}^3(M_3,\CS_n)&\,,\qquad \bar{\eta}_{\dot\alpha}\in H_{\CQ}^0(M_3,\CS_n)\,, \\
\psi_\alpha\in H_{\CQ}^1(M_3,\CS_n)&\,,\qquad \bar{\psi}_{\dot\alpha}\in H_{\CQ}^2(M_3,\CS_n)\,.
\ea\ee
The 4d chiralities of the zero modes align with the $\Z_2$ grading of the exterior algebra whereby the 4d chiral index of the representation ${\bf R}_n$ is given by the Euler characteristic of $\CQ$ restricted to the subbundle $\CS_n$. 

Alternatively one can characterize the zero mode spectrum in terms of approximate zero modes and their non-perturbative corrections. Approximate zero modes are Lie algebra valued 1-forms on $M_3$
\be\label{eq:ApproxZero}
\tn{Approximate Zero Mode\,:}\qquad \chi\in\Omega^*(M_3,\mathcal{S}_n)
\ee
which are annihilated by the Laplacian $H=\frac{1}{2}\lbbb \CQ,\CQ^\dagger\rbbb$ to all orders in perturbation theory. The 7d SYM gives following mass matrix for these modes
\be\label{eq:OverLapIntegralsMass}
\tn{Mass Matrix\,:}\qquad M_{AB}=\int_{M_3}\braket{\chi_A,\CQ\chi_B}\,, 
\ee
where the bracket is anti-linear in the first argument and contracts the Riemannian and Lie algebra indices using the metric on $M_3$ and Killing form of the Lie algebra $\mathfrak{g}_{\tn{ADE}}$ respectively. Generators for the cohomologies \eqref{eq:Cohomologies} are then determined by the kernel of the matrix \eqref{eq:OverLapIntegralsMass}. The SYM also gives the 4d Yukawa couplings as the overlap integral
\be\label{eq:OverLapIntegralsYukawa}
\tn{Yukawa Couplings\,:}\qquad Y_{ABC}=\int_{M_3}\braket{ \chi_{C\,},\lbb \chi_{A\,}\wedge\,,\chi_B\rbb}\,,
\ee
between three approximate zero modes labelled by $A,B,C$. Zero modes are determined by \eqref{eq:OverLapIntegralsMass} to linear combinations of approximate zero modes whereby \eqref{eq:OverLapIntegralsYukawa} also sets the Yukawa couplings between these.

\subsection{Higgs Bundles and ALE Geometry}
\label{sec:FlatHigssBundles}

\begin{figure}
  \centering
  \includegraphics[width=14cm]{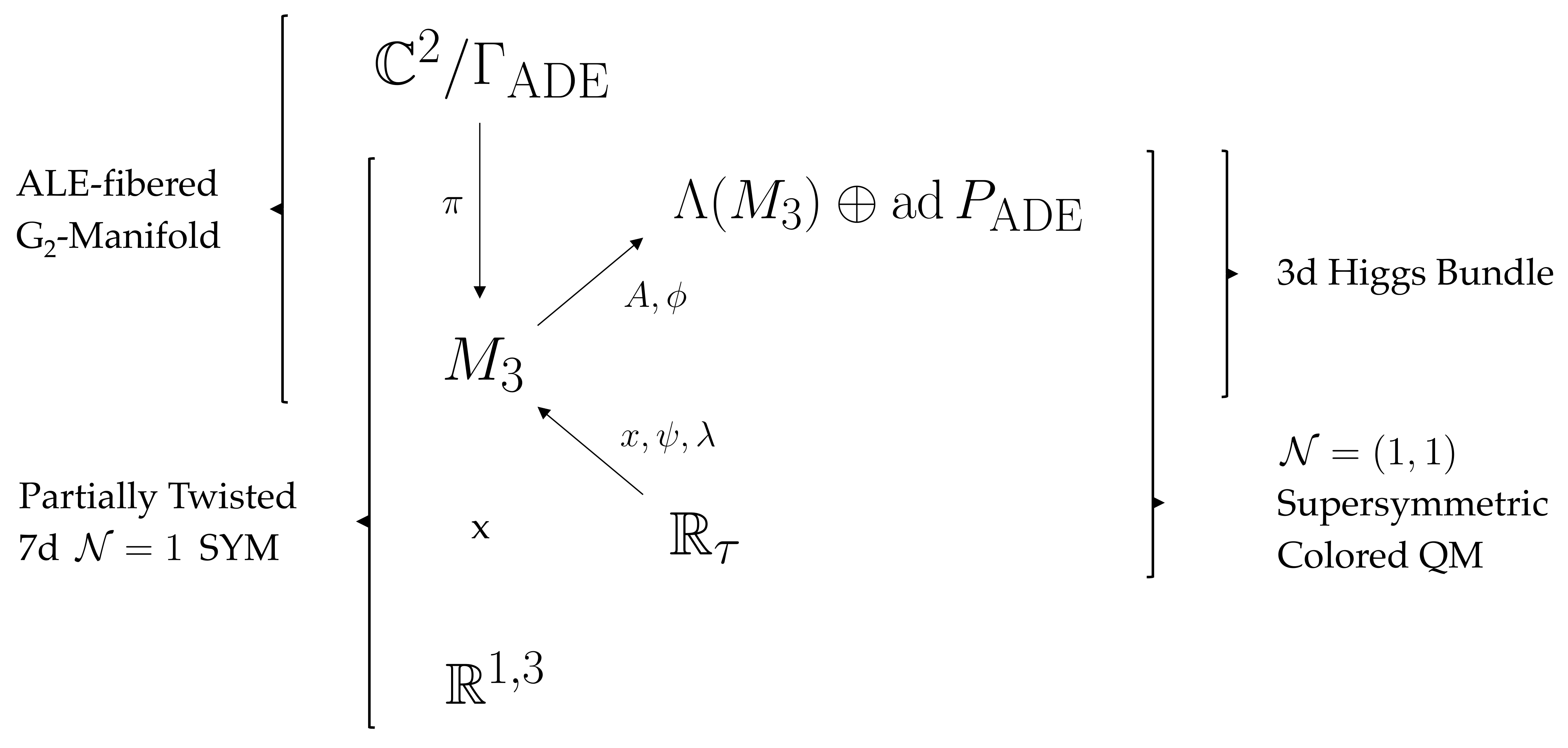}
    \caption{We sketch the relation between the ALE-fibered $G_2$-manifold, the partially twised 7d SYM, the Higgs bundle defined via its BPS system \eqref{eq:BPS} and the colored SQM probing the Higgs bundle. }\label{fig:SetUp}
\end{figure}

The Cartan components of the gauge field $A$ and Higgs field $\phi$ in the partially twisted 7d SYM originate from the supergravity 3-form and 11d metric in M-theory. Solutions to the BPS equations \eqref{eq:BPS} with flat abelian connections therefore lift to the local ALE-fibered $G_2$-manifolds described in section \ref{sec:LocalG2}. This is precisely the setting in which the BPS equations reduce to equations \eqref{eq:HiggsHarmonicity} encoding $G_2$ holonomy. Abelian solutions to the BPS-equations are given by a flat connection $A=A_IH^I$ and harmonic Higgs field $\phi=\phi_IH^I$. Here $H^I$ with $I=1,\dots, R$ denote the Cartan generators of the Lie algebra $\mathfrak{g}_{\tn{ADE}}$ and we refer to these solutions as Higgs bundles on $M_3$. With respect to a suitable basis of the Cartan subalgebra the non-vanishing 1-forms $\phi_I$ can be identified with those defined geometrically in \eqref{eq:HiggsComponents}. For this class of solutions we sketched how the structures introduced so far relate in figure \ref{fig:SetUp}.

We further restrict to set-ups where the eigenvalue 1-forms $\Lambda_K$ of the Higgs field $\phi$ and integral combinations thereof are Morse, that is the zeros of these 1-forms are isolated and their graph intersect the zero section of the cotangent bundle $T^*M_3$ transversely. The approximate zero modes setting the mass matrix \eqref{eq:OverLapIntegralsMass} and Yukawa couplings \eqref{eq:OverLapIntegralsYukawa} of the 4d theory are then in correspondence with the codimension 7 singularities \eqref{eq:SingEnhance} and their profiles sharply localize at the degeneration loci on $M_3$. These modes originate from M2-branes wrapping vanishing cycles above the marked points in figure \ref{fig:Relevant3Spheres}. As this locus consists of isolated points in $M_3$ both overlap integrals only receive non-perturbative contributions. These contributions originate from M2-branes wrapped on the 3-spheres depicted in figure \ref{fig:Relevant3Spheres}. In the effective 7d gauge theory description the contributions are associated with the tree-like graphs given by the projection of these 3-spheres.

The Higgs bundles can be further distinguished by their spectral cover. The spectral cover of a diagonal Higgs field $\phi$ is given by
\be\label{eq:GeneralSC}
\CC=\lbbb\tn{det}\,(s-\phi)=0\rbbb=\lbbb (x,\Lambda_K(x))\,|\, x\in M_3\rbbb=\cup_k \CC_k\subset T^*M_3 \,,
\ee 
which is the union of graphs of the eigenvalues $\Lambda_K$. The eigenvalues $\Lambda_K$ can be globally defined across $M_3$ or connected by branch sheets and respectively the spectral cover is fully reducible or not. We let $k$ run over irreducible components of $\CC$. We refer to the first configuration as split and the second configuration as non-split, the latter are a common occurrence in F-theory constructions \cite{Marsano:2009gv,Marsano:2009wr, Marsano:2011hv, Donagi:2009ra}. The irreducible components of split and non-split spectral covers are $n:1$ coverings 
\be
\CC_k~ \xrightarrow[]{\pi} ~ M_3 
\ee
of $M_3$. We have $n=1$ for all components of split spectral covers and $n\geq2$ for at least one component in the case of non-split spectral covers. The geometry of the adjoint bundle is determined by the flat connection $A$ to
\be
\tn{ad}\,P_{\tn{ADE}}=M_3\times\mathfrak{h}_{\tn{ADE}}\oplus \bigoplus_{\alpha}L_\alpha\,,
\ee
where the sum runs over all roots $\alpha$ of the Lie algebra $\mathfrak{g}_{\tn{ADE}}$ and $L_\alpha$ are line bundles on $M_3$ with connection $\alpha^IA_I\in \Omega^1(M_3)$. When the connection $A$ vanishes the adjoint bundle reduces to the direct product $\tn{ad}\,P_{\tn{ADE}}=M_3\times\mathfrak{g}_{\tn{ADE}}$.
 
We enlarge the space of solutions of the BPS equations \eqref{eq:HiggsHarmonicity} by allowing for source terms. The motivation for particular source terms is taken from the corresponding IIA string theory set-up for gauge algebras $\mathfrak{g}_{\tn{ADE}}=\mathfrak{su}(n)$ which is given by space-time filling D6-branes on $\R^{1,3}\times T^*M_3$ wrapping a special Lagrangian submanifold in $T^*M_3$. In the M-theory reduction to IIA string theory KK-monopoles reduce to D6-branes and the spectral cover is expected to flow to a special Lagrangian submanifold \cite{Thomas:2001vf, Pantev:2009de}. Here sources of codimension 2 and 3 lead to singularities in the Higgs field and D6-branes associated with the corresponding eigenvalues are non-compact. Embedding the local model into a compact geometry these would simply describe D6-branes extending beyond the approximated region. Concretely the BPS equations are altered by sources $j_I$ and $\rho_I$ of co-dimension two or three
\be\label{eq:MagSource}
d\phi_I=*j_I\,, \qquad *\,d*\phi_I=\rho_I\,.
\ee
When these sources are supported on knots this represents the world volume perspective of D6-branes intersecting along the knot which have recombined due to a condensation of the bifundamental chiral superfields localized at their intersection \cite{Erdmenger:2003kn}. In \cite{Pantev:2009de, Braun:2018vhk} sources with $j_I=0$ supported on graphs $\Gamma\subset M_3\subset T^*M_3$ were considered and leveraged to engineer chiral 4d gauge theories. In both set-ups the spectral covers associated to the set-up are split due to the absence of co-dimension one sources. 

Given a Higgs bundle and a Hermitian Lie-algebra valued function $f\in\Omega^0\lb M_3,\tn{ad}\,P_{\text{ADE}}\rb$ a one-parameter family of Higgs bundles is obtained via the deformation
\be\label{eq:DeformationOfHiggs}
\phi\rightarrow \phi_t=\phi+td_Af\,, \qquad t\in \R\,.
\ee
This deforms the operator $\CQ$ of \eqref{eq:ComplexifiedConnection} to $\CQ_t=d_A+[\phi_{t\,}\wedge\,,\cdot\,]$ but leaves the cohomologies \eqref{eq:Cohomologies} and therefore the particle content of the 4d physics unaltered. Indeed we have
\be\ba
\exp\lb -\lbb tf,\cdot\, \rbb \rb \CQ \exp\lb \lbb tf,\cdot\, \rbb \rb =d+\lbb \lb \phi+td_Af+iA\rb\wedge\,,\cdot\, \rbb= \CQ_t\,,
\ea\ee
which gives rise to the isomorphism
\be
H_\CQ^*(M_3,\CS)\cong H_{\CQ_t}^*(M_3,\CS)\,.
\ee
A second kind of deformation is simply given by rescaling the Higgs field 
\be\label{eq:Deformation}
\phi\rightarrow t\phi\,, \qquad t\in \R\,.
\ee
In the sourced set-ups of \eqref{eq:MagSource} this is equivalent to an overall scaling of the source terms. For exact Higgs fields $\phi=d_Af$ these two kinds of deformations agree. In the limit $t\rightarrow \infty$ the perturbative modes localize to the zeros of the Higgs field and the overlap integral \eqref{eq:OverLapIntegralsMass} and \eqref{eq:OverLapIntegralsYukawa} receive their dominant contributions from M2-brane instantons.

\section{Higgs Bundles with Non-split Spectral Covers}
\label{sec:Harmonic}

Higgs fields with non-split spectral covers form the most general class of solutions to the equations \eqref{eq:MagSource}. Their key feature are the branch cut loci of the Higgs field eigenvalues which are given by a collection of knots with a specified monodromy action interchanging the eigenvalues whenever a component of the branch cut locus is circled. A choice of Seifert surfaces for each knot determines a decomposition of the cover into simply connected sheets. The physics of non-split covers can then be understood as that of each such component subject to constraints imposed by the monodromy action. Here we expand on the topology of non-split spectral covers and give simple examples of Higgs fields with such covers for the base manifold $M_3=S^3$. In lower dimension non-split configurations have been discussed in \cite{Gaiotto:2009hg, Xie:2012hs, Wang:2015mra, Wang:2018gvb, Cecotti:2013lda}. The presented analysis extends the approach of \cite{Pantev:2009de, Braun:2018vhk} where split spectral covers were considered.

\subsection{Branch Cuts, Seifert Surfaces and Sources}

An irreducible, non-split spectral cover $\CC$ defined in \eqref{eq:GeneralSC} traced out by a diagonal Higgs field $\phi\in\Omega^1(S^3,\mathfrak{g}_{\tn{ADE}})$ constitutes an $n$-fold covering \eqref{eq:GeneralSC} of $S^3$ away from singularities of the Higgs field. The $n$ eigenvalues $\Lambda_K\in \Omega^1(S^3)$ of the Higgs field are not globally defined, but exhibit a one-dimensional branch locus. These branch loci lie along closed submanifolds of the base $S^3$ and therefore realize a collection of interlinked circles $K_{ik}$ which are embedded into $S^3$ as knots. We collect all linked knots $K_{ik}$, labelled by $i,k$, into a total of $l$ links $L_i$ and the branch locus becomes
\be\label{eq:BranchCutLocus}
\tn{Branch Locus\,:}\qquad L_i= \bigcup_k K_{ik}\,, \qquad K_{ik}\cong S^1\subset S^3\,,\qquad i=1,\dots,l\,.
\ee
The eigenvalues $\Lambda_K$ of the Higgs field $\phi=\tn{diag}(\Lambda_K)$ are only well-defined on a simply connected neighbourhood of the link complement $S^3\setminus \cup_i L_i$ and are acted on by a monodromy action when encircling any component of the branch locus. Equivalently, when encircling the branch locus the Higgs field $\phi$ returns to its original value up to a gauge transformation implementing the action the Weyl group 
\be\label{eq:Monodromy2}
\tn{Monodromy action}\,:\qquad \phi\quad \rightarrow \quad g_i \phi g_i^{-1}\,, \qquad g_i\in G_{\tn{ADE}}\,. 
\ee
We denote by $s_i\in \tn{Weyl}(\mathfrak{g}_{\tn{ADE}})$ the monodromy element associated to components of the links $L_i$. For $\mathfrak{g}_{\tn{ADE}}=\mathfrak{su}{(n)}$ we have for example $\tn{Weyl}(\mathfrak{g}_{\tn{ADE}})=S_n$ where $S_n$ is the symmetric group on $n$ letters.

To every link $L_i$ there exists and orientable two-dimensional surface $F_i$, called the Seifert surface of the link $L_i$  \cite{knots}, bounded by the link
\be\label{eq:SeifertSurfaces}
\del F_i=L_i\,.
\ee
We refer to the two sides of the Seifert surface $F_i$ as its positive $F_i^+$ and negative $F_i^-$ side. Any circle linking the collection of knots $L_i$ intersect its associated Seifert surface $F_i$. The eigenvalues $\Lambda_K$ of the Higgs field are therefore well-defined on $S^3\setminus \cup_i F_i$ above which the sheets of the spectral cover can be distinguished. 

The Higgs field $\phi$ is constrained by the BPS equations and consequently its eigenvalues $\Lambda_K\in \Omega^1(S^3)$ are closed and coclosed on $S^3\setminus \cup_i F_i$. The graphs of these 1-forms in the cotangent space $T^*S^3$ join above the Seifert surfaces $F_i$ to form the spectral cover $\CC\subset T^*S^3$. We refer to the graphs of $\Lambda_K$ as the $K$-th sheet of this cover with respect to a choice of Seifert surfaces $\cup_iF_i$. The BPS-equations descend to each sheet up to surface sources given by a one-form current $j_K$ and a zero-form density $\rho_K$ support on the Seifert surfaces $\cup_iF_i$
\be\label{eq:AmmendedBPS}
d\Lambda_K=*j_K\,, \qquad *\:d*\Lambda_K=\rho_K\,,  \qquad \tn{supp}\,j_K=\tn{supp}\,\rho_K= \bigcup_iF_i\subset S^3\,.
\ee
These are subject to two sets of consistency conditions, the first set of which are between sheets of the cover and read
\be\label{eq:Consistency}
\sum_{K=1}^n \rho_K=\sum_{K=1}^n j_K=0\,, \qquad \Lambda_K\big|_{F_i^+}=\Lambda_L\big|_{F_i^-}\,.
\ee
These require all sources to cancel between sheets and further constrain these to have profiles compatible with gluing the $K$-th sheet to $L$-th sheet along the two sides $F_i^\pm$ of the Seifert surface. In the gluing condition $K,L$ run over pairs such that both indices exhaust all sheets. The second set of conditions are between sources for the same sheet and follow from the compactness of $S^3$. The equations \eqref{eq:AmmendedBPS} can only be solved when the integrated sources $\rho_K$ vanish on each sheet
\be\label{eq:Laplace}
  \int_{S^3}*\rho_K =\sum_{i} \int_{S^3}*\rho_K\big|_{F_i}=0\,.
\ee

In this way the sources \eqref{eq:AmmendedBPS}, which are subject to \eqref{eq:Consistency} and \eqref{eq:Laplace}, determine the boundary conditions for the eigenvalues $\Lambda_K$ when decomposing the cover $\CC$ into sheets. The cancellation of sources between sheets ensures that the Higgs field $\phi$ is harmonic across the Seifert surfaces $F_i$ and traceless. The gluing condition encodes the monodromy action $s_i$ around the Links $L_i$ as each sheet is glued along the two sides $F^\pm_i$ to two other sheets. Equation \eqref{eq:Laplace} is a tadpole cancellation constraint.

\subsection{Example: Unknots, Disks and Surface Charge}
\label{sec:Example}

\begin{figure}
  \centering
  \includegraphics[width=10cm]{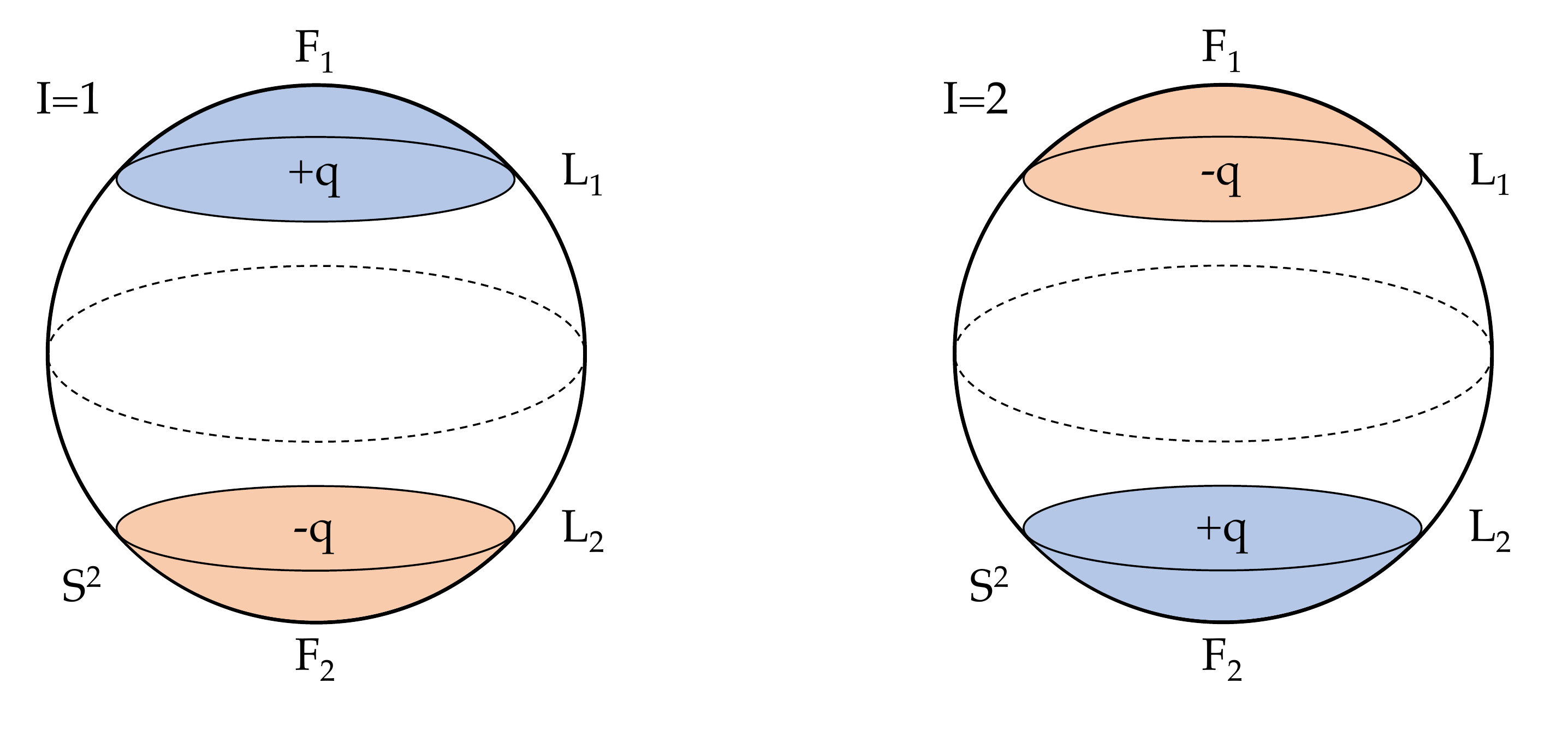}
    \caption{The picture shows a pair of unknots $L_i$ and their Seifert surfaces $F_i\subset S^2\subset S^3$ along with the sources $\rho_I,j_I$ these support with respect to each sheet $I=1,2$. They are supported on the 2-sphere $S^2_{\pi/2}$ which projects onto $\theta=\pi/2$ in \eqref{eq:Fibration}. The sourced Higgs field \eqref{eq:AmmendedBPS} realizes a branched double cover of $S^3$.}
    \label{fig:TheExample}
\end{figure}

We give a simple example of sources $\rho_K,j_K$ satisfying the conditions \eqref{eq:Consistency} and \eqref{eq:Laplace} with $K=1,2$ realizing non-compact, branched, double covers of the 3-sphere with a collection of circles removed. Consider the 3-sphere $S^3$ as a fibration of 2-spheres over an interval which we parametrize by $\theta\in [0,\pi]$
\be\label{eq:Fibration}
S^2\xhookrightarrow{}S^3\rightarrow [0,\pi]\,.
\ee 
At $\theta=0,\pi$ the fibral 2-sphere collapses. The 3-sphere $S^3$ is equipped with the round metric such that the geometry is symmetric under a reflection $\theta\rightarrow \pi-\theta$ fixing the central 2-sphere fiber $S^2_{\pi/2}$ projecting to $\theta=\pi/2$. On this 2-sphere we consider a total of $l$ separated unknots $S_i^1$ each bounding a disk $D_i$
\be\label{eq:SetUpExample}
L_i=S^1_i\subset S^2_{\pi/2}\,, \qquad F_i=D_i\subset S^2_{\pi/2}\,,\qquad i=1,\dots,l\,,
\ee
which function as the links and Seifert surfaces of \eqref{eq:BranchCutLocus} and \eqref{eq:SeifertSurfaces} respectively. 

We now consider source profiles $\rho_K,j_K$ with $K=1,2$ supported on the surfaces $D_i$ realizing non-compact double covers of $S^3$ away from the unknots $S_i^1$. These are constructed electrostatically by setting $j_K=0$ and declaring the disks $D_i$ to be perfect conductors for the electric source $\rho_K$. The eigenvalues $\Lambda_K$ of the Higgs field $\phi=\tn{diag}(\Lambda_1,\Lambda_2)$ are then identified with the electric field of the configurations in each sheet. In the first sheet $K=1$ the disk $D_i$ is assigned the electric charge $q_i$, while in the second sheet $K=2$ it is assigned the opposing charge $-q_i$. The distributed charge must further sum to vanish on each sheet $\sum_i q_i=0$. This manifestly satisfies two of the conditions in \eqref{eq:Consistency} and \eqref{eq:Laplace}. The gluing condition across the surfaces $D_i$ is then satisfied as the source distribution in both sheets is symmetric under reflection about $\theta=\pi/2$. This realizes an irreducible double cover
\be\label{eq:CoverSu(2)}
\CC~\rightarrow~ S^3 \setminus \cup_iL_i\,. 
\ee
We have depicted the set-up in the case of $l=2$ unknots and disks in figure \ref{fig:TheExample}.

The charge distributions $\rho_K$ diverges to the boundary and consequently so do the eigenvalues $\Lambda_K$. In a local normal coordinate system $(z,x)\in\C\times \R$ where one of the unknots is centered at $z=0$ and its associated disk $D_i$ stretches along $\R_- \times \R$, where $\R_-\subset \C$ is the negative real axis, we have
\be\label{eq:AsymptoticsBranchSheet}
\Lambda_K = c_K \lb \frac{dz}{\sqrt{z}}+ \frac{d\bar z}{\sqrt{\bar z}}\rb +\dots\,,
\ee
approaching the unknot with some real constant $c_K$. The omitted terms are regular in the $z\rightarrow 0$ limit and the branch cut of the square root stretches along $\R_-$. These asymptotics follow from the closure and co-closure of the Higgs field away from the branch locus and the discontinuity across the charged Seifert surface.  

The homology groups of the constructed double cover $\CC$ are computed using the Mayer-Vietoris sequence and read
\be
H_0(\CC,\Z)=\Z\,, \qquad H_1(\CC,\Z)=\Z^{2l-1}\,, \qquad  H_2(\CC,\Z)=\Z^{l-1}\,, \qquad H_3(\CC,\Z)=0\,.
\ee
The supersymmetric deformations of the cover are given by altering the charges $q_i$ assigned to each disk $D_i$ with respect to one of the sheets. The constraints \eqref{eq:Consistency} determine the associated opposite deformations on the second sheet. The condition \eqref{eq:Laplace} removes one degree of freedom yielding an $l-1$ dimensional deformation space. 

The zeros of the Higgs field eigenvalues $\Lambda_{1},\Lambda_2$ lie on $S^2_{\pi/2}$. They come in pairs as $\Lambda_{1}+\Lambda_2=0$ and there are a total of $2l-4$ zeros. Each eigenvalue derives from an electrostatic potential $f$ such that $df=\Lambda_1=-\Lambda_2$. For generic charge set-ups the potential $f$ is a Morse function. The zeros of the eigenvalues are critical points of this function and they can be distinguished according to their Morse-index. Let $N_\mu$ be the number of critical points of Morse-index $\mu$, here we have
\be\label{eq:zerosHiggs}
N_1=N_2=l-2\,,\qquad N_0=N_3=0\,,
\ee
The Morse index characterizes topological properties of the Higgs field zeros and determines the matter localized at these.

\subsection{Example: Twisted Connected Sum $G_2$-Manifolds}
\label{sec:TCS}

Non-split spectral covers feature in the local models associated twisted connected sum (TCS) $G_2$-manifolds \cite{MR2024648, MR3109862, Corti:2012kd}. These geometries have been discussed in the physics literature \cite{Halverson:2014tya, Braun:2016igl, Braun:2017uku, Braun:2019wnj} and constitute a landscape of $10^6$ compact $G_2$-manifolds. We give an overview of their construction and discuss the Higgs bundles and spectral covers of their local models.

\begin{figure}
  \centering
  \includegraphics[width=11cm]{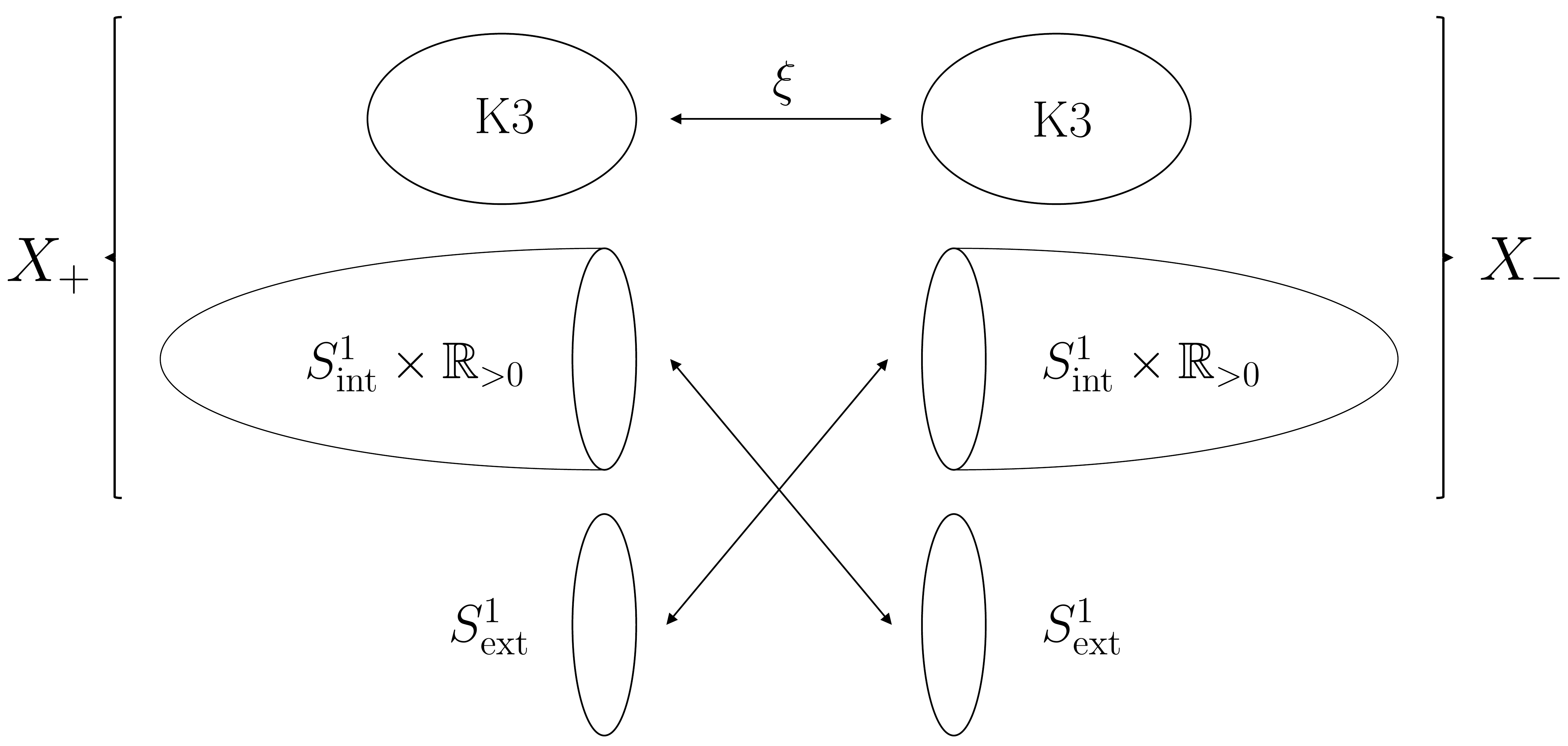}
    \caption{The figure shows a TCS $G_2$-manifold. It is glued from $X_+\times S^1_{\tn{ext}}$ and $X_-\times S^1_{\tn{ext}}$. Internal and external circles are interchanged by the gluing diffeomorphism. The hyperk\"ahler rotation or Donaldson gluing is implemented by $\xi:S^0_+\rightarrow S^0_-$.}\label{fig:TCS}
\end{figure}

TCS $G_2$-manifolds are constructed from a pair of building blocks $Z_{\pm}$ which are in turn constructed using methods from algebraic geometry to satisfy a list of topological constraints \cite[Section 3]{Corti:2012kd}. The building blocks $Z_\pm$ are algebraic 3-folds admitting a K3 fibration
\be
S_{\pm} ~ \xhookrightarrow{} ~ Z_\pm ~ \rightarrow ~ \P^1\,,
\ee
where $S_{\pm}$ is a generic K3 fiber. The fiber class $[S_\pm]=c_1(Z_\pm)$ is required to be given by first Chern class of its building block. From the building blocks $Z_\pm$ one constructs asymptotically cylindrical (aCyl) Calabi-Yau 3-folds $X_\pm=Z_\pm\setminus S_\pm^0 $ by excising a generic fiber $S_\pm^0$. The aCyl Calabi-Yau 3-folds $X_\pm$ are therefore fibered as
\be
S_{\pm} ~ \xhookrightarrow{} ~ X_\pm ~ \rightarrow ~ \C\,.
\ee
The boundaries of $X_\pm$ are given by $S^1_{\tn{int}}\times S^0_\pm$. Approaching the boundaries the Calabi-Yau structure $(\omega_\pm,\Omega_\pm)$ of $X_\pm$ asymptotes to that of the cylinder $\R_{>0}\times S^1_{\tn{int}}\times S_\pm^0$. The latter is given by $(\omega_{\infty,\pm},\Omega_{\infty,\pm})$ where $\omega_{\infty,\pm}=dt\wedge d\theta+\omega_{S,\pm}$ and $\Omega_{\infty,\pm}=\lb d\theta -idt\rb \wedge \Omega_{S,\pm}$. Here $(t,\theta)$ are coordinates on the cylinder $\R_{>0}\times S^1$ and $\omega_{S,\pm},\Omega_{S,\pm}$ are the Calabi-Yau structure of the K3 surface $S^0_\pm$. Alternatively, we can express the data of the K3 fiber $S^0_\pm$ using its Hyperk\"ahler triplet $(\omega_{1,\pm},\omega_{2,\pm},\omega_{3,\pm})$ which gives $\omega_{S,\pm}=\omega_{1,\pm}$ and $\Omega_{S,\pm}=\omega_{2,\pm}+i\omega_{3,\pm}$.

Each aCyl Calabi-Yau $X_\pm$ is extended to a 7-manifold $X_\pm\times S^1_{\tn{ext}}$ whose boundary is given by $S^1_{\tn{int}}\times S^1_{\tn{ext}}\times S^0_\pm$ by trivially adding an external circle. A compact 7-manifold is now constructed by gluing this pair along their boundaries. The gluing diffeomorphism interchanges the external and internal circles and identifies the K3 surfaces $S^0_\pm$ via the map $\xi:S^0_+\rightarrow S^0_-$. The diffeomorphism $\xi$ is referred to as a hyper-K\"ahler rotation, often called a Donaldson matching, and acts on the hyper-K\"ahler triplets as
\be\label{eq:DonaldsonMatching}
\xi^*\lb \omega_{1,-}\rb =\omega_{2,+}\,, \qquad \xi^*\lb \omega_{2,-}\rb=\omega_{1,+}\,, \qquad \xi^*\lb \omega_{3,-}\rb =-\omega_{3,+}\,.
\ee
The base and fibers are glued separately and thereby the resulting smooth, compact 7-manifold is K3-fibered over a 3-sphere. Furthermore, it admits a metric with $G_2$ holonomy \cite{MR2024648, Corti:2012kd}. We sketch the gluing construction in figure \ref{fig:TCS}.

Local models for TCS $G_2$-manifolds are now obtained by replacing the K3 fibers with ALE fibers, i.e. they are given by the fibrations \eqref{eq:ALE101} with $M_3=S^3$. The spectral cover of this local model results from gluing the spectral covers associated with $X_\pm\times S^1_{\tn{ext}}$ and its topology is fixed by the Donaldson matching \eqref{eq:DonaldsonMatching}. To expand on this consider the restriction map
\be
\rho_\pm:H^2(Z_\pm,\Z)\rightarrow H^2(S^0_\pm,\Z)\cong (-E_8^{\oplus 2})\oplus U^{\oplus 3}\equiv L,
\ee
where $L$ is the K3 lattice of signature $(3,19)$. The Donaldson matching gives an isomorphism $H^2(S^0_+,\Z)\cong H^2(S^0_-,\Z)$ and the 2-forms in the intersection 
\be\label{eq:Extendable2cycles}
\mathfrak{g}=\tn{Im}\,\rho_+ \cap \tn{Im}\,\rho_-\,,
\ee 
are dual to 2-cycles in the fibers which sweep out 5-cycles across the base 3-sphere. These are in turn dual to $\tn{rank}(\mathfrak{g})$ independent harmonic 2-forms on the TCS $G_2$-manifold. In a KK-reduction of the supergravity 3-form this give rise to $\tn{rank}(\mathfrak{g})$ abelian gauge fields in 4d. To understand the relation to the spectral cover of the local model we begin with the spectral covers of $X_\pm\times S^1_{\tn{ext}}$. These are constructed by replacing the K3 fibers with ALE fibers and collecting the Hyperk\"ahler periods of the 2-cycles $\sigma_I\in H_2(S^0_\pm,\Z)$ as in \eqref{eq:HiggsComponents}. The Donaldson matching then prescribes the gluing of the Higgs fields $\phi^\pm$ as
\be\ba\label{eq:SpecCovGlue}
\phi_{I,1}^-&=\int_{\sigma_I}\omega_{1,-}=\int_{\sigma_I}\omega_{2,+}=\phi_{I,2}^+\,,\\
\phi_{I,2}^-&=\int_{\sigma_I}\omega_{2,-}=\int_{\sigma_I}\omega_{1,+}=\phi_{I,1}^+\,,\\
\phi_{I,3}^-&=\int_{\sigma_I}\omega_{3,-}=-\int_{\sigma_I}\omega_{3,+}=\phi_{I,3}^+\,.\\
\ea\ee
The gluing requires the ALE fibers on both sides to be of the same type. The resulting spectral cover $\CC=\cup_{k=1}^N\CC_k$ is traced out by the glued forms \eqref{eq:SpecCovGlue}. The difference $\CC_k-\CC_l$ of two irreducible components lifts to a 5-cycle in the ALE-fibration of which $N-1$ are independent and again give dual 2-forms which determine the number of 4d gauge fields in a KK-reduction of the supergravity 3-form on the ALE-fibration. We therefore find the number $N$ of irreducible components of the spectral cover $\CC$ to be given by
\be
N=\tn{rank}\,(\mathfrak{g})+1\,.
\ee

We now touch on the discussion to singular ALE-fibrations for which some of the 2-cycles $\sigma_I$ are collapsed throughout the base 3-sphere. These were argued in \cite{Braun:2017csz, Braun:2017uku} to describe the local geometry of 7-folds constructed from singular aCyl Calabi-Yau 3-folds $X_\pm$. The 3-fold $X_\pm$ have singular K3 fibers $S_\pm$ with a generic ADE singularity and $X_\pm\times S^1_{\tn{ext}}$ are expected to glue to singular, compact TCS $G_2$-manifolds. The details of the singular limit for $X_\pm$ are discussed for different ADE singularities in \cite{Lerche:1996an, Billo:1998yr}. The associated spectral covers have Higgs fields where some of the Cartan components $\phi_I$ vanish identically, i.e. the singularities specifies the number of zero sections contributing sheets to the cover $\CC$. 

The landscape of TCS $G_2$-manifolds realize via their local models a large class of examples of split and non-split spectral covers over a base 3-sphere. The source loci of \eqref{eq:AmmendedBPS} are left implicit in these constructions, this was already found to be the case for split spectral covers discussed in \cite{Braun:2018vhk}.  

\subsection{Topology of Cyclically Branched Covers and Recombined Higgs Fields}
\label{sec:GeneralizedNonSplit}

\begin{table}
\renewcommand{\arraystretch}{1.2}
\begin{center}
 \begin{tabular}{|| c | c | c |  c |  c | c ||}
 \hline
Knot Name & Sketch & 2-fold $\CC$ & 3-fold $\CC$ & 4-fold $\CC$ & 5-fold $\CC$ \\ [0.5ex] 
 \hline\hline
\parbox[b][1.9cm][c]{0.5cm}{ $0_1$ } & \includegraphics[width=2cm]{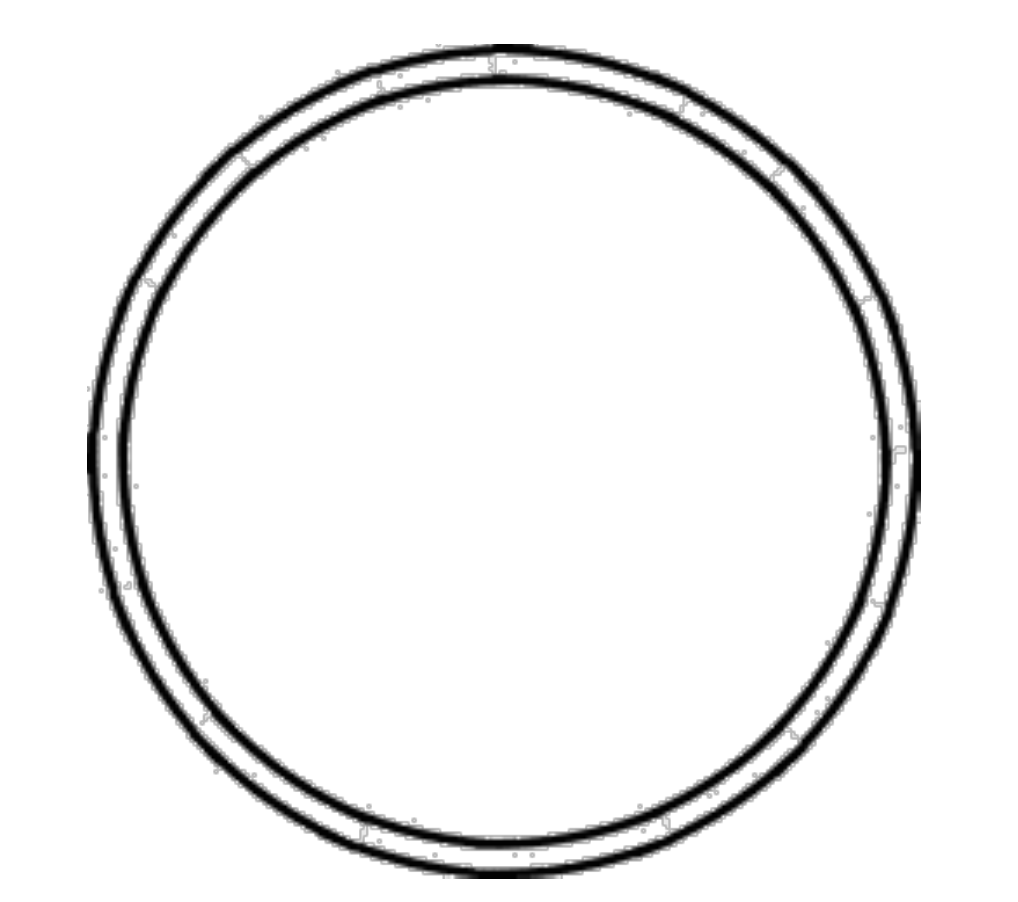} & \parbox[b][2cm][c]{1.5cm}{ \centering $1$}  &  \parbox[b][2cm][c]{1.5cm}{ \centering $1$}  &  \parbox[b][2cm][c]{1.5cm}{ \centering $1$} &  \parbox[b][2cm][c]{1.5cm}{ \centering $1$} \\ 
 \hline
\parbox[b][2cm][c]{0.5cm}{ $3_1$ } & \includegraphics[width=2cm]{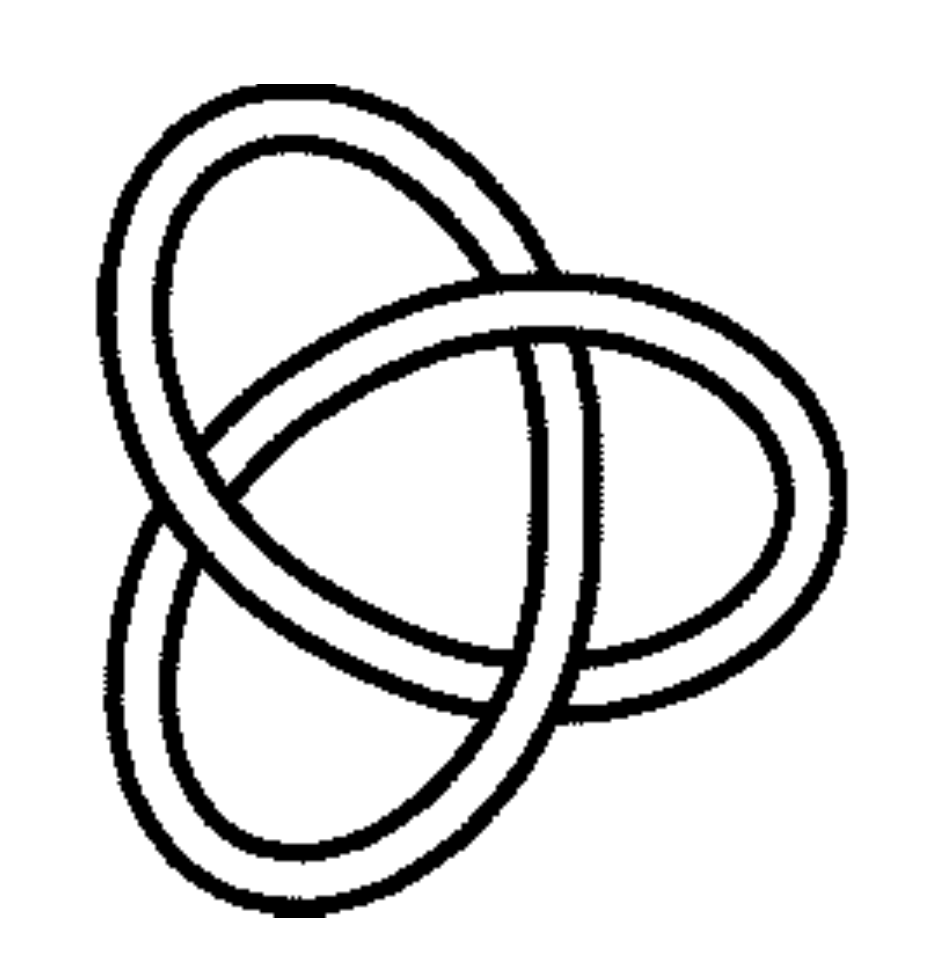} & \parbox[b][2cm][c]{0.5cm}{ $\Z_3$}  & \parbox[b][2cm][c]{1.5cm}{\centering $\Z_{2}\times \Z_2 $}  & \parbox[b][2cm][c]{0.5cm}{$\Z_3$} &  \parbox[b][2cm][c]{1.5cm}{ \centering $1$} \\
 \hline
\parbox[b][2cm][c]{0.5cm}{ $4_1$ } & \includegraphics[width=2cm]{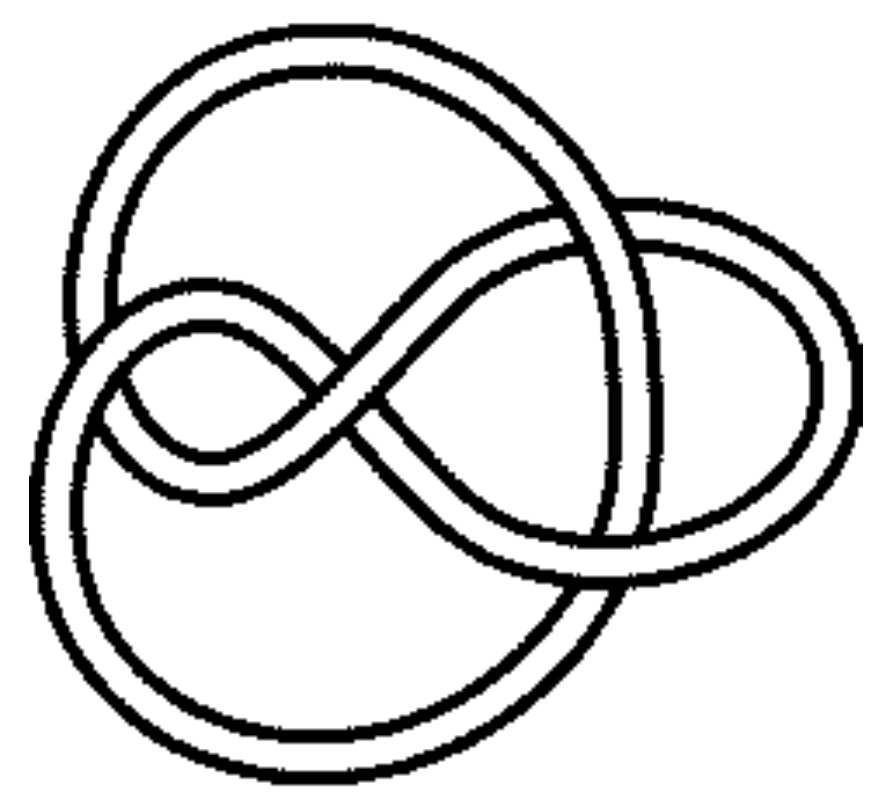} & \parbox[b][2cm][c]{1.5cm}{ \centering $\Z_5$} & \parbox[b][2cm][c]{1.5cm}{ \centering $\Z_4\times \Z_4 $}& \parbox[b][2cm][c]{1.5cm}{\centering $\Z_3\times \Z_{15} $}&  \parbox[b][2cm][c]{1.5cm}{ \centering $\Z_{11}\times \Z_{11}$}\\
 \hline
\parbox[b][1.9cm][c]{0.5cm}{ $5_1$ } & \includegraphics[width=2cm]{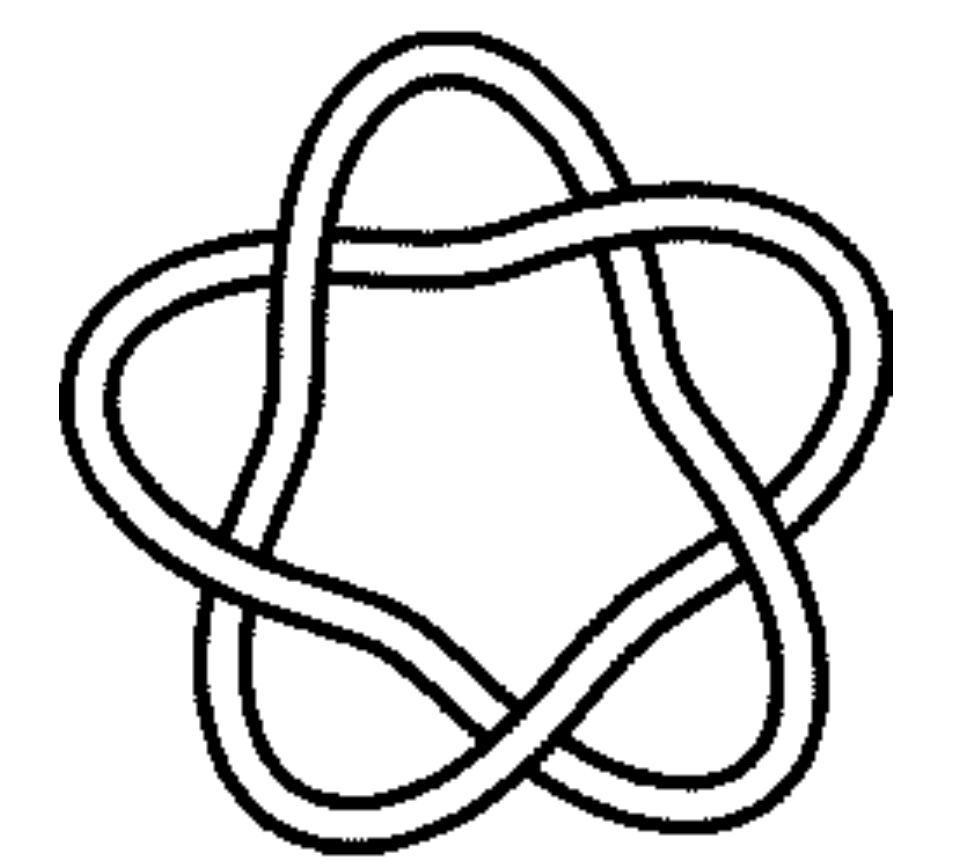} & \parbox[b][1.9cm][c]{1.5cm}{ \centering $\Z_5$} & \parbox[b][1.9cm][c]{1.5cm}{ \centering $1$}  & \parbox[b][1.9cm][c]{1.5cm}{ \centering $ \Z_5$} &  \parbox[b][2cm][c]{3.5cm}{ \centering $\Z_2\times\Z_2\times\Z_2\times\Z_2$}  \\
 \hline
\end{tabular}
\end{center}
\caption{We table examples of knots. Each column list the torsion component of the first homology of the $n$-fold covering space $\CC \rightarrow S^3$ of the knot \cite{knots}. The torsion numbers are tabled in \cite{KnotTable}. The pictures are taken from \cite{KnotAtlas}.}
\label{tab:ExampleKnots}
\end{table}

The data of the 4d theory engineered by a geometry with a split or non-split spectral cover can be extracted from a particle probing $S^3$ with a potential set by the Higgs field. Of interest here is in part the topology of the spectral cover, as we explain in section \ref{sec:Split} and \ref{sec:NonSplit}. In this section we discuss the topology of a simple class of non-split spectral covers. For concreteness we consider spectral covers associated with the Lie algebra $\mathfrak{g}_{\tn{ADE}}=\mathfrak{su}(n)$.

We focus on irreducible spectral covers with a single component, more general covers are given by unions of these irreducible covers. Further we restrict to set-ups for which the monodromy elements $s_i=s\in S_n$ are identical for all components of the branch locus and of order $n$ for $n$-sheeted coverings. In this setting the topology in the vicinity of branch link $L_i$ is that of the branched multi-covering studied in knot theory \cite{knots}, from which we excise the links $L_i$ along which the Higgs field diverges. We refer to these covers as irreducible, cyclically branched $n$-sheeted coverings. The example of section \ref{sec:Example} realizes such a cover for $n=2$ and $\mathfrak{g}_{\tn{ADE}}=\mathfrak{su}(2)$ with $L_i=S^1$ and $s=-1\in S_2$.

We start with a solution to \eqref{eq:AmmendedBPS} for eigenvalue 1-forms $\Lambda_K$ where $K=1,\dots, n$. The eigenvalues $\Lambda_K$ sweep out a $n:1$ cover $\CC\rightarrow S^3\setminus \cup_iL_i$ away from the branch locus and picking Seifert surfaces for each link $(L_i,F_i)$ the spectral cover $\CC$ can be written as
\be\label{eq:CC}
\CC=\tilde\CC \setminus  \cup_{\substack{i}} L_i \,,
\ee
where the covering $\tilde \CC$ is glued from $n$ copies of the base with the Seifert surfaces removed 
\be\label{eq:GluedCoverSec3}
\tilde\CC=\lb S^3\setminus \cup_i F_i  \rb_1 \# \dots \# \lb S^3\setminus \cup_i F_i \rb_n\,.
\ee
The cut out $S^3\setminus \cup_i F_i $ contains two copies of the Seifert surfaces $F^\pm_i$ corresponding to its positive and negative sides which intersect along the links $L_i$. The gluing in \eqref{eq:GluedCoverSec3} is performed by identifying $F_i^+$ in the $i$-th gluing factor with $F_i^-$ in the $(i+1)$-th factor and finally gluing $F_i^+$ in the $n$-th gluing factor to $F_i^-$ in the first. Each gluing factor is in correspondence with a sheet of the spectral cover. For further details we refer to \cite{knots,Cecotti:2011iy}.

The homology groups of the cover \eqref{eq:GluedCoverSec3} are computed by an application of the Mayer-Vietoris sequence to a decomposition of the cover $\tilde\CC$ into patches whose projection to the base contain at most a single Seifert surface $F_i$. The homology groups of the spectral cover \eqref{eq:CC} are then computed by another application of the Mayer-Vietoris sequence to the covering $\tilde\CC=\CC\cup T$ where $T$ is tubular neighbourhood of the links $\cup_iL_i\subset T$. We restrict to the case in which the links $L_i=K_i$ are simply knots and $T$ thus becomes a collection of $l$ solid tori. We give further details in appendix \ref{app:Homology}. For an $n$-sheeted cover with $l$ knots $K_i$ the result reads 
\be\label{eq:Homologies}
H_1(\CC,\Z)=\Z^{(n-1)(l-1)+l}\oplus  \bigoplus_{i=1}^l H_1^{(n)}(K_i)\,, \qquad H_2(\CC,\Z)=\Z^{(n-1)(l-1)}\,,
\ee
together with $H_0(\CC,\Z)=\Z$ and $H_3(\CC,\Z)=0$. Each knot contributes a torsion factor to the first homolog group while the number of links and sheets determines the free factor in \eqref{eq:Homologies}. In table \ref{tab:ExampleKnots} we list the group $ H_1^{(n)}(K_i)$ for some low component coverings, a substantially more extensive list of examples is given in \cite{KnotTable}.

The cover \eqref{eq:CC} inherits a natural metric from its gluing factors. The eigenvalues $\Lambda_K$ of the Higgs field then combine to a harmonic 1-form on the spectral cover
\be\label{eq:EffectiveOneForm}
\Lambda\in \Omega^1(\CC)\,, \qquad \Lambda\big|_{\lbb S^3\setminus \cup_iF_i \rbb_K}=\Lambda_K\,, \qquad K=1,\dots,n\,,
\ee
which by constructions restricts on each gluing factor to one of the local 1-form eigenvalues $\Lambda_K$ of the Higgs field. Supersymmetric deformations of the cover $\CC$ are now described by harmonic perturbations $\Lambda\rightarrow \Lambda+\delta\Lambda$ or equivalently $n$ harmonic perturbations $\Lambda_K\rightarrow \Lambda_K+\delta\Lambda_K$ which glue consistently across the branch sheets $\cup_iF_i$.

Finally note that we can equip the cotangent bundle $T^*S^3$ with an auxiliary Calabi-Yau structure whose symplectic 2-form $\omega$ and holomorphic 3-form $\Omega$ are given by
\be
\omega=\frac{i}{2}\sum_{i=1}^3 dz^i\wedge d\bar z^i\,, \qquad \Omega=dz_1 \wedge dz_2 \wedge dz_3\,,
\ee
where $dz^i=dx^i+idy^i$ with $x^i, y^i$ being local coordinates on $S^3,T^*_xS^3$ respectively. With respect to this auxiliary Calabi-Yau geometry the spectral cover $\CC$ is an immersed, non-compact Lagrangian submanifold, which follows from $\omega|_{\CC}=d\phi=0$.

\section{Colored SQMs probing Higgs Bundles}
\label{sec:GeneralHiggsBundleSQM}

Given a vacuum of the 7d SYM in terms of a complex flat connection \eqref{eq:ComplexifiedConnection} the massless modes in 4d are determined by the mass matrix \eqref{eq:OverLapIntegralsMass} and their interactions are set by the Yukawa integral \eqref{eq:OverLapIntegralsYukawa}. These overlap integrals can be interpreted as amplitudes of a colored $\CN=2$ supersymmetric quantum mechanics. The relevant structures of the SQM for this identification are its physical Hilbert space  $\CH_{\tn{phys.}}$ and supercharge $\CQ$ which are given by
\be\label{eq:HilbertSpaceSQM}
\CH_{\tn{phys.}}=\Lambda\lb M_3,\tn{ad}\,P_{\text{ADE}}\rb\,, \qquad \CQ=d+[\lb\phi+iA\rb\wedge\,,\,\cdot\,].
\ee
Here we present this new $\CN=2=(1,1)$ supersymmetric quantum mechanics. In \cite{ALVAREZGAUME1984269, Rietdijk:1992jp} similar quantum mechanical systems with less supersymmetry have been considered. We refer to the SQM as `colored' due to the presence of additional fermions over the SQM considered in \cite{Witten:1982im} which extend the Hilbert space by color degrees of freedom associated with the Lie algebra $\mathfrak{g}_{\tn{ADE}}$. The colored SQM is constructed working backwards from \eqref{eq:HilbertSpaceSQM}.
 
\subsection{Set-up and Conventions}

\label{sec:SetUp}
We consider the manifold $M_3$ with metric $g$ and a principal bundle $P_{\tn{ADE}}\rightarrow M_3$ with gauge group $G_{\tn{ADE}}$ over it. The corresponding Lie algebra is denoted $\mathfrak{g}_{\tn{ADE}}$\,. This gives rise to the associated adjoint vector bundle $\tn{ad}\,P_{\tn{ADE}}\rightarrow M_3$\,. Both are naturally complexified. Greek indices run as $\alpha,\beta,\gamma=1,\dots,\dim G_{\tn{ADE}}$ and are associated to the fiber while latin indices run as $i,j,k=1,2,3$ and are associated to the base. The Killing form $\kappa_{\alpha\beta}$ gives rise to a non-degenerate pairing on the fibers of $\tn{ad}\,P_{\tn{ADE}}\rightarrow M_3$ which is used to raise and lower greek indices. Latin indices are raised and lowered with the metric $g_{ij}$\,. The generators of the Lie algebra $\mathfrak{g}_{\tn{ADE}}$ are denoted by $T_\alpha$ and are taken to satisfy
\be 
[T_\alpha,T_\beta]=ic_{\alpha\beta\gamma}T^{\gamma} \,.
\ee
We probe the geometry $\tn{ad}\,P_{\tn{ADE}}\rightarrow M_3$ with a non-linear supersymmetric sigma model. We denote the flat worldline by $\R_\tau$ and take $\tau$ to denote the time coordinate on it. The bosonic and fermionic fields are given by the maps $x:\R_\tau\rightarrow M_3$ and sections $\psi:\R_\tau\rightarrow x^{*}(TM_3)$ respectively. Further we add a color field given by sections $\lambda:\R_\tau\rightarrow x^{*}(\tn{ad}\,P_{\tn{ADE}})$\,. The dynamics of the model are governed by a non-dynamical background connection $A\in\Omega^1(M_3,\tn{ad}\,P_{\tn{ADE}})$ and Higgs field $\phi\in\Omega^1(M_3,\tn{ad}\,P_{\tn{ADE}})$ on the target manifold $M_3$\,. These are real Lie algebra valued 1-forms on the target manifold $M_3$\,. The connection $A_{i\alpha}$ and Higgs field $\phi_{i\alpha}$ are required to satisfy the BPS equations \eqref{eq:BPS}\,.

The sigma model can thus be summarized as\medskip
\be\label{eq:SQMSetUp}
\begin{tikzcd}[row sep=large]
x^*(TM_3\oplus \tn{ad}\,P_{\tn{ADE}})\otimes\C \arrow[d,swap,shift left=1,"{\psi,\,\lambda~~}"]  & (TM_3\oplus \tn{ad}\,P_{\tn{ADE}})\otimes\C \arrow[l,"x^*"]   \arrow[d,swap,shift right=1,"\pi\,"] \\
\R_{\tau}\arrow[r,"x"] \arrow[u,swap,shift left=1,"{~~\pi_\tau}"] & M_3 \arrow[u,swap,shift right=1,"{\,A,\,\phi}"]
\end{tikzcd}\medskip
\ee 
where $\pi,\pi_\tau$ denote the canonical projections. Expanded in components the fields $\psi,\lambda$ read
\be\label{eq:FermionsOfSQM}
\psi(\tau)=\psi^i(\tau)\frac{\del}{\del x^i}\bigg|_{x(\tau)}\,,\qquad \lambda(\tau)=\lambda^\alpha(\tau) e_\alpha\big|_{x(\tau)}\,,
\ee
where $e_\alpha$ are fiber coordinates induced by a local trivialisation of $\tn{ad}\,P_{\tn{ADE}}$\,. Both $\psi,\bar\psi$ and $\lambda,\bar\lambda$ are taken to be anti-commuting fermionic fields. The latter we package into bilinears
\be\label{eq:FermionGenerator}
\tilde T=-\lbb \bar\lambda, \lambda\rbb=\tilde T^\alpha e_\alpha=-ic^\alpha_{~\,\beta\gamma}\bar\lambda^\beta \lambda^\gamma e_\alpha \,, \qquad \tilde T_\alpha^\dagger =\tilde T_\alpha\,,
\ee
which we pair with the connection $A_{i\alpha}$ and Higgs field $\phi_{i\alpha}$ to form the color contracted 1-forms 
\be\ba\label{eq:BilinearBackground}
A_\lambda&=\lb A_\lambda\rb_{i}dx^i= A_{i}^\alpha \tilde T_\alpha dx^i=\kappa\lb\bar\lambda,[A_i,\lambda]\rb dx^i \,,\\
\phi_\lambda&=\lb \phi_\lambda\rb_{i}dx^i=\phi_{i}^\alpha \tilde T_\alpha dx^i=\kappa\lb\bar\lambda,[\phi_i,\lambda]\rb dx^i\,. 
\ea\ee
The bilinears $\tilde T$ quantize to the Lie algebra generators $T$\,. To remind of this contraction we introduce a subscript $\lambda$ as in \eqref{eq:BilinearBackground}\,.

We combine the connection $A_{i\alpha}$ and Higgs field $\phi_{i\alpha}$ into a complex Lie algebra valued 1-form $\varphi$ with components
\be
\varphi_{i\alpha}=\phi_{i\alpha}+iA_{i\alpha}\,.
\ee
There are now three connections on $M_3$ given by the natural connection $D$ on $\tn{ad}\,P_{\tn{ADE}}$ and its complexification $\CQ$ which read
\be\label{eq:ConnectionsSQM}
D=d+i[A\wedge\,,\cdot\,]\,, \qquad \CQ=d+[\varphi\wedge\,,\cdot\,]\,,
\ee
together with the Levi-Civita connection $\nabla$ of the metric $g_{ij}$\,. Each of these pulls back to the world line $\R_{\tau}$ in \eqref{eq:SQMSetUp} and acts on the fermions $\psi,\bar\psi,\lambda,\bar\lambda$ of \eqref{eq:FermionsOfSQM} as
\be\ba\label{eq:pullbackconnections}
\nabla_\tau \psi^i&=\partial_{\tau} \psi^{i}+\Gamma_{jk}^{i} \dot x^{j}\psi^k \,, \\
D_\tau \lambda_\alpha&=\del_\tau \lambda_\alpha +c_{\alpha\beta\gamma}\dot{x}^iA_i^\beta\lambda^\gamma\,,\\
\CQ_\tau \lambda_\alpha&=\del_\tau \lambda_\alpha -ic_{\alpha\beta\gamma}\dot{x}^i\varphi_i^\beta\lambda^\gamma\,.
\ea\ee 
The pullback is referenced by adding the world line parameter $\tau$ as an index to the respective connections.

\subsection{Colored $\CN=(1,1)$ Supersymmetric QM}
\label{sec:SQM}

The dynamics of the sigma model described in section \ref{sec:SetUp} is governed by the Lagrangian
\be\ba\label{eq:SQMLagrangian}
\CL&= \frac{1}{2}\dot{x}^i\dot{x}_i+i\bar{\psi}^i\nabla_\tau \psi_i+i\bar{\lambda}^\alpha D_\tau\lambda_\alpha+\frac{i}{2}\lb F_{ij}\rb_\lambda \bar{\psi}^i\psi^j-\frac{1}{2} R_{ijkl} \psi^{i} \bar{\psi}^{j} \psi^{k} \bar{\psi}^{l} \\
&~~~\,-\lb D_{(i}\phi_{j)}\rb_\lambda \bar{\psi}^i\psi^j-\frac{1}{2} \phi_\lambda^i \phi_{\lambda,i}-\frac{1}{2}[\phi_i,\phi_j]_\lambda\bar{\psi}^i\psi^j+\zeta \lb \bar{\lambda}^\alpha\lambda_\alpha-n\rb\,.
\ea\ee
Here $R_{ijkl}$ denotes the Riemann curvature tensor, the bracket notation $D_{(i}\phi_{j)}$ denotes a symmetrisation of indices, the integer $n$ is set to $n=1$ and $\zeta$ is a Lagrange multiplier. The action \eqref{eq:SQMLagrangian} is invariant under
\be\ba\label{eq:SQMVariations}
\delta x^i&=\epsilon\bar{\psi}^i-\bar{\epsilon}\psi^i\,, \\
\delta\psi^i&=i\epsilon\dot{x}^i+\epsilon \phi_\lambda^i-\epsilon\Gamma_{j k}^{i} \bar{\psi}^{j} \psi^{k}\,,\\
\delta\bar{\psi}^i&=-i\bar{\epsilon}\dot{x}^i+\bar\epsilon \phi_\lambda^i-\bar{\epsilon}\Gamma_{jk}^{i} \bar{\psi}^{j} \psi^{k}\,,\\
\delta\lambda^\alpha&=-i\epsilon c^{\alpha}_{\,~\beta\gamma} \bar{\psi}^i \varphi_i^\beta\lambda^\gamma-i\bar{\epsilon} c^{\alpha}_{\,~\beta\gamma} \psi^i   \bar\varphi_i^\beta \lambda^\gamma \,, \\
\delta\bar{\lambda}^\alpha&= -i\epsilon c^{\alpha}_{\,~\beta\gamma}\bar{\psi}^i  \varphi_i^\beta \bar{\lambda}^\gamma -i\bar{\epsilon } c^{\alpha}_{\,~\beta\gamma}\psi^i \bar\varphi_i^\beta\bar{\lambda}^\gamma\,.
\ea\ee 
The supercharges associated to the variations \eqref{eq:SQMVariations} are given by
\be\ba\label{eq:SQMSuperscharges}
\CQ=\bar{\psi}^i\lb i\dot{x}_i+\phi_\lambda^i\rb \,, \qquad \CQ^\dagger=\psi^i\lb -i\dot{x}_i+\phi_\lambda^i \rb \,.
\ea\ee
There is no R-symmetry rotating the supercharges. We check the supersymmetric variations \eqref{eq:SQMVariations} and provide a derivation of \eqref{eq:SQMSuperscharges} in Appendix \ref{app:SUSY}.

The physics of the quantum mechanics \eqref{eq:SQMLagrangian} is that of a particle moving in the target space $M_3$\,. In addition to its position, its state is characterized by its fermion and color content which are given by vectors in the pullback of the exterior algebra $\Lambda M_3$ and adjoint bundle $\tn{ad}\,P_{\tn{ADE}}$ to the world line respectively. The latter are the fermions $\lambda,\bar{\lambda}$ and determine the color contracted Higgs field $\phi_\lambda$ setting the potential for the particle via \eqref{eq:BilinearBackground}.

Quantization of the SQM \eqref{eq:SQMLagrangian} leads to the physical Hilbert space
\be\label{eq:Hilberspace}
\CH_{\tn{phys.}}=\Lambda\lb M_3,\tn{ad}\,P_{\text{ADE}}\rb\,,
\ee
consisting of Lie algebra valued forms on $M_3$. The Lagrange multiplier in \eqref{eq:SQMLagrangian} gives rise to the constraint that only states with a single $\bar\lambda$ excitations are considered physical which precludes states in higher powers of the adjoint representation of $\mathfrak{g}_{\tn{ADE}}$ from contributing to the spectrum. States of even, odd degrees are bosonic, fermionic respectively. The supercharge is realized on $\CH_{\tn{phys.}}$ as the operator 
\be\label{eq:SuperchargeSQM}
\CQ=d+[\lb\phi+iA\rb\wedge\,,\,\cdot\,]\,.
\ee
We give further detail on the quantisation procedure in appendix \ref{sec:canonicalquant}.

\subsection{Perturbative Ground States and Instantons}
\label{sec:PerturbativeGroundStates}

Perturbative ground states of the quantized SQM are given by Lie-algebra valued forms $\chi\in\Omega^p\lb M_3,\tn{ad}\,P_{\text{ADE}}\rb$ annihilated by the Hamiltonian $H=\frac{1}{2}\lbbb \CQ,\CQ^\dagger\rbbb$ or equivalently by the two supercharges $\CQ,\CQ^\dagger$ to all orders in perturbation theory
\be\label{eq:PBFSCharacterization}
H\chi=0\quad \leftrightarrow \quad \CQ\chi=0\,, ~~ \CQ^\dagger\chi=0\,.
\ee
In the path integral formulation of the SQM perturbative ground states correspond to constant maps fixed by the Euclidean fermionic supersymmetry variations $\delta^E \psi^i,\delta^E\bar\psi^i$ which emphasizes the second condition given in \eqref{eq:PBFSCharacterization}. We give the Euclidean versions of the Lagrangian \eqref{eq:SQMLagrangian} and variations \eqref{eq:SQMVariations} together with the Hamiltonian of the SQM in appendix \ref{sec:HamEu}. A characterization of the perturbative ground states already follows from inspection of the unquantized supercharges \eqref{eq:SQMSuperscharges}, constant maps annihilated by the supercharges necessarily map to points at which the Higgs field $\phi_\lambda$ vanishes. We conclude that perturbative ground states are labelled by pairs
\be\ba\label{eq:PGSSQM}
(x_A,\lambda_A)\in M_3\times\mathfrak{g}_{\tn{ADE}}\,,
\ea\ee
which are such that the color contracted Higgs field at $x_A$ with respect to $\lambda_A$ vanishes
\be\label{eq:EffHiggs}
\phi_{\lambda_A}(x_A)=\kappa\lb \bar\lambda_A,[\phi_i(x_A) ,\lambda_A]\rb dx^i=0\,.
\ee
Here we have introduced capital latin indices $A,B,C$ which label pairs in $M_3\times \mathfrak{g}_{\tn{ADE}}$\,. Further we assume that $\phi_{\lambda_A}$ has simple isolated zeros or equivalently that it is a Morse 1-form. 

To fully determine a perturbative ground state \eqref{eq:PGSSQM} we further need to specify its $\psi,\bar\psi$ fermion content. This however is already fixed by a given pair $(x_A,\lambda_A)$ by considering how the 1-form $\phi_{\lambda_A}\in\Omega^1(M_3)$ vanishes at $x_A\in M_3$\,. Consider a small sphere $S^2_\epsilon \subset M_3$ on which the color contracted Higgs field $\phi_{\lambda_A}$ does not vanish and which encloses the point $x_A\in M_3$\,. Then we have a map of spheres
\be
\frac{\phi_{\lambda_A}}{||\phi_{\lambda_A}||}:S^2_\epsilon \rightarrow S^2\,.
\ee
The degree $\mu(x_A,\lambda_A)$ of this map topological characterizes the vanishing of the 1-form $\phi_{\lambda_A}$ at $x_A\in M_3$\,. The number of $\bar\psi$ excitations of the perturbative ground state, or equivalently its degree $p$ as a differential form, is given by $p=\mu(x_A,\lambda_A)$\,. This generlizes the notion of Morse index as introduced in \cite{Witten:1982im} and explained in \cite{Braun:2018vhk}. The pairs \eqref{eq:PGSSQM} thus fully label perturbative ground states\footnote{Here we have discussed generic localized perturbative ground states. To a given Higgs field background $\phi$ there also exist color vectors $\lambda$ such that the color contracted Higgs field $\phi_\lambda\equiv 0$ vanishes identically. We say that these color vectors and associated ground states of $\CQ,\CQ^\dagger$ live in the bulk. Whenever $\phi_\lambda\neq 0$ we refer to the color vectors and their associated perturbative ground states as localized. Generically the local 1-form $\phi_\lambda$ will have isolated simple zeros, this is the case discussed here. We do not discuss higher dimensional zero loci of the color contracted Higgs field $\phi_\lambda$.}. In Dirac notation we denote these by
\be\ba\label{eq:PG}
\chi_A=\ket{x_A,\lambda_A,\mu_A}\in \Omega^{\,\mu_A}(M_3,\tn{ad}\,P_{\tn{ADE}}) \,.
\ea\ee

Given two perturbative ground states $\chi_A,\chi_B$ we construct a third perturbative ground state $\chi_{AB}=\lbb\chi_{A\,}\wedge\,,\chi_B\rbb$ as, if $\chi_A,\chi_B$ are annihilated to all orders in perturbation theory by $\CQ$, then so is $\chi_{AB}$ by
\be\ba\label{eq:GenerateExtraGroundState}
\CQ\lbb\chi_{A\,}\wedge\,,\chi_B\rbb&=[\CQ\chi_A\wedge\,,\chi_B]+(-1)^{\mu_A}[\chi_A\wedge\,,\CQ\chi_B]\,.
\ea\ee
It is also annihilated to all orders in perturbation theory by an analogous relation for $\CQ^\dagger$ proving it a perturbative ground state itself. Perturbative ground states are thus seen to come in families, the above procedure can be repeated with either of the pairs $(\chi_{A,B},\chi_{AB})$\,. However $\chi_{AB}\neq 0$ is not necessarily true, the terms in \eqref{eq:GenerateExtraGroundState} may potentially cancel or more trivially the degree of $\chi_{AB}$ may exceed the dimension of the target space $M_3$. 

Half-BPS instantons are field configurations minimizing the Euclidean Lagrangian and are annihilated by half of the supercharges \eqref{eq:SQMSuperscharges} in Euclidean time. They are distinguished by boundary conditions fixing the initial and final position of the particle. Field configuration may only converge to stationary points on $M_3$ allowing for $\dot x=0$, i.e. instantons necessarily connect perturbative ground states. From the Euclidean Lagrangian we obtain the flow and parallel transport equations
\be\label{eq:GradientFlow}
 \dot x^i \pm\phi_\lambda^i =\dot x^i \pm i c_{\alpha\beta\gamma} g^{ij}\bar\lambda^\alpha\phi_j^\beta\lambda^\gamma= 0\,, \qquad D_\tau\lambda_\alpha=0\,,
\ee
supplemented with the constraint $\bar\lambda\lambda=1$ enforced by the Lagrange multiplier. An instanton of the colored SQM solves \eqref{eq:GradientFlow} piecewise and connects multiple perturbative ground states. We refer to instantons of the SQM as generalized instantons whenever they connect more than two perturbative ground states, this more general class of instantons is absent in SQMs without $\lambda,\bar\lambda$ color degrees of freedom.

Instanton connecting two perturbative ground states, as familiar from Witten's SQM or Morse theory, start out at a point $(x_A,\lambda_A)\in\tn{ad}\,P_{\tn{ADE}}$ satisfying $\phi_{\lambda_A}(x_A)=0$ where the color contracted Higgs field $\phi_{\lambda_A}$ is given in \eqref{eq:EffHiggs}. From this initial configuration the instanton flows on $M_3$ along a path $\gamma$ determined by the 1-form $\phi_{\lambda(\tau)}$ where $\lambda(\tau)$ is the parallel transport of $\lambda_A$ along the path $\gamma$ with respect to the background connection $A$ on $M_3$\,. The flow can end at a point $(x_B,\lambda_B)\in\tn{ad}\,P_{\tn{ADE}}$ satisfying $\phi_{\lambda_B}(x_B)=0$. Summarizing we have
\be
(x_A,\lambda_A)\,, ~\phi_{\lambda_A}(x_A)=0 \qquad \xrightarrow[D_\tau\lambda(\tau)^{\,}=^{\,}0]{\quad \dot{x}(\tau)^{\,}=^{\,}\pm \phi_{\lambda(\tau)}\quad}\qquad (x_B,\lambda_B)\,,~\phi_{\lambda_B}(x_B)=0\,,
\ee
where $\tau$ runs from $-\infty$ to $+\infty$ from left to right. Completing the square in the Euclidean Lagrangian, instanton effects are found to be suppressed by
\be
S_{\tn{inst}}=\mp \int_{-\infty}^{+\infty}d\tau  \dot{x}^i\phi_{\lambda,i}>0\,,
\ee
where the sign depends on whether ascending or descending flows are considered in \eqref{eq:GradientFlow}. 

\begin{figure}
  \centering
  \includegraphics[width=11cm]{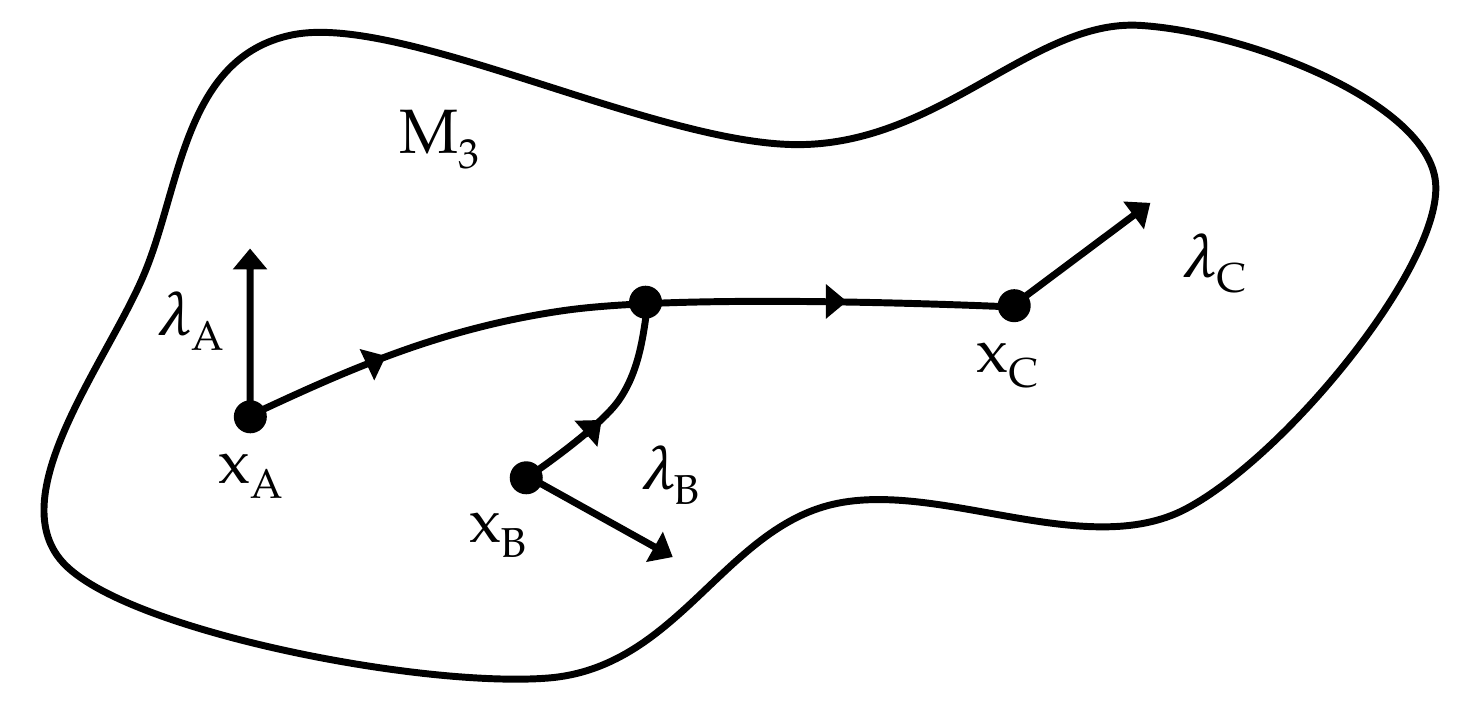}
    \caption{Sketch of an instanton connecting three perturbative ground states labeled by $(x_A,\lambda_A)$, $(x_B,\lambda_B)$ and $(x_C,\lambda_C)$. The color degrees of freedom are valued in the pull back bundle $x^*(\tn{ad}\,P_{\tn{ADE}})$ and are depicted as internal vectors attached to the localization site of the perturbative ground states. The three legs of the instanton are piecewise determined by the flow equations \eqref{eq:GradientFlow}.}\label{fig:Instanton}
\end{figure}

Generalized instantons connecting three perturbative ground states are pieced together from flows parametrized by half-lines where $\tau$ runs from $-\infty$ to 0 or from $0$ to $+\infty$ on each segment. We depict such a generalized instantons connecting three perturbative ground states labelled by $(x_A,\lambda_A), (x_B,\lambda_B)$ and $(x_C,\lambda_C)$ in figure \ref{fig:Instanton}. Along each leg the instanton is determined by the flow equations \eqref{eq:GradientFlow} and boundary conditions imposed at the junction and perturbative ground states. We discuss these generalized instantons in greater detail in section \ref{sec:OrgPerturbGS}.

\subsection{SYM and SQM}
\label{sec:ThePointOfSQM}
The colored SQM is a powerful computational and organisational tool when applied to the compactification of the partially twisted 7d SYM on $M_3$, we briefly discuss the dictionary between the SQM and SYM which follows from \eqref{eq:HilbertSpaceSQM}. 

The perturbative ground states of the SQM \eqref{eq:PG} are to be identified with the approximate zero modes \eqref{eq:ApproxZero} of the partially twisted 7d SYM. As a consequence the matrix elements of the supercharge $\CQ$ with respect to the perturbative ground states is given by the mass matrix \eqref{eq:OverLapIntegralsMass} of the 4d modes associated with the approximate zero modes. The ground states of the SQM then determine the massless spectrum in 4d \eqref{eq:Cohomologies}. The identification of perturbative ground states and approximate zero modes allows for an interpretation of the Yukawa overlap integral \eqref{eq:OverLapIntegralsYukawa} as a tunneling amplitude. States occupying two perturbative ground states $\chi_A,\chi_B$ can tunnel to a third $\chi_C$ and the overlap $Y_{ABC}$ then gives the amplitude for this process. We give a summary of these relations in appendix \ref{sec:Dictionary}.

The non-perturbative effects of the SYM derived from an ALE-geometry are understood to originate from M2-brane instantons wrapping supersymmetric 3-cycles. In the SYM these effects are in correspondence with flow trees of the Higgs field which are given by the projection of the supersymmetric 3-cycle to the base $M_3$, see e.g. figure \ref{fig:Relevant3Spheres}. These flow trees are precisely piecewise solutions to the flow equations \eqref{eq:GradientFlow} and thereby in one to one correspondence with the generalized instantons of the SQM. Along these graphs the approximate zero modes and perturbative ground states have maximal overlap and consequently these give the dominant contributions to the two integrals \eqref{eq:OverLapIntegralsMass} and \eqref{eq:OverLapIntegralsYukawa}.

\section{Higgs Bundles with Split Spectral Covers}
\label{sec:Split}

The simplest backgrounds to study the correspondence between non-perturbative effects in the 7d SYM, which originate from M2-brane instantons in M-theory, and generalized instantons of the colored SQM are abelian solutions to the BPS-equations with split spectral covers. These backgrounds have previously been studied in \cite{Pantev:2009de, Braun:2018vhk, Heckman:2018mxl, Barbosa:2019bgh, Cvetic:2020piw} and serve as a precursor to studying abelian solutions to the BPS equations with non-split spectral covers. Configurations with split spectral covers already display many features relevant for model building. Further, the cohomologies \eqref{eq:Cohomologies} characterizing the 4d massless matter content are computable in many cases and are easily engineered to give a chiral spectrum \cite{Pantev:2009de, Braun:2018vhk}.

Here we find that the single particle sector of the colored SQM decomposes into a direct sum of Witten SQMs \cite{Witten:1982im}, one for each generator of the Lie algebra $\mathfrak{g}_{\tn{ADE}}$. These interact via multi-particle effects encoded in higher order operations on the Morse-Witten complex of the colored SQM. They originate from M2-branes associated with the Y-shaped instantons as shown in figure \ref{fig:Relevant3Spheres} and higher-point generalized instantons. We quantify these effects by computing Yukawa overlap integrals \eqref{eq:OverLapIntegralsYukawa} via supersymmetric localization.

\subsection{Colored SQM and Witten's SQM}
\label{sec:ColorAndWitten}

We consider backgrounds characterized by a vanishing connection $A=0$ and a diagonal Higgs field $\phi=\phi_IH^I$. The Cartan components $\phi_I\in \Omega^1(M_3)$ are singular 1-forms on $M_3$ solving the sourced equations \eqref{eq:MagSource}. The color contracted Higgsfield $\phi_\lambda$, introduced in \eqref{eq:BilinearBackground}, now becomes
\be\label{eq:FlatEffHiggs}
\phi_\lambda=\kappa\lb \bar\lambda , [\phi, \lambda]\rb=\sum_{\alpha}\alpha^I\phi_{I}\bar\lambda^\alpha\lambda_\alpha\,,
\ee
where the sum runs over all roots $\alpha$ of the Lie algebra $\mathfrak{g}_{\tn{ADE}}$. The Lagrangian of the SQM probing the Higgs bundle simplifies from \eqref{eq:SQMLagrangian} to 
\be\ba\label{eq:SQMLagrangian2}
\CL&= \frac{1}{2}\dot{x}^i\dot{x}_i+i\bar{\psi}^i\nabla_\tau \psi_i+i\bar{\lambda}^\beta \dot\lambda_\beta-\lb \nabla_{(i}\phi_{j)}\rb_\lambda \bar{\psi}^i\psi^j-\frac{1}{2} \phi_\lambda^i \phi_{\lambda,i} \\
&~~~\,-\frac{1}{2} R_{ijkl} \psi^{i} \bar{\psi}^{j} \psi^{k} \bar{\psi}^{l}
+\zeta \lb \bar{\lambda}^\beta\lambda_\beta-n\rb,
\ea\ee
where $\beta=1,\dots, \dim\mathfrak{g}_{\tn{ADE}}$ runs over all generators $T^\beta$ of the Lie algebra $\mathfrak{g}_{\tn{ADE}}$. The bundle geometry is $\tn{ad}\,P_{\text{ADE}}= M_3\times \mathfrak{g}_{\text{ADE}}$ and as a consequence the Hilbert space \eqref{eq:Hilberspace} which is now given by Lie algebra valued forms $\CH_{\tn{phys.}}=\Lambda\lb M_3,\mathfrak{g}_{\text{ADE}}\rb$ decomposes into the direct sum
\be\ba\label{eq:HilberspaceDecomp}
\CH_{\tn{phys.}}&=\bigoplus_\beta \CH_{\tn{phys.}}^{(\beta)}\,, \medskip \\  \CH_{\tn{phys.}}^{(\beta)}&=\Lambda\lb M_3\rb \otimes T^\beta\,,
\ea\ee
paralleling the decomposition of $\tn{ad}\,P_{\text{ADE}}$ into a sum of line bundles. States in $\CH_{\tn{phys.}}^{(\beta)}$ are $p$-forms oriented along the generator $T^\beta$ in $\Omega^p(M_3,\mathfrak{g}_{\text{ADE}})$. Specializing to a Cartan-Weyl Basis $\lbbb H^I, E^\alpha \rbbb$ of the Lie algebra $\mathfrak{g}_{\text{ADE}}$ we can sharpen the decomposition \eqref{eq:HilberspaceDecomp} to
\be\ba\label{eq:HilberspaceDecomp2}
\CH_{\tn{phys.}}&= \bigoplus_\alpha \CH_{\tn{phys.}}^{(\alpha)}  \oplus \, \bigoplus_I \CH_{\tn{phys.}}^{(I)}  \,, 
\ea\ee
and refer to the first summand  $\oplus_\alpha \CH_{\tn{phys.}}^{(\alpha)}$ as the localized sector and to the second summand $\oplus_I \CH_{\tn{phys.}}^{(I)}$ as the bulk sector of this SQM. They are built from
\be\label{eq:HilbertSubspaces}
\CH^{(\alpha)}_{\tn{phys.}}=\Lambda(M_3)\otimes E^\alpha\,,  \qquad \CH_{\tn{phys.}}^{(I)} =\Lambda\lb M_3\rb \otimes H^I\,.
\ee
The supercharge $\CQ$ respects this decomposition as all component functions $\phi_i$ of the Higgs field $\phi=\phi_idx^i$ are valued in the Cartan subalgebra, i.e. it restricts to operators on the subspaces \eqref{eq:HilbertSubspaces}
\be\ba\label{eq:DecompQ}
\CQ^{(\alpha)}\,&:\qquad  \CH_{\tn{phys.}}^{(\alpha)}~\rightarrow ~ \CH_{\tn{phys.}}^{(\alpha)}\,, \qquad \chi \otimes E^\alpha~\mapsto~\lb d\chi+\alpha^I\phi_I\wedge \chi\rb\otimes  E^\alpha\,, \\
\CQ^{(I)}\,&:\qquad  \CH_{\tn{phys.}}^{(I)}~\rightarrow~  \CH_{\tn{phys.}}^{(I)}\,, \qquad \chi \otimes H^I~\mapsto~ d\chi\otimes  H^I\,,
\ea\ee
where $\chi\in\Omega^p(M_3)$ is a $p$-form on $M_3$. The Hamiltonian $H=\frac{1}{2}\lbbb \CQ,\CQ^\dagger \rbbb$ decomposes similarly into restrictions as \eqref{eq:DecompQ} which govern the time evolution of states of definite color
\be\label{eq:Hamiltonians}
H^{(\alpha)}=\frac{1}{2}\lbbb \CQ^{(\alpha)},\CQ^{(\alpha)\dagger} \rbbb\,, \qquad H^{(I)}=\frac{1}{2}\lbbb \CQ^{(I)},\CQ^{(I)\dagger}\rbbb\,,
\ee
Stripping off the trivial Lie algebra generator in each sector we obtain Hamiltonians acting on the exterior algebra $\Lambda(M_3)$. We thus find a copy of Witten's SQM for every Lie algebra generator and more precisely obtain the correspondences
\be\ba\label{eq:WittenSQMs}
E^\alpha\in\mathfrak{g}_{\tn{ADE}}\qquad  &\leftrightarrow \qquad \tn{Witten's SQM with supercharge } \CQ=d+\alpha^I\phi_I\wedge\,,\\
H^I\in\mathfrak{g}_{\tn{ADE}}\qquad  &\leftrightarrow \qquad \tn{Witten's SQM with supercharge } \CQ=d\,.\\
\ea\ee
The study of colored SQMs with split Higgs fields thus equates to studying the interaction between the family of uncolored SQMs \eqref{eq:WittenSQMs} embedded within it. In appendix \ref{sec:R1} we study the above from view point of the Lagrangian and make contact with the analysis presented in \cite{Braun:2018vhk}.

\subsection{Organizing Perturbative Ground States}
\label{sec:OrgPerturbGS}

The linear combinations of perturbative ground states \eqref{eq:PG} which constitute true ground states of the SQM, and therefore zero modes along $M_3$ in the compactification of the 7d SYM, are determined by the cohomology groups of the Morse-Witten complex. The Morse-Witten complex of the colored SQM collects the Morse-Witten complexes of each copy of Witten's SQM in \eqref{eq:WittenSQMs} into a single complex. 

The Morse-Witten complex is built from the free abelian groups $C^\mu(M_3,\phi)$ generated by the perturbative ground states \eqref{eq:PG} over the complex numbers
\be\ba\label{eq:GeneralChains}
C^\mu(M_3,\phi)&=\bigoplus_{\alpha}  C^{\mu,\alpha}(M_3,\alpha^I\phi_I)\,, \\ C^{\mu,\alpha}(M_3,\alpha^I\phi_I)&=\bigoplus_{a} \C\,  \ket{x_a,\lambda^\alpha,\mu_a}\equiv C_\alpha^\mu(M_3,\phi)\,,
\ea\ee
where $\mu$ fixes the degree of the perturbative ground state as a differential form. It is graded by the fermion number operators associated with the fermions $\psi,\bar\psi$ and $\lambda,\bar\lambda$. The supercharge gives rise to the boundary map on the complex \eqref{eq:GeneralChains} and as a consequence of the decomposition \eqref{eq:DecompQ} the colored Morse-Witten complex is found to decompose into multiple standard Morse-Witten complexes whose chain groups are $C_\alpha^\mu(M_3,\phi)$ for fixed color $\alpha$. We take capital latin indices to run over generic perturbative ground states of the colored SQM and decapitalized latin indices to run over all perturbative ground states of a fixed color, i.e. or equivalently over all perturbative ground states of a subcomplex of the SQMs in \eqref{eq:WittenSQMs}.
 
The color restricted supercharge $\CQ^{(\alpha)}$ of \eqref{eq:DecompQ} now gives rise to the standard boundary map \cite{Witten:1982im, Hori:2000kt, Gaiotto:2015aoa} generated by oriented flow lines \eqref{eq:GradientFlow} of $\alpha^I\phi_I$ we have 
\be\label{eq:MWcomplex22}
\begin{tikzcd}[row sep=0.25in]
 C_\alpha^3(M_3,\phi)  &  C_\alpha^2(M_3,\phi) \arrow[l,swap,yshift=2,"~\CQ^{(\alpha)}"]  &  C_\alpha^1(M_3,\phi) \arrow[l,swap,yshift=2,"~\CQ^{(\alpha)}"] &  C_\alpha^0(M_3,\phi) \arrow[l,swap,yshift=2,"\CQ^{(\alpha)}"]\,.
\end{tikzcd}
\ee 
The adjoint of the supercharge $\CQ^{(\alpha)}$ maps in the opposite direction. There is no such complex for colors in the bulk of the SQM. Each of the complexes \eqref{eq:MWcomplex22} can be analyzed separately and its cohomologies are the Novikov/Lichnerowicz cohomologies \cite{Lichnerowicz, Novikov_multivaluedfunctions, dur4050} with respect to the closed 1-form $\alpha^I\phi_I$ on $M_3$. The cohomology groups of the supercharge $\CQ$ of the colored SQM thus decomposes into a direct sum
\be\label{eq:CohoNonSplit}
H_{\CQ}^*(M_3,\mathfrak{g}_{\tn{ADE}})\cong \lb \bigoplus_{I=1}^R H_{\tn{dR}}^*(M_3)\rb \oplus \lb \bigoplus_{E^\alpha \,\in\, \mathfrak{g}_{\tn{ADE}}} H_{\tn{Nov.}}^*(M_3,\alpha^I\phi_I) \rb\,,
\ee
where each summand is in correspondence with an SQM of \eqref{eq:WittenSQMs}. For exact 1-forms $\alpha^I\phi_I=\alpha^Idf_I$ derived from Morse functions $\alpha^If_I$ the complex \eqref{eq:MWcomplex22} is that of Morse theory on a manifold with boundary. The boundary is generated by excising the source terms (more generally supported on graphs) as introduced in \eqref{eq:MagSource} and for purely electrically sources Higgs fields $j_I=0$ the Novikov cohomoloiges in \eqref{eq:CohoNonSplit} reduce to relative cohomologies and are readily computed \cite{Pantev:2009de, Braun:2018vhk}.

The complexes \eqref{eq:MWcomplex22} of different color can interact via a cup product originiating from \eqref{eq:GenerateExtraGroundState} and mediated by Y-shaped instantons. These multi-particle effects are absent in ordinary SQMs. Consider three perturbative ground states
\be\ba\label{eq:PerturbGS}
\chi_a&=\ket{x_a,\lambda^\alpha,\mu_a}\in \Omega^{\mu_a}(M_3)\otimes E^\alpha\,,\\
\chi_b&=\ket{x_b,\lambda^\beta,\mu_b}\in\Omega^{\mu_b}(M_3)\otimes E^\beta\,,\\
\chi_c&=\ket{x_c,\lambda^\gamma,\mu_c}\in\Omega^{\mu_c}(M_3)\otimes E^\gamma\,,
\ea\ee
which we assume to be energy eigenstates with energies $E_{0,r}$ of the Hamiltonian $H=\frac{1}{2}\{ \CQ,\CQ^\dagger \}$. In general energy eigenstates will be linear combinations of the perturbative ground states to which the arguments below extend naturally. We further restrict to cases which allow for the normalisation $\kappa(T_a,T_b)=\delta_{ab}$ of generators to simplify exposition. 

The Y-shaped instantons determine the leading order contribution to the overlap integral \eqref{eq:OverLapIntegralsYukawa}. The integral vanishes unless three selection rules are satisfied
\be\label{eq:SelectionRules}
\mu_a+\mu_b=\mu_c\,, \qquad {\alpha+\beta}=\gamma\,, \qquad E_{0,a}+E_{0,b}=E_{0,c}\,.
\ee
If these are satisfied the Yukawa integral can be simplified to
\be\label{eq:OverlapAppendix}
Y_{abc}=\int_{M_3}\braket{ \chi_{c\,},\lbb \chi_{a\,}\wedge\,,\chi_b\rbb}=\int_{M_3}*\,\overline{\chi}^{\,(\gamma)}_c\wedge \chi^{(\alpha)}_a\wedge \chi^{(\beta)}_b\,.
\ee
where we took the trace over the Lie algebra generators in the second equality and made the complex conjugation in the first factor explicit. Here the raised indices $(\alpha,\beta,\gamma)$ refer to the differential form part of the perturbative ground stated stripped of its Lie algebra generator.

We evaluate this integral in three steps. The first step consists of rewriting the perturbative ground states as projections of profiles which are highly localized at the point $x_r\in M_3$ associated to the perturbative ground state with $r=a,b,c$. We then rewrite the overlap integral as a path integral of the colored SQM in which the unprojected localized profiles go over into boundary conditions. This path integral then splits into three pieces each associated with a definite color which we evaluate via supersymmetric localization. 

To begin note that the operator creating a perturbative ground state can be rewritten as
\be\label{eq:RewritePerturbativeGS}
\chi_{r\,}= \lim_{T\rightarrow -\infty(1+i\delta) } \frac{e^{-iHT}\Psi_{r\,} e^{iHT}}{e^{-iE_{0,r}T}\braket{\chi_r|\Psi_r}}\equiv \Psi_r\big|_{-\infty}\,,\qquad 0<\delta \ll 1\,,
\ee
where $r=a,b,c$. The Hamiltonian $H$ is the Legendre transform of the Lagrangian given in \eqref{eq:SQMLagrangian2} and is given explicitly in \eqref{eq:SQMHamiltonian}. Here $\Psi_r= \Psi_r^{(\alpha)} \bar\lambda^\alpha$ (no sum) creates a Lie algebra valued $\mu_{r}$-form oriented along the generator $E^\alpha$ whose support only contains the point $x_{r}\in M_3$ and no other points at which perturbative ground states localize. The slightly imaginary limit projects $\Psi_r$ onto the state of lowest energy with non-trivial overlap, this state is $\chi_r$. Using the basis \eqref{eq:HilberSpaceBasis} we extract the component functions as
\be\ba\label{eq:ComponentFunctionPerturb}
\lb \Psi_r\big|_{-\infty}\rb_{i_1\dots i_{\mu_r}}^{(\alpha)}(x)&=  \bra{x}\lambda^\alpha \psi_{i_1}\dots \psi_{i_{\mu_r}} \Psi_r\big|_{-\infty}\ket{0}\\
&= \lim_{T\rightarrow -\infty(1+i\delta) }  \bra{x}\lambda^\alpha \psi_{i_1}\dots \psi_{i_{\mu_r}} \lb e^{-iHT}\Psi_{r}\rb \lb e^{-iE_{0,r}T}\braket{\chi_r|\Psi_r}\rb^{-1}\ket{0}\\
&=\lim_{T\rightarrow -\infty(1+i\delta) } \lb e^{-iE_{0,r}T}\braket{\chi_r|\Psi_r}\rb^{-1} \bra{x}\lambda^\alpha \psi_{i_1}\dots \psi_{i_{\mu_r}} e^{-iHT}\ket{\Psi_{r} }\\
&=\lim_{T\rightarrow -\infty(1+i\delta) } \lb e^{-iE_{0,r}T}\braket{\chi_r|\Psi_r}\rb^{-1} \\
&~~~~\times \bra{x}\lambda^\alpha \psi_{i_1}\dots \psi_{i_{\mu_r}} e^{-iHT}\lb \sum_s \ket{\chi_s}\bra{\chi_s}+\sum_{n}\ket{n}\bra{n}\rb\ket{\Psi_{r}} \\
&=\lim_{T\rightarrow -\infty(1+i\delta) } \lb e^{-iE_{0,r}T}\braket{\chi_r|\Psi_r}\rb^{-1} \bra{x}\lambda^\alpha_{(r)}\psi_{i_1}\dots \psi_{i_{\mu_r}} e^{-iHT}\ket{\chi_r}\braket{\chi_r|\Psi_r}\\
&=\bra{x}\lambda^\alpha\psi_{i_1}\dots \psi_{i_{\mu_r}} \ket{\chi_r}\\
&=\chi_{r,i_1\dots i_{\mu_r}}^{(\alpha)}(x)
\,,
\ea\ee
which proves \eqref{eq:RewritePerturbativeGS}. Here the sum $\sum_s$ runs over all perturbative ground states while the sum $\sum_n$ runs over all higher energy eigenstates in the physical Hilbertspace $\CH_{\tn{phys.}}$ of \eqref{eq:HilberspaceDecomp2}. The support of the states $\Psi_r$ is localized at $x_r$ and excludes the sites of localization of all other perturbative ground states. Consequentially $\braket{\chi_s|\Psi_r}=\delta_{sr}\braket{\chi_r|\Psi_r}$ holds. Note further that we can anticommute the color fermions $\lambda,\bar\lambda$ past another in \eqref{eq:ComponentFunctionPerturb} which results in a simplification of the Hamiltonian evolving the states. We have
\be\label{eq:KeyEq}
\chi_{r,i_1\dots i_{\mu_r}}^{(\alpha)}(x)= \lim_{T\rightarrow -\infty(1+i\delta) }  \bra{x}\psi_{i_1}\dots \psi_{i_{\mu_r}} \lb e^{-iH^{(\alpha)}T}\Psi_{r}^{(\alpha)}\rb \lb e^{-iE_{0,r}T}\braket{\chi_r|\Psi_r}\rb^{-1}\ket{0}\,,
\ee
where $H^{(\alpha)}$ is the Hamiltonian given in \eqref{eq:Hamiltonians}.

Next we rewrite the overlap integral \eqref{eq:OverlapAppendix} using the 
expression \eqref{eq:KeyEq} for the profile of the perturabtive ground states
\be\ba\label{eq:KeyEq2}
Y_{abc}&=\lim_{T\rightarrow -\infty(1+i\delta)} \int_{M_3} d^3x \lb e^{-i\lb E_{0,a}+E_{0,b}-E_{0,c}\rb T}  \braket{\chi_a|\Psi_a} \braket{\chi_b|\Psi_b} \overline{\braket{\chi_c|\Psi_c}}\rb^{-1}  \\ &~~~~  \times\bra{x}\psi_{i_1}\dots \psi_{i_{\mu_a}} \lb e^{-iH^{(\alpha)} T}\Psi_{a}^{(\alpha)}\rb \ket{0}\\ &~~~~  \times\bra{x}\psi_{j_1}\dots \psi_{j_{\mu_b}} \lb e^{-iH^{(\beta)} T}\Psi_{b}^{(\beta)}\rb \ket{0}\\
&~~~~\times\bra{x}\psi_{k_1}\dots \psi_{k_{3-\mu_c}} \lb e^{iH^{(\gamma)} T}*\overline{\Psi}_{c}^{\,(\gamma)}\rb \ket{0}\\ &~~~~ \times\,\epsilon^{i_1\dots i_{\mu_a}j_1\dots j_{\mu_b} k_1\dots k_{3-\mu_c}}
\ea\ee
We take $\Psi_{r}$ to be $\delta$-function like supported at $x_{r}$, rescale the Higgs field $\phi\rightarrow t\phi$ and from now on work to leading order in $1/t$. In the $t\rightarrow \infty$ limit the profile of the normalized perturbative ground states $\chi_{r}$ increasingly localizes at $x_{r}$\,. To leading order we thus have 
\be\label{eq:LeadingOrderApprox}
\braket{\chi_r|\Psi_r}=1+\CO(1/t)\,.
\ee
The energies cancel by \eqref{eq:SelectionRules} and together with \eqref{eq:LeadingOrderApprox} we find \eqref{eq:KeyEq2} to simplify to
\be\ba
Y_{abc}&=\lim_{T\rightarrow -\infty(1+i\delta)} \int_{M_3} d^3x\, \epsilon^{i_1\dots i_{\mu_a}j_1\dots j_{\mu_b} k_1\dots k_{3-\mu_c}}
\\ &~~~~\times\bra{x}\psi_{i_1}\dots \psi_{i_{\mu_a}} \lb e^{-iH^{(\alpha)} T}\Psi_{a}^{(\alpha)}\rb \ket{0}
\\ &~~~~ \times\bra{x}\psi_{i_1}\dots \psi_{i_{\mu_b}} \lb e^{-iH^{(\beta)} T}\Psi_{b}^{(\beta)}\rb \ket{0}
\\ &~~~~ \times\bra{x}\psi_{i_1}\dots \psi_{i_{3-\mu_c}} \lb e^{iH^{(\gamma)} T}*\overline{\Psi}_{c}^{\,(\gamma)}\rb \ket{0} \\
&~~~~+\CO(1/t)\,.
\ea\ee
We now transition to the path integral representation by rewriting each matrix element above as a separate path integral. These are associated to paths with time intervals $(T,0]_{a,b}$ and $[0,-T)_c$\,. The profiles $\Psi_{r}$ are supported at $x_{r}$ and give rise to boundary conditions for the path integral at infinite times. All in all we have
\be\ba\label{eq:PIYukawa}
Y_{abc}&=\int_{M_3}d^3 x_0\int \prod_{\substack{ -\infty <\, \tau\, < 0 \\ x_{a,-\infty}\,=\,x_a\\ x_{b,-\infty}\,=\,x_b }} dx_{a,\tau\,} d\psi_{a,\tau\,} d\bar\psi_{a,\tau\,} dx_{b,\tau\,} d\psi_{b,\tau\,} d\bar\psi_{b,\tau\,}  \prod_{\substack{ 0 < \,\tau\, < \infty \\ x_{c,\infty}\,=\,x_c}} dx_{c,\tau\,} d\psi_{c,\tau\,} d\bar\psi_{c,\tau}\\
&~~~~ \exp\lbb {i\lb S^{(\alpha)}[x_a,\psi_a,\bar\psi_a]+S^{(\beta)}[x_b,\psi_b,\bar\psi_b]+S^{(\gamma)}[x_c,\psi_c,\bar\psi_c]\rb}\rbb~+~\CO(1/t)\,.
\ea\ee
Here we have introduced the half line actions 
\be\ba\label{eq:ActionLorentz}
S^{(\alpha)}=\int_{-\infty}^0 d\tau~\CL^{(\alpha)} \,,\qquad S^{(\beta)}=\int_{-\infty}^0 d\tau~\CL^{(\beta)}\,, \qquad  S^{(\gamma)}=\int^{\infty}_0  d\tau~ \CL^{(\gamma)} \,,
\ea\ee
where the color restricted Lagrangians $\CL^{(\alpha)}$ are the Legendre transformation of the color restricted Hamiltonians \eqref{eq:Hamiltonians}. These actions are associated with the time intervals $(T,0]_{a,b}$ and $[0,-T)_c$ in the $T\rightarrow -\infty$ limit. The slightly imaginary limit makes the Feynman propagator the relevant propagator here. Further we have written $x(\tau)=x_\tau$ and denoted the three paths generated by insertions of the identity operator by the labels $a,b,c$. Note that these are only defined on half of the real line. These are constrained to start or end at the points $x_a,x_b,x_c\in M_3$ where the perturbative ground states localizes at infinite time and join at a common point $x_0\in M_3$. 

\begin{figure}
  \centering
  \includegraphics[width=8cm]{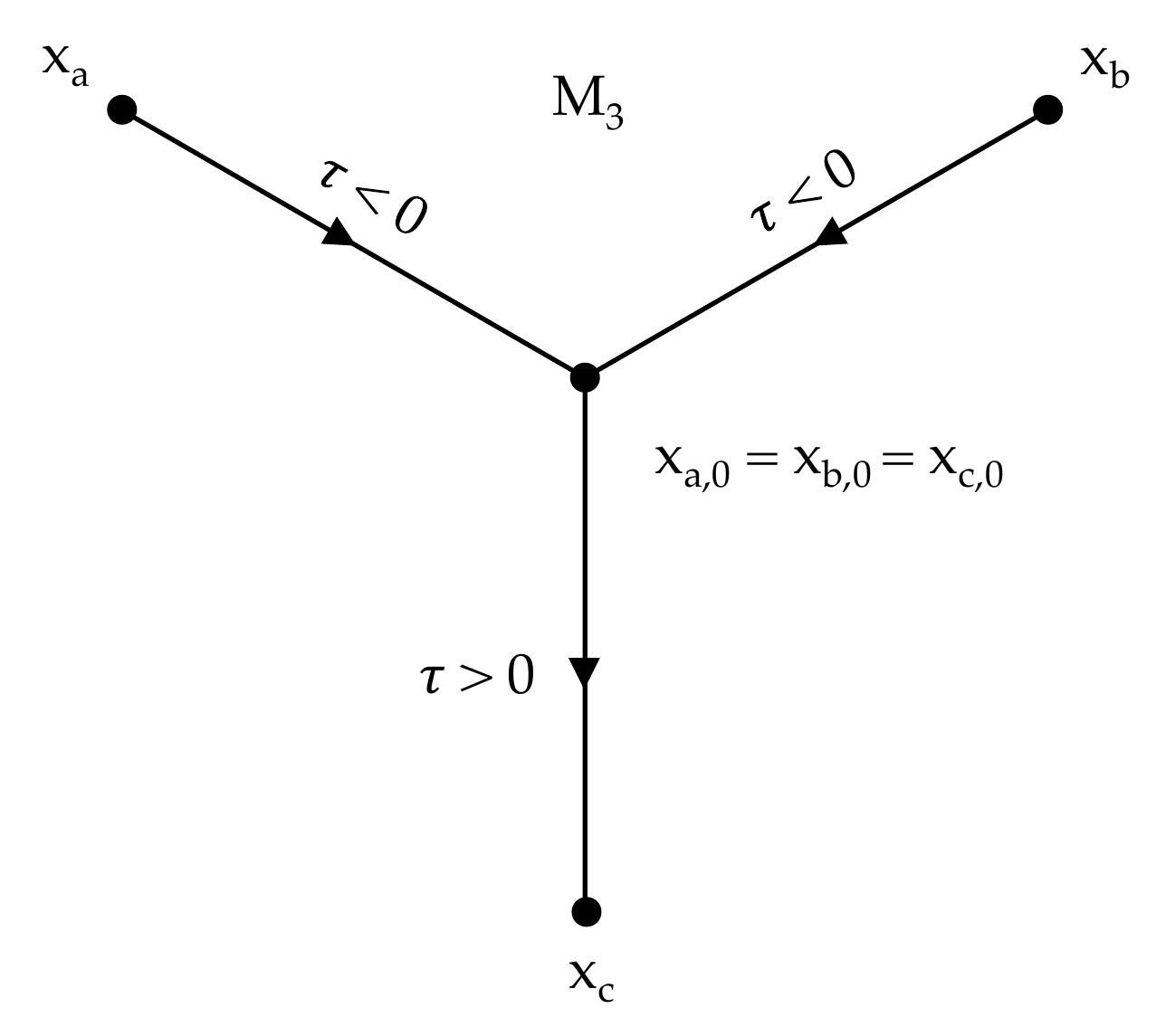}
    \caption{The figure shows an example of a Y-shaped graph whose end points are fixed at the points $x_{a,b,c}$\,. It is parametrized by two copies of $\R_-$ and one copy of $\R_+$\,. The set of Y-shaped graphs constrained in this manner constitute the configuration space the path integral in \eqref{eq:PIYukawa} localizes to. These Y-shaped flow trees are examples of generalized instantons which are a novel phenomenon of colored SQMs.}\label{fig:YPath}
\end{figure}

The expression \eqref{eq:PIYukawa} is technically not a path integral, the space of field configurations integrated over is that of all Y-shaped graphs whose end points are given by $x_{a,b,c}$. We depict such a configuration in figure \ref{fig:YPath}. In the SQM $Y_{abc}$ is to be identified with the tunneling amplitude of two particles of color $\lambda_a,\lambda_b$ located at $x_a,x_b$ respectively combing to a particle of color $\lambda_c$ located at $x_c$.

As the final step we now evaluate the integral \eqref{eq:PIYukawa} via supersymmetric localization. We rotate to euclidean time $\tau\rightarrow -i\tau$ and denote the resulting actions with a subscript, we have
\be\label{eq:FactoringYukawa}
Y_{abc}=\int_{M_3}d^3 x_0\prod_{r=a,b,c}\int Dx_{r} D\psi_r D\bar\psi_r \,e^{-S^{(\alpha_r)}_E[x_r,\psi_r,\bar\psi_r]}~+\CO(1/t)\,.
\ee
The total action
\be
S_E=S_E^{(\alpha)}+S_E^{(\beta)}+S_E^{(\gamma)}\,,
\ee
is not invariant under the supersymmetries derived from \eqref{eq:SQMVariations}. Half of the supersymmetry is broken by the boundaries of the actions \eqref{eq:ActionLorentz}, explicit computation yields
\be\label{eq:BPScond}
\delta S_E= \bar\epsilon\lb \psi^i_{a\,}\dot x_{a,i}+\psi^i_{b\,}\dot x_{b,i}+\psi^i_{c\,}\dot x_{c,i}\rb_{\tau=0}\,,
\ee
whereby only the symmetry generated by $\epsilon$ is unbroken. Considering the factors of \eqref{eq:FactoringYukawa} separately we see that the path integral thus localizes to ascending flow-lines of the 1-forms $\alpha^I\phi_{I},\beta^I\phi_{I}$ on each leg emanating from $x_{a,b}$ and to ascending flow lines of $\gamma^I\phi_{I}$ on the leg ending at $x_{c}$ of the Y-shaped configuration depicted in figure \ref{fig:YPath}. These flow-lines are required to meet at a common point $x_0\in M_3$ at time $\tau=0$. We refer to such a BPS configuration as a flow tree $\Gamma_{abc}$. In a three dimensional set-up the only relevant triplet of perturbative ground state have degrees $\mu_a=\mu_b=1$ and $\mu_{c}=2$ as the D-term constraint excludes perturbative ground states of degree 0 or 3. The moduli space of such flow trees is generically zero dimensional which follows by dimension count. Ascending, descending flow lines emanating from a point of Morse index $\mu=1,2$ sweep out a manifold of codimension 1 respectively. A common point of these flows is obtained upon intersecting these submanifolds whose expected codimension is 3. Due to the common center point $x_0$ there is no zero mode associated to time translations. 

The BPS locus of localization are thus Y-shaped flow trees as depicted in figure \ref{fig:YPath}. The localization computation then gives the result
\be\ba\label{eq:YukawaEvaluatedText}
Y_{abc}&=\sum_{\Gamma_{abc}} (\pm)_{\Gamma_{abc}} \exp\lb -t\int_{\Gamma_\alpha} \alpha^I\phi_{I}  -t\int_{\Gamma_\beta} \beta^I\phi_{I}+t\int_{\Gamma_\gamma}\gamma^I\phi_{I}  \rb\\&~~~\,+\CO(1/t)\,,
\ea\ee
where $\Gamma_{\sigma}$ with $\sigma=\alpha,\beta,\gamma$ are flow lines of the 1-form $\sigma^I\phi_I$ originating and ending at the respective perturbative ground states at $x_{a,b,c}$ and $x_0$. They glue to the flow tree $\Gamma_{abc}$ over which the sum runs. The sign $(\pm)_{\Gamma_{abc}}$ denotes a fermion determinant. When $\sigma^I\phi_I=df^{(\sigma)}$ is exact this simplifies to
\be\label{eq:YukawaEvaluated2Text2}
Y_{abc}=\sum_{\gamma_{abc}} (\pm)_{\gamma_{abc}} \exp\lb -tf^{(\alpha)}(x_a)-tf^{(\beta)}(x_b)+tf^{(\gamma)}(x_c) \rb\,+\CO(1/t)\,.
\ee
This fixes the proportionality constant which we could not determine in \cite{Braun:2018vhk}. For exact Higgs fields $df^{(\sigma)}$ the moduli space of Y-shaped flow trees has been described in \cite{Fukaya_morsehomotopy}, where it is shown to be an oriented 0d manifold, the relative signs $(\pm)_{\gamma_{abc}}$ are then a choice of orientation on this moduli space. 

The overlap integral $Y_{abc}$ therefore gives rise to a map between the chain groups of the embedded Morse-Witten complexes
\be\label{eq:YukawaMaptext}
Y=[\,\cdot\, \wedge\,, \cdot\,]\,: \quad C_\alpha^{\mu_a}(M_3)\times C_\beta^{\mu_b}(M_3) ~~ \rightarrow ~~ C_{\alpha+\beta}^{\mu_a+\mu_b}(M_3)\,,
\ee
which maps pairs of perturbative ground states according to the Y-shaped flow trees
\be
(\chi_a,\chi_b) ~~ \mapsto ~~ \sum_cY_{abc}\chi_c\,,
\ee
where
$\chi_a,\chi_b,\chi_c$ are given in \eqref{eq:PerturbGS}. Ground states of the colored SQM are linear combination of perturbative ground states and thus the map $Y$ descends to the cohomology of the colored SQM complex \eqref{eq:GeneralChains} where it describes a cup product.

\begin{figure}
  \centering
  \includegraphics[width=15cm]{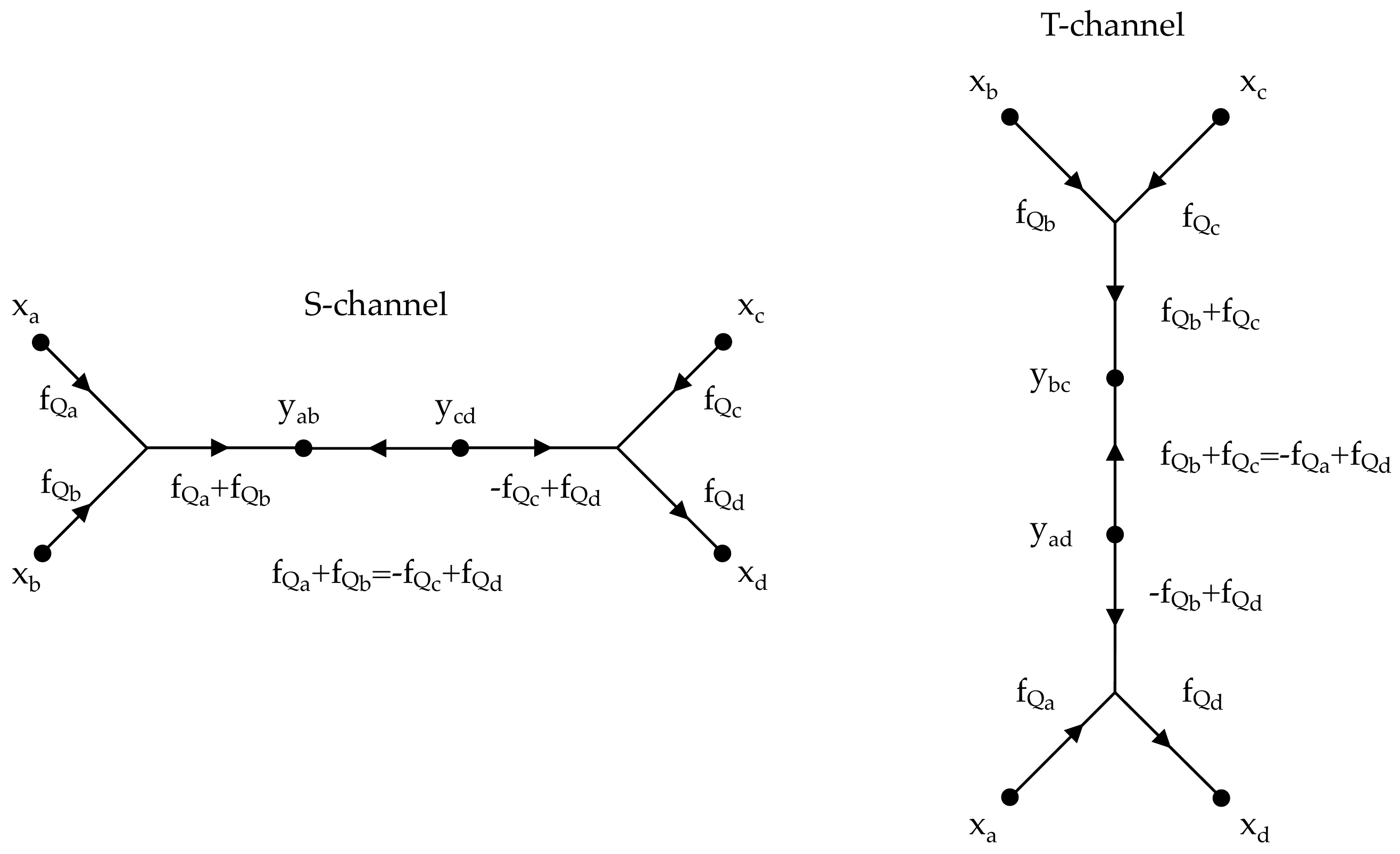}
    \caption{Picture of the flow trees contributing to the Massey product of length 3\,. Two summands marked with $S,T$ respectively contribute to the Massey product $m_3$ in \eqref{eq:Massey3Action}. Pictorially these are given by S-channel and T-channel like contributions. The Massey product $m_3$ maps perturbative ground states localized at the points $x_a,x_b,x_c$ to one localized at $x_d$\,. Both channels are a given by two Y-shaped gradient flow trees connect by a gradient flow line. This structure descend from \eqref{eq:Massey3Action} which involves two cup products $Y$ and a single boundary operator $\CQ$. Here we assume globally exact Higgs fields $\phi_{Q_r}=df_{Q_r}$. In the picture we mark the functions governing the gradient flows. Two intermediate perturbative ground states are label by $y_{rs}$ for each channel.}
 \label{fig:M3Prod}
\end{figure}

The Massey products $m_n$ of length $n$ generalize the cup product $Y$\,. These are realized by gradient flow trees connecting $n+1$ perturbative ground states and are associated to a collection of Y-shaped gradient flow trees and gradient flow lines. We restrict our discussion to the Massey products of length $3$ which are given by the map 
\be\ba
m_3\,: \quad C_\alpha^{\mu_a}(M_3)\times C_\beta^{\mu_b}(M_3)\times C_\gamma^{\mu_c}(M_3)~~\rightarrow ~~ C^{{\mu_{a}}+{\mu_{b}}+{\mu_{c}}-1}_{\alpha+\beta+\gamma}(M_3)
\ea\ee
and is defined by
\be\ba\label{eq:Massey3Action}
\Big( \ket{x_a, \lambda^\alpha,\mu_a} , \ket{x_b, \lambda^\beta,\mu_b}, \ket{x_c, \lambda^\gamma,\mu_c}\Big)\quad \mapsto \qquad &(-1)^{\mu_a+\mu_b-1\,}Y(S,\ket{x_c, \lambda^\gamma,\mu_c})\\+\,&(-1)^{\mu_b+\mu_c-1\,}Y(\ket{x_a, \lambda^\alpha,\mu_a},T)\,,
\ea\ee
where $S,T$ are perturbative ground states determined by the reverse flows
\be\ba
QS&=(-1)^{\mu_a\,} Y(\ket{x_a, \lambda^\alpha,\mu_a},\ket{x_b, \lambda^\beta,\mu_b}) \,, \\
QT&=(-1)^{\mu_b\,} Y(\ket{x_b, \lambda^\beta,\mu_b},\ket{x_c, \lambda^\gamma,\mu_c}) \,.
\ea\ee
Up to signs these maps are easily understood as concatenations of gradient flow lines and $Y$-shaped flow trees. The quantities $Y(S,\ket{x_c, \lambda_c,\mu_c})$ and  $Y(\ket{x_a, \lambda_a,\mu_a},T)$  corresponds to s,t-channel like graphs respectively. They are depicted in figure \ref{fig:M3Prod}\,. General Massey products of length $n$ are described similarly. By linear extension all Massey products $m_n$ descend to the cohomolgies of the complexes of \eqref{eq:MWcomplex22}.

Summarizing we note that the set of perturbative ground states of the colored SQM can be organized into separate Morse-Witten complexes whose boundary maps are given by the color restrictions $\CQ^{(\alpha)}$ of the supercharge $\CQ$. The supercharge $\CQ$ giving rise to boundary maps. These complexes interact via the cup product $Y$ and Massey products $m_n$ with $n\geq 3$ which give rise to $3$-point and $(n+1)$-point tunneling amplitutdes among the perturbative ground states. We summarize the corresponding geometrical and field theoretic structures in appendix \ref{sec:Dictionary}. Generalizing from \eqref{eq:OverLapIntegralsMass} and \eqref{eq:OverLapIntegralsYukawa} the massey products are in correspondence with irrelevant couplings in the 4d $\CN=1$ gauge theory. We therefore focus on the flow lines and Y-shaped flow trees related to the 3-spheres shown in figure \ref{fig:Relevant3Spheres} going forward.

\subsection{Partial Higgsing}
\label{sec:RankK}

When the group $G_{\tn{ADE}}$ is only partially Higgsed the correspondence \eqref{eq:WittenSQMs} degenerates. Consider the rank $n$ Higgsing
\be\ba\label{eq:GeneralSplitting}
G_{\tn{ADE}}\quad&\rightarrow\quad G_{\tn{GUT}}\times U(1)^n\,,\\
\tn{Ad}\, G_{\tn{ADE}}\quad&\rightarrow\quad \lb \tn{Ad}\,G_{\tn{GUT}} \otimes {\bf 1}\rb \oplus \lb {\bf 1}\otimes \tn{Ad}\,U(1)^n\rb \oplus \sum_Q {\bf R}_Q\,,
\ea\ee
where $Q=(q_1,\dots,q_n)$ is a vector of $U(1)$ charges. Then for every generator $E^\alpha\in {\bf R}_Q$ the supercharge of the associated SQM reads $\CQ=d+Q^I\phi_I\wedge$. The correspondence \eqref{eq:WittenSQMs} can be rephrased as 
\be\ba\label{eq:WittenSQMs3}
 {\bf R}_Q \qquad  &\leftrightarrow \qquad \tn{Witten's SQM with supercharge } \CQ=d+Q^I\phi_I\wedge\,,
\ea\ee
making the degeneracy manifest. Representation not transforming under $U(1)^n$ correspond to a free SQM mapping into $M_3$ whose supercharge is the exterior derivative. Here the selection rule in \eqref{eq:SelectionRules} for the roots of the Lie algebra becomes the well known constraint on Yukawa interactions of requiring the sum of $U(1)$ charges to vanish.

\section{Higgs Bundles with Non-Split Spectral Covers}
\label{sec:NonSplit}

We now turn to colored SQMs probing Higgs bundles with non-split spectral covers. These covers are branched and were discussed in section \ref{sec:Harmonic}, they are the spectral covers generically encountered in F-theory constructions  \cite{Donagi1993SpectralC, Friedman:1997yq, Donagi:2008ca, Hayashi:2008ba, Blumenhagen:2009yv, Marsano:2009gv, Marsano:2009wr, Donagi:2009ra, Hayashi:2009ge, Hayashi:2010zp, Marsano:2011hv}. Here we explore the Morse-theoretic consequences of the presence of branch sheets and find that previously distinct copies of Witten's SQM combine into a single SQM whose target space is now topologically an irreducible component of the spectral cover. Consequently the cohomology of the supercharge $\CQ$ on $M_3$ computes topological properties for the spectral cover components $\CC_k$ rather than those of the base manifold $M_3$. We discuss how to count zero modes in these models and determine the gauge symmetry of the associated 4d physics. We further comment on turning on flat abelian connections $A$ and how these generically lift zero modes. As in section \ref{sec:Harmonic} we specialize to $M_3=S^3$.

\subsection{Combination of Witten SQMs}
\label{sec:Fusion}

We consider the Lagrangian \eqref{eq:SQMLagrangian2} with a Higgs field $\phi=\tn{diag}\,(\Lambda_K)\in \Omega^1(S^3,\mathfrak{g}_{\tn{ADE}})$ solving the sourced BPS equations \eqref{eq:AmmendedBPS} whose associated $n$-sheeted spectral cover is irreducible and cyclically branched as described in section \ref{sec:GeneralizedNonSplit}. For concreteness we furthermore restrict to Lie algebras $\mathfrak{g}_{\tn{ADE}}=\mathfrak{su}(n)$. The topology of such covers is fixed by the pairs $(L_i,F_i)$ where $L_i=\del F_i\subset M_3$ denotes the links of the branch locus and $F_i\subset M_3$ a choice of Seifert surfaces together with a cyclic monodromy action $s\in S_n$.

\begin{figure}
  \centering
  \includegraphics[width=6cm]{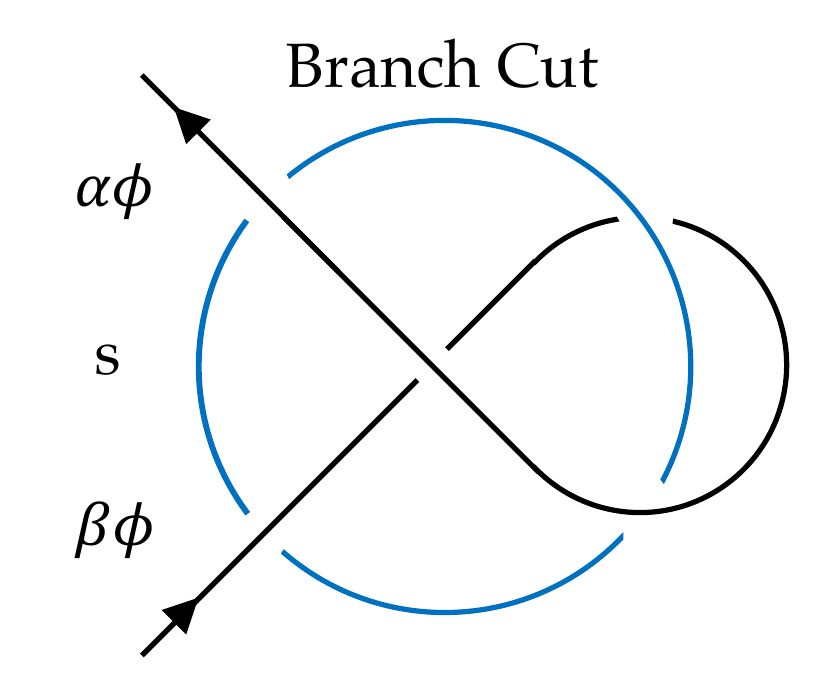}
    \caption{Sketch of a flow line (black) encircling a branch cut (blue). The incoming flow is determined by the Morse 1-form $\alpha^I\phi_I$ and the monodromy action associated to the branch circle is given by $s=-1$, then the outgoing flow is determined by $\beta^I\phi_I$. The two Witten SQMs associated to the roots $\alpha,\beta$ are coupled and combine to a single SQM.}
 \label{fig:BranchCut}
\end{figure}

We begin analysing the 1-particle sector of the colored SQM. The notion of perturbative ground states and the flow equations between these are identical to the case of non-split spectral covers, but the global structure of flow lines is altered. Along a path linking the branch locus the eigenvalues of the Higgs field are interchanged according to the monodromy action which is given by
\be\label{eq:Monodromy}
\phi\quad \rightarrow \quad g \phi g^{-1}\,, \qquad g\in SU(n)\,, 
\ee
where the element $g$ is determined by the monodromy element $s$. A particle following the flow line set by a sum of Higgs field eigenvalues $\alpha^I\phi_I$ follows a different combination of eigenvalues $\beta^I\phi_I$ after circling the branch locus and changes color. We have depicted this process in figure \ref{fig:BranchCut}. The color change is determined by the monodromy action
\be\label{eq:MonodromyColor}
E^\alpha\quad \rightarrow \quad g E^\alpha g^{-1}\,, 
\ee
and looping around the branch locus multiple times we find an orbit of generators
\be\label{eq:MonodromyOrbit}
E^{[\alpha]}=\lbbb g^k E^\alpha g^{-k}\,|\, k=0,\dots,n-1 \rbbb\,.
\ee
For a standard choice of Cartan-Weyl basis $E^\alpha$ conjugation by $g^k$ acts as a permutation of the roots $\alpha$ and we find an associated orbit of colors $[\alpha]$ to the action \eqref{eq:MonodromyOrbit}. 

The eigenvalue 1-forms of the Higgs field can be distinguished on the simply connected subspace $S^3\setminus \cup_iF_i$ and while flowing in $S^3\setminus \cup_iF_i$ the particle is of definite color. Traversing the Seifert surfaces $F_i$ the particle changes color according to \eqref{eq:MonodromyColor}. This leads to an interpretation of the Seifert surfaces as defects in the colored SQM. The wave functions of particles of definite color need not extend smoothly across the Seifert surfaces in $S^3\setminus \cup_iF_i$ but rather they are required to glue smoothly to a wave function profile on $S^3\setminus \cup_iF_i$ associated with a color prescribed by the monodromy action \eqref{eq:MonodromyColor}. Equivalently, they must glue exactly as the eigenvalues of the Higgs field in \eqref{eq:EffectiveOneForm}. By this effect particles of color $\alpha$ evolve identically to an uncolored particle probing $n$ copies of $S^3\setminus \cup_iL_i$. Each copy is associated with a color $\beta\in[\alpha]$ and the potential governing the particle is determined in the respective copy by the 1-form $\beta^I\phi_I$. Due to \eqref{eq:EffectiveOneForm} this gives a well-defined potential on the $n$-fold glued space \eqref{eq:CC} which is topologically the spectral cover $\CC$. With this the correspondence \eqref{eq:WittenSQMs} is altered to\smallskip
\be\ba\label{eq:WittenSQMs2}
E^{[\alpha]}\subset \mathfrak{su}{(n)}\qquad  &\leftrightarrow \qquad \tn{Witten's SQM on } \CC \tn{ with supercharge } \CQ=d+\Phi_{[\alpha]\,}\wedge\,,\\
H^I\in\mathfrak{su}{(n)}\qquad  &\leftrightarrow \qquad \tn{Witten's SQM on } M_3 \tn{ with supercharge } \CQ=d \,.\\ 
\ea\ee
Here the 1-form $\Phi_{[\alpha]}\in \Omega^1(\CC)$ is defined by gluing the 1-forms $\beta^I\phi_I$ across the gluing factors given in \eqref{eq:CC}.

The branch cuts of the Higgs field or equivalently its Seifert surface defects break the gauge symmetry to the stabilizer $\tn{Stab}(\phi)$ which consists of gauge transformations leaving $\phi$ invariant. They are generated by the generators $H$ of the maximal torus of the gauge group which satisfy
\be\label{eq:SymmetryRestrict}
gHg^{-1}=H\,.
\ee
For the $n$-sheeted irreducible coverings discussed in this section all of the gauge symmetry is broken. More general Higgs fields whose spectral covers have $N+1$ irreducible components have their gauge group broken to $U(1)^N$. This may enhance to include factors of $SU(k)$ if $k$ eigenvalues of the Higgs field take the same value. 

\subsection{Monodromies and Partial Higgsing}
\label{sec:GluingMonodromies}

We are interested in preserving some of the gauge symmetry and non-split Higgs field backgrounds whose spectral cover \eqref{eq:GeneralSC} has multiple components. The eigenvalues $\Lambda_K$ associated with each irreducible component of the cover can be activated successively whereby we can focus on Higgs fields where $n$ eigenvalues have been set to vanish and $m$ have been activated to trace out an irreducible $m$-sheeted cover.

We begin by consider a Higgs field background valued in the Lie algebra $\mathfrak{su}(n+m)$ for which $m$ eigenvalues are turned on as described in section \ref{sec:GeneralizedNonSplit}. This naively realizes a partial Higgsing of the gauge symmetry from $SU(n+m)$ to $SU(n)\times U(1)^m$ as discussed in section \ref{sec:RankK}. The adjoint representation breaks into representations of $SU(n)\times U(1)^m$ as
\be\ba\label{eq:braking}
\tn{Ad}\, SU(n+m)\quad\rightarrow\quad &\lb \tn{Ad}\,SU(n) \otimes {\bf 1}\rb \oplus \lb {\bf 1}\otimes \tn{Ad}\,U(1)^m\rb  \\
& \oplus \sum_{i=1}^m \lb {\bf n}_{Q_i}\oplus \overline{\bf n}_{-Q_i}\rb \oplus\sum_{j=1}^{m^2-m} {\bf 1}_{Q_j} \,.
\ea\ee
Here we denote the fundamental representation of $SU(n)$ by ${\bf n}$ and both $Q_i,Q_j$ are charge vectors of $U(1)^k$. There are $m$ pairs of the fundamental representation of $SU(n)$ and $m(m-1)$ trivial representations charged under $U(1)^k$. 

Monodromy effects \eqref{eq:SymmetryRestrict} now break the gauge symmetry to $SU(n)\times U(1)$ and the colored SQM now groups the representations in \eqref{eq:braking} into representations of this reduced gauge symmetry. The $m$ pairs of fundamental representations ${\bf n}_{Q_i}, \overline{\bf n}_{-Q_i}$ belong to the same monodromy orbit of colors with length $m$ \eqref{eq:MonodromyOrbit} and combine to a single pair of fundamental representations ${\bf n}_+,\overline{\bf n}_-$ of the gauge symmetry $SU(n)\times U(1)$. Similarly the $m(m-1)$ trivial representations are grouped into $(m-1)$ trivial representations which are uncharged under the new gauge group. The $m$ representations $\tn{Ad}\,U(1)^m$ combine to $\tn{Ad}\,U(1)$. The latter follows from the common geometric origin of the Higgs field $\phi$ and the connection $A$. The gauge fields valued in $\tn{Ad}\,U(1)^m$ are in correspondence with the $m$ activated Higgs field eigenvalues. They are constrained to glue in the same way as the eigenvalues \eqref{eq:EffectiveOneForm} across the branch sheets and are not independent. Summarizing we find that the monodromy effects 
lead to following representation content
\be\ba\label{eq:braking2}
&\lb \tn{Ad}\,SU(n) \otimes {\bf 1}\rb \oplus \lb {\bf 1}\otimes \tn{Ad}\,U(1)^m\rb  \oplus\sum_{j=1}^{m^2-m} {\bf 1}_{Q_j} \oplus \sum_{i=1}^m \lb {\bf n}_{Q_i}\oplus \overline{\bf n}_{-Q_i}\rb\\
 \rightarrow \qquad & \lb \tn{Ad}\,SU(n) \otimes {\bf 1}\rb \oplus \lb {\bf 1}\otimes \tn{Ad}\,U(1)\rb  \oplus\sum_{k=1}^{m-1} {\bf 1}_0^{(k)} \oplus \lb {\bf n}_{+}\oplus \overline{\bf n}_{-}\rb
\ea\ee
of the reduced gauge symmetry group $SU(n)\times U(1)$. The raised superscript ${\bf 1}^{(k)}_0$ is introduced to distinguish the $m-1$ uncharged trivial representation.

We check these results by considering the circle reduction of M-theory on the ALE geometry set by the Higgs field background to the IIA set-up. This is given by $n+m$ D6-branes of which $m$ have been Higgsed leaving a stack of $n$ coincident branes. The $m$ D6-branes recombine into a single D6-brane which explains the gauge symmetry reduction to $SU(n)\times U(1)$. Further this interpretations explains the single pair of fundamental representations ${\bf n}_+,\overline{\bf n}_-$ which correspond to the open string sector between the stack of $n$ D6-branes and the recombined, Higgsed D6-brane. The modes in the uncharged trivial representations originate from the self-intersection of the Higgsed D6-brane.

The colored SQM now further determines a simplification of the cohomology groups $H_{\CQ}^*(S^3,\mathfrak{g}_{\tn{ADE}})$ with $\mathfrak{g}_{\tn{ADE}}=\mathfrak{su}(n+m)$ which determine the 4d spectrum \eqref{eq:Cohomologies}. The spectral cover is the union of $n$ copies of the zero section in $T^*S^3$ and the Higgsed eigenvalues which sweep out the irreducible 3-manifold $\CC\subset T^*S^3$ given topologically by
\be\label{eq:GluedCover23}
\CC=\lbb S^3\setminus\lb \bigcup_i F_i \rb \rbb_1 \# ~\dots~ \# \lbb S^3\setminus\lb \bigcup_i F_i \rb\rbb_m\setminus (\cup_i L_i)\,.
\ee
For further details we refer to section \ref{sec:GeneralizedNonSplit} and appendix \ref{app:Homology}. We discuss the zero mode counting for each summand of \eqref{eq:braking2} in turn. The fields transforming in $ \tn{Ad}\,SU(n) \otimes {\bf 1}$ are not effected by the Higgs field background and the relevant zero modes in the reduction on $M_3$ are counted by the de Rham cohomology groups $H_{\tn{dR}}^*(S^3,\R)$. The fields transforming in ${\bf 1}\otimes \tn{Ad}\,U(1) $ commute with the Higgs field, but, as explained above, zero modes are counted by the de Rham cohomology groups $H^*_{\tn{dR}}(\CC,\R)$. The fields transforming in the $m-1$ uncharged trivial representations ${\bf 1}^{(k)}$ are similarly effected by the branch cuts. Such representations resulted from combining $m$ charged representations ${\bf 1}_{Q_j}$ and the relevant Higgs field for each of these is given by $Q_j^I\phi_I$. The charge vectors $Q_j$ are nothing but the roots $\alpha_j$ of $\mathfrak{su}(m)$ and the glued representations ${\bf 1}_{Q_j}$ precisely fit into a color orbit of the monodromy action \eqref{eq:MonodromyColor}. The sum $\sum_{k=1}^{m-1} {\bf 1}^{(k)}=\sum_{[\alpha]} {\bf 1}^{[\alpha]}$ in \eqref{eq:braking2} is equivalently expressed as a sum over color orbits. 
The $m$ 1-forms $Q_j^I\phi_I$ associated with the color orbit $[\alpha]$ glue across the $m$ factors in \eqref{eq:GluedCover23} to the 1-form $\Phi_{[\alpha]}$ on the gluing space $\CC$. As a consequence zero modes are counted by the Novikov cohomology groups $H^*_{\tn{Nov.}}(\CC,\Phi_{[\alpha]})$. The fields transforming in ${\bf n}_+$ are identically argued to be counted by $H^*_{\tn{Nov.}}(\CC,\Phi_{[\beta]})$ where $\beta$ is a positive root of $\mathfrak{su}(n+m)$ that is neither a root of the subalgebras $\mathfrak{su}(n)$ or $\mathfrak{su}(m)$. Of course there are many different (precisely $nm$) such roots but due to the degeneracy explained in section \ref{sec:RankK} all such roots yield the same 1-form $\Phi_{[\beta]}$. Zero modes transforming in $\overline{\bf n}_-$ are simply counted by the groups $H^*_{\tn{Nov.}}(\CC,-\Phi_{[\beta]})$. Due to its distinguished role we denote $\Phi_{[\beta]}$ simply by $\Phi$. 

We can now generalize \eqref{eq:CohoNonSplit} for partial Higgsings with non-split spectral covers. For the Lie algebra $\mathfrak{g}=\mathfrak{su}(n+m)$ and monodromy orbits $[\alpha]$ of $\mathfrak{su}(m)$ we have, counting with multiplicities,
\be\ba\label{eq:ZeroModeCount}
H^*_{\CQ}(S^3,\mathfrak{g})&=\lb \bigoplus_{i=1}^{n^2-1} H^*_{\tn{dR}}(S^3,\R)\rb \oplus H^*_{\tn{dR}}(\CC,\R) \\ 
&~~~\,\oplus \lb \bigoplus_{[\alpha]} H^*_{\tn{Nov.}}(\CC,\Phi_{[\alpha]}) \rb \oplus \lb   \bigoplus_{\,k=1}^{m}  \lbb H^*_{\tn{Nov.}}(\CC,\Phi) \oplus H^{*}_{\tn{Nov.}}(\CC,-\Phi) \rbb \rb\,.
\ea\ee

More generally we can consider other ADE gauge groups and turn on Higgs fields similarly as above. Consider for example the gauge symmetry breaking $E_8\rightarrow SU(5)_{\tn{GUT}}\times SU(5)_\perp$ where the Higgs field is turned on along $SU(5)_\perp$. Such a breaking is described by a five sheeted spectral cover of $SU(5)_\perp$ traced out by the non-vanishing eigenvalues of the Higgs field. The Higgsing is a special case of
\be\ba\label{eq:RepsE8}
E_8\quad &\rightarrow \quad SU(5)_{\tn{GUT}}\times SU(5)_\perp\\
{\bf 248}\quad &\rightarrow \quad \lb {\bf 24}, {\bf 1}\rb \oplus  \lb{\bf 1}, {\bf 24}\rb \oplus \lb {\bf 10}, {\bf 5}\rb \oplus\lb\overline{\bf 5}, {\bf 10}\rb\oplus \lb\overline{\bf 10}, \overline{\bf 5}\rb \oplus \lb{\bf 5}, \overline{\bf 10}\rb\,,
\ea\ee
for which $SU(5)_\perp$ is further reduced to $U(1)$ when taking the eigenvalues of the Higgs field to trace out an irreducible 5-fold covering. The representations of $SU(5)_{\tn{GUT}}$ follow from orbits of the Weyl group action $S_5$ on the representations in \eqref{eq:RepsE8} of $SU(5)_\perp$. The Higgs field breaks $SU(5)_\perp$ naively to $S[U(1)^5]$, the spectrum transforming under $SU(5)_{\tn{GUT}}\times U(1)$ then follows as in \eqref{eq:braking2}. We normalize the $U(1)$ charge of the fundamental representations ${\bf 5}$ to unity and then find following spectrum transforming under the gauge symmetry $SU(5)_{\tn{GUT}}\times U(1)$
\be\ba\label{eq:RepsE2}
 {\bf 24}_0 \oplus  \lb {\bf 1}_0 \oplus \sum_{k=1}^4 {\bf 1}_0^{(k)} \rb \oplus  {\bf 10}_{+1} \oplus \overline{\bf 10}_{-1}  \oplus 2\times \overline{\bf 5}_{+2}\oplus 2\times {\bf 5}_{-2}\,.
\ea\ee
The zero modes of $\CQ$ transforming in each representation are again characterized by a Higgs field on the space \eqref{eq:GluedCover23} constructed via gluing. For example the matter curves (here points) of $({\bf 10,5})$ in \eqref{eq:RepsE8} give matter transforming in the anti-symmetric representation of $SU(5)_{\tn{GUT}}$ which localizes at $\Lambda_K=0$ for the $K=1,\dots,5$ eigenvalues of the Higgs field. The eigenvalues $\Lambda_K$ glue to a 1-form $\Lambda$ on $\CC$ as in \eqref{eq:EffectiveOneForm}. The massless matter transforming in ${\bf 10}_{+1}$ of \eqref{eq:RepsE2} is therefore counted by $H^*_{\tn{Nov.}}(\CC,\Lambda)$. Similarly the massless matter in $(\overline{\bf 5},{\bf 10})$ localizes at $\Lambda_K+\Lambda_L=0$ with $K>L$. The monodromy action groups these ten 1-forms into two groups of five 1-forms which glue to the 1-forms $\Lambda_{\tn{as}}^{(1)},\Lambda_{\tn{as}}^{(2)}$ on $\CC$. The matter transforming in the two representations $\overline{ \bf 5}_{+2}$ are therefore counted by $H^*_{\tn{Nov.}}(\CC,\Lambda_{\tn{as}}^{(i)})$ with $i=1,2$.

\subsection{Example: 2-sheeted Covers and Monodromy}

\begin{figure}
  \centering
  \includegraphics[width=15.5cm]{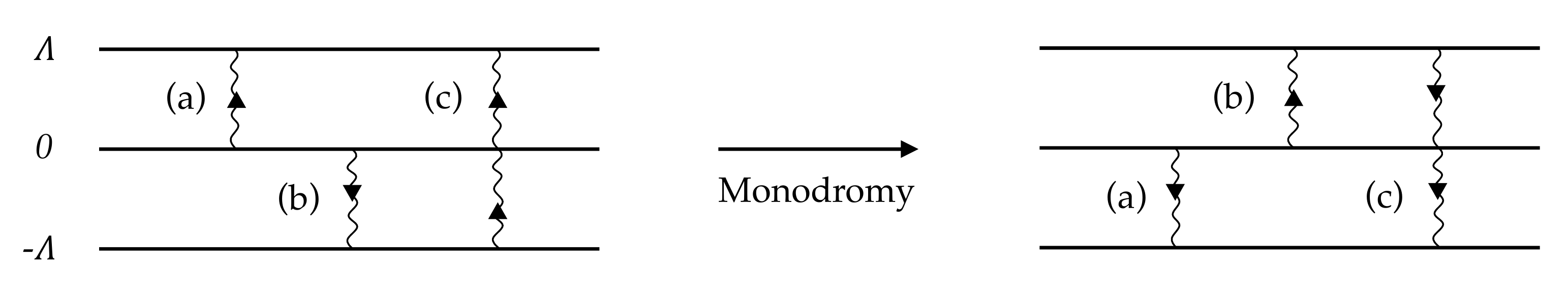}
    \caption{The picture locally shows the four D6-branes which are IIA realization of the Higgs field $\eqref{eq:SimpleHiggsCut}$. The D6-branes labeled by $\pm\Lambda$ are connected by branch sheets and along a closed path linking a branch cut locus $S_i^1$ the components of the combined D6-brane interchange. By (a,b,c) we label three open string sectors and their image when transported around the branch locus. The Chan-Paton factors of the string determine to which root of the Lie algebra $\mathfrak{su}(4)$ it is associated. The pairs of roots \eqref{eq:BulkSQM}-\eqref{eq:UnchargedSQM} are now understood as the open string sectors which are mapped onto another by the monodromy action.}
 \label{fig:BranchCut3}
\end{figure}

We give an explicit example of the effects discussed in the previous sections. Consider the family of two-sheeted covers \eqref{eq:CoverSu(2)} constructed in section \ref{sec:Example} and embed these two sheets $(\Lambda,-\Lambda)$ into an $\mathfrak{su}(4)$ valued Higgs field $\phi$ on $M_3=S^3$ as
\be\label{eq:SimpleHiggsCut}
\phi=\tn{diag}(0,0,\Lambda,-\Lambda)\,. 
\ee
Here $\Lambda$ is a 1-form with branch loci along a collection of circles $\cup_iS^1_i$ defined on $S^3\setminus \cup_i D_i$ where $D_i$ are disks realizing the branch sheets and bound by the branch locus $\del D_i=S^1_i$. With respect to the Cartan basis 
\be
H^1=\tn{diag}(1,-1,0,0)\,, \qquad H^2=\tn{diag}(0,1,-1,0)\,, \qquad H^3=\tn{diag}(0,0,1,-1)\,,
\ee
consider the six positive roots
\begin{align}\label{eq:Roots}
\alpha_1&=(2,-1,0)\,, \qquad &&\alpha_2=(-1,2,-1)\,, \qquad  && \alpha_3=(0,-1,2)\,,\\
\alpha_4&=(1,1,-1)\,, \qquad &&\alpha_5=(-1,1,1)\,, \qquad &&\alpha_6=(1,0,1)\,. \label{eq:Roots2}
\end{align}

When traversing a closed path linking one of the circles $S_i^1$ the third and fourth sheet of the spectral cover are interchanged, i.e. the Higgs field $\phi$ returns to \eqref{eq:Monodromy}
\renewcommand{\arraystretch}{0.9}
\be\label{eq:Monodromyaction}
\phi\rightarrow -\phi=g\phi g^{-1}\,, \qquad g=\lb \begin{matrix} 1 & 0 & 0 & 0 \\  0 & 1 & 0 & 0 \\ 0 & 0 & 0 & 1 \\ 0 & 0 & -1 & 0 \end{matrix}\rb\in SU(4)\,,
\ee
which realizes a $\Z_2$ monodromy action. The gauge group is broken to $SU(2)\times U(1)$. The supercharge $\CQ=d+[\phi\wedge\,,\cdot\,]$ preserves the standard complexified Lie algebra generators $E^{\alpha_i}$ associated with the roots \eqref{eq:Roots} and restricts to each of the respective subspaces, in the notation of \eqref{eq:DecompQ}, to
\begin{align}\label{eq:RestrictedSC}
\CQ^{(\alpha_1)}&=d\,,\qquad && \CQ^{(\alpha_2)}=d-\Lambda\,\wedge\,,\qquad && \CQ^{(\alpha_3)}=d+2\Lambda\,\wedge\,, \\
\CQ^{(\alpha_4)}&=d-\Lambda\,\wedge\,,\qquad &&\CQ^{(\alpha_5)}=d+\Lambda\,\wedge\,,\qquad  &&\CQ^{(\alpha_6)}=d+\Lambda\,\wedge\,. \label{eq:RestrictedSC2}
\end{align}
The gauge transformation \eqref{eq:Monodromyaction} determines which copies of Witten's SQM associated with different roots of $\mathfrak{su}(4)$ combine across the branch sheets. The conjugation of \eqref{eq:Monodromyaction} acts on the positive generators of $\mathfrak{su}(4)$ as
\begin{align}\label{eq:FusionRules}
g E^{\alpha_1} g^{-1}&=E^{\alpha_1}\,, \qquad &&gE^{\alpha_2}g^{-1}=-E^{\alpha_5}\,, \qquad && gE^{\alpha_3} g^{-1}=-(E^{\alpha_3})^T=-E^{-\alpha_3}\,, \\
g E^{\alpha_4} g^{-1}&=-E^{\alpha_6}\,, \qquad &&gE^{\alpha_5}g^{-1}=E^{\alpha_2}\,, \qquad &&gE^{\alpha_6} g^{-1}=E^{\alpha_4}\,,
\end{align}
and the roots \eqref{eq:Roots} and \eqref{eq:Roots2} together with their negative copies are grouped into the color orbits 
\begin{align}
[\alpha_1]&=\lbbb \alpha_1\rbbb\,,\quad  && [-\alpha_1]=\lbbb-\alpha_1\rbbb\,, \qquad &&\big(\tn{Ad}\,SU(2)\big)\label{eq:BulkSQM}\\
[\alpha_2]&=\lbbb\alpha_2,\alpha_5\rbbb\,, \quad  &&~\:\:[\alpha_4]=\lbbb\alpha_4,\alpha_6\rbbb\,,\qquad  &&({\bf n}_+) \label{eq:fundSQM}\\
[-\alpha_2]&=\lbbb-\alpha_2,-\alpha_5\rbbb\,, \qquad && [-\alpha_4]=\lbbb-\alpha_4,-\alpha_6\rbbb\,, \qquad \qquad &&(\overline{\bf n}_-)\label{eq:antifundSQM}\\
[\alpha_3]&=\lbbb\alpha_3,-\alpha_3\rbbb  && \qquad &&({\bf 1}_0)\,. \label{eq:UnchargedSQM}
\end{align}
The twelve SQMs naively associated with the roots of $\mathfrak{su}(4)$ in \eqref{eq:WittenSQMs} consequently combine across the branch sheets to SQMs associated with the color orbits \eqref{eq:BulkSQM}-\eqref{eq:UnchargedSQM}. The generators $E^{\pm\alpha_1}$ commute with the Higgs field and give free SQMs mapping into $S^3$. The remaining color orbits contain two roots and are over \eqref{eq:WittenSQMs2} in correspondence with SQMs mapping into the target space
\be\label{eq:NewTargetSpace}
\CC=\lb S^3\setminus D\rb_1\# \lb S^3\setminus D\rb_2\setminus L\,, \qquad D=\bigcup_{i=1}^N D_i\,,\qquad L=\bigcup_{i=1}^N S^1_i\,,
\ee 
whose metric is inherited from the gluing factors. Each gluing component is associated with one of the roots in of the pairs \eqref{eq:fundSQM}-\eqref{eq:UnchargedSQM}. The 1-forms $\Lambda,-\Lambda$ glue to a single harmonic 1-form $\Phi$ on $\CC$ and consequently the supercharges \eqref{eq:RestrictedSC}, \eqref{eq:RestrictedSC2} combine in pairs to give the supercharges of the SQMs mapping into \eqref{eq:NewTargetSpace}. 

We briefly comment on the IIA string theory interpretation of the above effects. In the type IIA set-up associated with the Higgs field \eqref{eq:SimpleHiggsCut} we locally have four D6-branes of which two have combined to a connected object corresponding to the spectral cover component $\CC$. The transformations \eqref{eq:FusionRules} are then understood as open string sectors identified by the monodromy action. For instance, an open strings locally connecting the first and third D6-branes are found to connect the first and fourth D6-brane when transported around the branch locus. We depict this interpretation in figure \ref{fig:BranchCut3}. 

The monodromy orbits already fix the representation content \eqref{eq:braking2} transforming under $SU(2)\times U(1)$ which here reads
\be
\tn{Ad}\,SU(2)_0 \oplus \tn{Ad}\,U(1)  \oplus {\bf 1}_0 \oplus  {\bf 2}_{+}\oplus {\bf 2}_{-}\,,
\ee 
where the roots associated with each representation are as given in \eqref{eq:BulkSQM}-\eqref{eq:UnchargedSQM}. In figure \ref{fig:BranchCut3} the open string sectors corresponding to ${\bf 2}_{+}, {\bf 2}_{-}, {\bf 1}_0 $ are marked with $(a,b,c)$ respectively. The reflective symmetry $\theta\rightarrow \pi-\theta$ of the set-up, we refer to section \ref{sec:Example}, requires all instanton effects potentially lifting perturbative ground states to come in pairs and cancel. All perturbative ground states are therefore ground states of the colored SQM and their count determines the cohomologies in \eqref{eq:ZeroModeCount}. These are localized at the zero of the Higgs field, counted in \eqref{eq:zerosHiggs}, and therefore the Novikov cohomology groups evaluate to 
\be\label{eq:NovKovGroups}
H^*_{\tn{Nov.}}(\CC,\Phi)=\lbbb 0, \R^{l-2}, \R^{l-2}, 0 \rbbb\,,
\ee
where $l$ is the number of disks participating in the gluing construction \eqref{eq:NewTargetSpace}. The Novikov groups for the 1-form $\Phi_{[\alpha_3]}=2\Phi$ on $\CC$ of the color orbit $[\alpha_3]$ also evaluate to \eqref{eq:NovKovGroups}.

\subsection{Flat Abelian Higgs bundles with Split Covers}
\label{sec:Connection}

The most general supersymmetric vacua of the 7d SYM solving \eqref{eq:BPS} are solutions with non-vanishing connections. As a precursor to analysing these we consider split and non-split Higgs bundles for which a flat abelian connection $A$ along the Cartan subalgebra $\mathfrak{h}_{\tn{ADE}}$ of the gauge algebra $\mathfrak{g}_{\tn{ADE}}$ has been turned on. Properties of these configurations have been explored in \cite{PhDBarbosa, Barbosa:2019hts}. The Higgs field $\phi$ and connection $A$ have a common geometric origin whereby we restrict to abelian configurations for $A$ which do not break the gauge symmetry further from the initial Higgsing. We have $A=A_IH^I$ where the Cartan components $A_I$ are only non-vanishing if the Cartan components $\phi_I$ is also non-vanishing. Further, branch cut structures of the Higgs field $\phi$ and connection $A$ agree when present. The two fields commute $[\phi,A]=0$ whereby the BPS equations are unaltered. The gauge bundle $\tn{ad}\,P_{\tn{ADE}}$ decomposes into line bundles 
\be
\tn{ad}\,P_{\tn{ADE}}=M_3\times \mathfrak{h}_{\tn{ADE}} \oplus \lb \bigoplus_{\alpha} L_\alpha\rb\,,
\ee
where $\alpha$ runs over the roots of the Lie algebra $\mathfrak{g}_{\tn{ADE}}$.

The supercharge associated to the Lie algebra generator $E^\alpha$ now takes the form $Q^{(\alpha)}=d+\alpha^I(t\phi+iA)_{I}\wedge$ and perturbative ground states are approximated well by
\be\label{eq:PerturbGSWithAbConnection}
\ket{x_A,\lambda^\alpha,|K_A|}=\exp\lbb - t|c^{(\alpha)}_k| (x^k)^2 + i(-1)^{k_\alpha} \alpha^IA^k_I(x_A)x_k\rbb dx^{K_A} \otimes E^\alpha
\ee
Here we have written locally $\alpha^I\phi_I=df^{(\alpha)}$ and expanded $f^{(\alpha)}=c^{(\alpha)}_k(x^k)^2+\CO(|x|^3)$ in normal coordinates centered at $x_A$\,. The gauge field is closed $dA=0$ but need not vanish at a zero of the Higgs field $\phi_Q$\,. We approximate it therefore locally as $\alpha^IA_I=dg^{(\alpha)}$ with $g^{(\alpha)}=\alpha^IA^k_I(x_A)x_k+\CO(|x|^2)$\,. The exponential in \eqref{eq:PerturbGSWithAbConnection} then subsumes these leading order approximations up to signs. Here $K_A\subset \lbbb 1,2,3\rbbb$ is subset of indices $k$ for which $c_k^{(\alpha)}<0$ and $k_\alpha=0$ if $k\in K_A$ and 1 otherwise. The index $A$ runs over the perturbative ground states of color $\alpha$. 

\begin{figure}
  \centering
  \includegraphics[width=5cm]{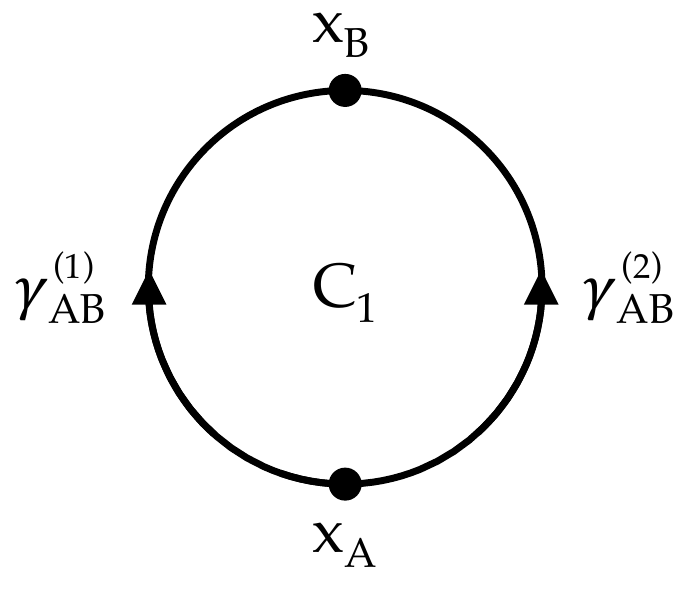}
    \caption{The picture shows two perturbative ground states localized at points $x_A,x_B\in M_3$ which are connected by two instantons along the paths $\gamma_{AB}^{(1)},\gamma_{AB}^{(2)}$. The paths wrap a 1-cycle $C_1$. In the absence of a connection the non-perturbative contributions to the matrix element of the supercharge between the perturbative ground states at $x_A,x_B$ cancel. If a connection is turned on then these contributions are shifted by a phase, no longer cancel and ground states are lifted.}\label{fig:WilsonLine}
\end{figure}

The supercharge acts on the perturbative ground states \eqref{eq:PerturbGSWithAbConnection} as 
\be\ba\label{eq:SuperchargeActionWithAbConnection}
\CQ\ket{x_A,\lambda^\alpha,|K_A|}&=\sum_{\gamma_{AB}}(\pm)_{\gamma_{AB}}\exp\lbb -\int_{\,x_A}^{\,x_B} \alpha^I\lb t\phi_{I}+iA_I\rb\big|_{\gamma_{AB}}\rbb\ket{x_B,\lambda^\beta,|K_A|+1}\\&~~~\, +\CO(1/t)\,,
\ea\ee
which follows from a computation similar to the one leading to the canoncial result \eqref{eq:SuperChargeAction} in the appendix. Here $\gamma_{AB}$ is a flow line of $\alpha^I\phi_I$ connecting the two sites of localization $x_A,x_B\in M_3$ and the signs $(\pm)_{\gamma_{AB}}$ again originates from fermion determinant in the path integral computation. If the Higgs field exhibits branch cuts then these can be looped by $\gamma_{AB}$ and ground states of different color $\lambda^\alpha \neq \lambda^\beta$ can connect. The tunnelling is again restricted to be between states with a relative fermion number of $1$. Note that compared to the result \eqref{eq:SuperChargeAction} for vanishing connections $A=0$ we have crucially picked up a Wilson line
\be\label{eq:WilsonLine}
W_\alpha(\gamma_{AB})=\exp\lb -i\int_{\gamma_{AB}}\alpha^IA_I\rb\,,
\ee
which shifts the boundary map of the associated Morse-Witten complex by a phase.

The Wilson line \eqref{eq:WilsonLine} obstructs the cancellation of instanton contributions. Consider a non-trivial 1-cycle $C_1$ contained in $M_3$ or more generally in the covering space $\CC$ with respect to which the connection $A$ has non-trivial holonomy $\int_{C_1}A\neq 0$. Assume that the Higgs field $\alpha^I\phi_I=df^{(\alpha)}$ is such that it has two vanishing points $x_A,x_B$  connected by two flow lines $\gamma_{AB}^{(1)},\gamma_{AB}^{(2)}$ which wrap $C_1$ as depicted in figure \ref{fig:WilsonLine} and have opposition signs, 
\be
(\pm)_{\gamma_{AB}^{(1)}}=+1\,, \qquad (\pm)_{\gamma_{AB}^{(2)}}=-1\,.
\ee
We sketch this set-up in figure \ref{fig:WilsonLine}. The action of the supercharge on the perturbative ground state $\ket{x_A,\lambda_A,|I|}$ localized at $x_A$ is computed to leading order by
\be\ba\label{eq:Superchargeaction}
\CQ\ket{x_A,\lambda_A,|K_A|}&=c\lbb \exp \lb  -i\int_{\gamma_{AB}^{(1)}}\alpha^IA_I\rb-\exp \lb  -i\int_{\gamma_{AB}^{(2)}}\alpha^IA_I\rb \rbb\ket{x_B,\lambda_B,|K_A|+1}\\
&=c\exp \lb  -i\int_{\gamma_{AB}^{(1)}}\alpha^IA_I\rb\lbb 1-\exp \lb  i\int_{C_1}\alpha^IA_I\rb \rbb\ket{x_B,\lambda_B,|K_A|+1}
\ea\ee
where we introduced the constant $c=\exp\lb {tf^{(\alpha)}(x_A)-tf^{(\alpha)}(x_B)}\rb $\,. Unless the holonomy $\int_{C_1}\alpha^IA_I=2\pi n$ is an integer multiple of $2\pi$ turning on $A$ will break the cancelation between the instanton effects and both $\ket{x_A,\lambda_A,|K_A|},\ket{x_B,\lambda_B,|K_B|}$ are no longer ground states. The prefactor $\exp \lb  -i\int_{\gamma_{AB}^{(1)}}A_Q\rb$ can be gauged to vanish.

\section{Effective 4d Physics}
\label{sec:4d}

So far we have focussed on understanding the zero modes of the operator $\CQ$ on $M_3$ given in \eqref{eq:ComplexifiedConnection} and identified an equivalent formulation of these as the ground states of a colored quantum mechanics. We now turn to discuss the KK-reduction on $M_3$ and the effective 4d gauge theory. The relevant scales for the reduction of the partially twisted 7d SYM are the volume $\tn{Vol}\,M_3$ setting the mass scale $M_{KK}$ and the mass scale $M_{\phi}$, set by an appropriate average of the Higgs field, marking the scale at which the gauge symmetry is broken. 

In addition to the massless sector we also find two sets of light modes. The M2-brane 
induce masses for perturbatively massless fields which are exponentially below $M_{KK}$ and the torsion factors \eqref{eq:Homologies} in the homology groups of the spectral covers yields modes below $M_{\phi}$ \cite{Wen:1985qj, freed1986,  Marchesano:2006ns, Camara:2011jg}. After discussing these two sets of light modes we briefly give a summary of the 4d effective physics adapting results in \cite{Braun:2018vhk} to the set-up at hand.

\subsection{Instantons, Torsion and Light Modes}
\label{sec:LightModes}

The light modes in the KK-reduction of the 7d SYM on $M_3$ are given by approximate zero modes, which receive non-perturbative contributions to their masses through M2-brane instantons, and modes resulting from an expansion of the 7d fields in the generators of the torsion cohomology classes of the sepectral cover. We discuss each in turn beginning with modes receiving instanton corrections.

The ADE singularity in the ALE fibration enhances at each zero of the Higgs background \eqref{eq:SingEnhance} and M2-branes wrapped on the collapsing vanishing cycle contribute an approximate zero mode $\chi_A$ on $M_3$. These transform in non-adjoint representation of the unbroken gauge symmetry and contribute matter in 4d. Their masses are set by the non-vanishing eigenvalues of the mass matrix \eqref{eq:OverLapIntegralsMass} which we reproduce for convenience here
\be\label{eq:RepMassMatterMatrix}
M_{AB}=\int_{M_3}\braket{\chi_A,\CQ\chi_B}\,, \qquad M_{AB}\sim M_{KK}\exp\lb-t\braket{\tn{Vol}\,S^3} \rb\,.
\ee
The associative 3-spheres $S^3$ are traced out by a single vanishing cycle in the ALE fibration and with $\braket{\tn{Vol}\,S^3}$ we denote their average volume. In our analysis we computed the non-perturbative effects of M2-branes wrapped on these 3-cycles to leading order in $1/t$, see e.g. \eqref{eq:YukawaEvaluatedText}. In this large $t$ limit the modes \eqref{eq:RepMassMatterMatrix} become increasingly light and should be integrated out effecting the running of the gauge coupling at low energies.

We now discuss the light modes originating from possible torsion factors in the 
in first homology group of a component $\CC$ of its spectral cover
\be\label{eq:TorHom}
\tn{Tor}\,H_1(\CC,\Z)=\Z_{m_1}\oplus \dots \oplus \Z_{m_p}\,.
\ee
These torsion factors generally have two origins. For split spectral covers they follow from those of the base manifolds as in this case each spectral cover component is one to one covering of the base away from the singularities of the Higgs field. On the other hand, for non-split spectral covers additional torsion factors can originate form the branch cut structure of the cover as demonstrated in \eqref{eq:Homologies}. We study the effect of torsion from the view point of the partially twisted 7d SYM from which the 4d physics follows via a KK reduction. Hodge theory does not give harmonic forms for each factor in \eqref{eq:TorHom} but it is still possible to associate $p$ 1-forms $\alpha_i$ and closed 2-forms $\beta_i$ with each torsion generator. These satisfy
\be\label{eq:Relationt}
d\alpha_i=(L^{-1})_i^{~j}\beta_j\,, \qquad (L^{-1})_i^{~j}\in \Z\,,\qquad \int_{\CC}\alpha_i\wedge \beta_j=\delta_{ij}\,, \qquad i,j=1,\dots,p\,.
\ee
where $L$ is the linking form on $H_1(\CC,\Z)$. We give further details in appendix \ref{sec:Torsion}. These $2p$ forms span eigenspaces of the Laplace operator and are characterized by a positive definite mass matrix $M$ as
\be
\Delta \alpha_i=-M_{i}^{~j}\alpha_j\,,\qquad\Delta \beta_i=-\widetilde M_{i}^{~j}\beta_j\equiv-(LML^{-1})_{i}^{~j}\beta_j\,. 
\ee
The forms $\alpha_i,\beta_j$ are the lightest massive eigenvectors of the Laplacian on $\CC$ with mass scales below the scale $M_\phi$ \cite{Marchesano:2006ns, Camara:2011jg}. In the case of non-split spectral covers the torsion groups \eqref{eq:TorHom} enter the reduction through \eqref{eq:ZeroModeCount} where we originally considered real coefficients to count zero modes. The cohomology group $\tn{Tor}\,H^1_{\tn{dR}}(\CC,\Z)$ now additionally contributes the forms $\alpha_i$ in \eqref{eq:Relationt} for the expansion of the $U(1)$ gauge field $A_{7d}$ of the 7d SYM associated with the irreducible spectral cover component $\CC$. The standard KK-expansion
\be
A_{7d}=A_{4d}+\sum_{i=1}^p\rho_i\alpha_i+\dots\,,
\ee
gives $p$ scalars $\rho_i$ which by supersymmetry complete into light chiral multiplets in 4d.

\subsection{4d Gauge Symmetry, Matter Content and Superpotential}

We summarize the structures determining the 4d gauge symmetry, matter content and superpotential. Instrumental in our analysis is the colored $\CN=2$ supersymmetric quantum mechanics \eqref{eq:SQMLagrangian} which fully determines the effective 4d physics from the partially twisted 7d SYM compactified on $M_3$. Here we consider a non-split Higgs field background on $S^3$ whose spectral cover $\CC=\cup_k\CC_{k}$ consists of $N+1$ components with a Higgs field breaking the gauge symmetry locally to $G_{\tn{GUT}}\times U(1)^m\subset G_{\tn{ADE}}$. We discussed set-ups of this kind in section \ref{sec:NonSplit} and for these we determined the following 4d physics.

\begin{enumerate}
\item Gauge Symmetry: The gauge symmetry is broken by monodromy effects \eqref{eq:SymmetryRestrict} to the stabilizer of the Higgs field $G_{\tn{GUT}}\times U(1)^N$. In the SQM picture the wave functions associated with the gauge fields of the naive gauge group $U(1)^m$ are constrained to glue consistently across the Seifert surfaces breaking the gauge symmetry to $U(1)^N$. In the local $G_2$ geometry the differences $\CC_{k}-\CC_{l}$ lift to 5-cycles in the ALE geometry which are Poincar\'e dual to $N$ independent 2-forms. Expanding the 11d supergravtiy 3-form in these yields $N$ abelian gauge fields. In the IIA set-up for $G_{\tn{ADE}}=SU(n)$ the Higgsed sheets of the spectral cover descend to $N$ independent D6-branes and each contributes an abelian factor to the gauge symmetry.
\item Massless Matter: 4d $\CN=1$ vector and chiral multiplets are counted by the cohomology groups \eqref{eq:Cohomologies}. For split and non-split spectral covers considered in this paper these are computed to \eqref{eq:CohoNonSplit} and \eqref{eq:ZeroModeCount} respectively. In the SQM picture these cohomologies count the ground states, alternatively they can be derived from the perturbative ground states corrected by flow line instanton effects and are computed by the kernel of the mass matrix \eqref{eq:OverLapIntegralsMass}. In the local $G_2$ geometry the perturbative ground states are in correspondence with M2-branes wrapping vanishing cycles. These yield light states in 4d and Euclidean M2-branes wrapping associative 3-spheres traced out by the vanishing cycles determine their masses. In a IIA set-up we have perturbatively massless states originating from open strings localized at the intersection of D6-branes and non-perturbative mass contributions via world sheet instantons.

For example, consider the breaking $E_8\rightarrow SU(5)_{\tn{GUT}}\times SU(5)_\perp$ which is further broken to $SU(5)_{\tn{GUT}}\times U(1)$ by a Higgs field with an irreducible 5-sheeted cover $\CC$ as considered in section \ref{sec:GluingMonodromies}. The eigenvalues of the Higgs field glue across the branch surfaces, along which the sheets of the cover mix, to a 1-form $\Lambda\in \Omega^1(\CC)$. For instance, massless matter transforming in ${\bf 10}_{+1}$ of \eqref{eq:RepsE2} is then counted by the Novikov cohomology groups
\be\ba\label{eq:MasslessMatter}
\tn{Chiral multiplets in ${\bf 10}_{+1}$\,:}&\qquad H^1_{\tn{Nov.}}(\CC,\Lambda)\,, \\
\tn{Conjugate-Chiral multiplets in $\overline{\bf 10}_{-1}$\,:}&\qquad H^2_{\tn{Nov.}}(\CC,\Lambda)\,.
\ea\ee
Other matter content is counted similarly. The chiral index for these representations is given by
\be
\chi({{\bf 10}_{+1}}\,\oplus\, \overline{\bf 10}_{-1})= H^1_{\tn{Nov.}}(\CC,\Lambda)-H^2_{\tn{Nov.}}(\CC,\Lambda)
\ee
and may be non-vanishing as the spectral cover $\CC$ is non-compact. In addition to these massless modes there are light modes below the two scales $M_{KK},M_{\phi}$ of the set-up, see section \ref{sec:LightModes}.

\item Superpotential: Yukawa couplings for massless matter are determined by the generalized Y-shaped instantons of the colored SQMs which connect three of its perturbative ground states. The contribution of a single such instanton explicitly reads \eqref{eq:YukawaEvaluatedText}. Globally the set of instantons is constrained to encode a cup product \eqref{eq:YukawaMaptext}. In the local $G_2$-manifold these interactions are due to euclidean M2-brane instantons wrapped on associative 3-spheres traced out by three linearly dependent vanishing cycles, as depicted on the right hand side in figure \ref{fig:Relevant3Spheres}. In the IIA set-up these descend again to world sheet instantons.
\end{enumerate}

\section{Conclusion and Outlook}
\label{sec:CONC}

In this paper we studied M-theory on local $G_2$-manifolds via an effective description as a partially twisted 7d SYM theory. We determined the gauge group of the resulting 4d $\CN=1$ gauge theory and its matter content. The 4d superpotential in this reduction is generated non-perturbatively by Euclidean M2-brane instantons. We characterized the low energy avatars of these M2-brane instantons in the 7d SYM theory and found a one to one correspondence with the instantons of a colored supersymmetric quantum mechanics. This correspondence allowed for the computation of superpotential contributions of individual M2-brane instantons from the perspective of the 7d SYM and constrained the set of all M2-brane instantons to encode topological structures. The latter can not be read off from the geometry and requires the colored SQM or equivalently the Higgs bundle of in the 7d SYM to make manifest. The SQM serves as a computational tool in the reduction to 4d which, given a Higgs bundle associated with local $G_2$-manifold, outputs the data of the 4d theory. This data is given in terms of cohomology groups and operations on these. The SQM is not straightforwardly suited for constructing interesting Higgs bundles or quantifying the relevant cohomological structures and it is in these two challenges which future research opportunities lie:
\begin{enumerate}
\item Construction of Higgs Bundles: The class of TCS $G_2$-manifolds suggests that large number of Higgs fields $\phi=\diag\lb \Lambda_K \rb$ with split spectral covers exist. These are solutions to the equations 
\be\label{eq:AmmendedBPS2}
d\Lambda_K=*j_K\,, \qquad *\:d*\Lambda_K=\rho_K\,,
\ee
and additionally satisfy gluing conditions in the presence of branch cuts, see section \ref{sec:Harmonic}. Here $j_K,\rho_K$ are sources supported on codimension $\geq1$ subloci of the base manifold. It would be interesting to classify possible source terms and therefore local models as they are considered in the physics literature.

\item Construction of Complex Flat Connections: The partially twisted 7d SYM with gauge group $G_{\tn{ADE}}$ allows for a larger class of vacua which are not necessarily solutions to \eqref{eq:AmmendedBPS2}. Rather they are solutions to the BPS equations
\be\label{eq:BPSConc}
iF_A+[\phi,\phi]=0\,, \quad d_A\phi=0\,,\qquad d_A* \phi=0\,,
\ee
possibly extended by source terms on the right hand side. These equations state that the operator $\CQ=d+\phi+iA$ is a complex flat $\mathfrak{g}_{\tn{ADE},\C}$ connection and solutions to \eqref{eq:BPSConc} are more loosely referred to as T-branes. The system \eqref{eq:BPSConc} also arises as the BPS equations of a partially twisted 5d $\CN=2$ SYM in the context of the 3d-3d correspondence \cite{Dimofte:2010tz, Dimofte:2011ju, Chung:2014qpa, Gukov:2016gkn}, for an overview see \cite{Dimofte:2014ija}. Here complex flat connections have been intensely studied. In a compactification program we are interested in cases in which the moduli space of complex flat connections is finite dimensional, as e.g. studied for \cite{Falbel2015DimensionOC, Abouzaid2017ASM} when $\mathfrak{g}_{\tn{ADE},\C}=\mathfrak{sl}(2,\C)$. Isolated vacua, as for example studied in the setting of knot complements, do not yield phenomenologically interesting 4d physics.

The problem of solving \eqref{eq:AmmendedBPS2} and \eqref{eq:BPSConc} for singular configurations with source terms can alternatively be formulated as a problem on a manifold $\CM_3$ with boundary $\del \CM_3$ where the boundary follows from excision of the source terms. Solutions are then characterized as complex flat connection on $\del M_3$ which can be extended throughout the bulk of the manifold. The 4d physics associated with these configurations would be worthwhile exploring.

\item Computation of Cohomology Groups: Given solutions to \eqref{eq:AmmendedBPS2} or more generally \eqref{eq:BPSConc} the 4d spectrum and superpotential are determined by the cohomology groups of the flat connection $\CQ=d+\phi+iA$. The colored SQM sets up a Morse-theoretic interpretation of these cohomology groups. This allowed for the evaluation of these cohomology groups, e.g. for purely electrically sourced split Higgs field brackgrounds (i.e. $j_K=0$ in \eqref{eq:AmmendedBPS2}) in \cite{Pantev:2009de,Braun:2018vhk} using \cite[Appendix D]{FarberClosed1forms}. It would be of great interest to enlarge the types of BPS configurations for which the cohomology groups of the connection $\CQ=d+\phi+iA$ can be computed.

\item Derivation of the colored SQM: the colored SQM captures all non-perturbative effects associated with M2-branes. It is natural to conjecture that it derives from an M2-brane probing the ALE-fibered $G_2$-manifold via a suitable reduction when wrapped on the vanishing cycles. The colored SQM in this work was constructed bottom-up by prescribing its supercharge and Hilbert space, a top-down derivation is currently not available in the literature.

\end{enumerate}

\noindent {\bf Acknowledgements}

It is a pleasure to thank Sebastjan Cizel, Cyril Closset, Julius Eckhard, Heeyeon Kim, Sakura Sch\"afer-Nameki for comments on the manuscript and insightful discussions. The author also benefitted from interesting discussions with Thomas Rochais and Ethan Torres. The author is supported by the Studienstiftung des Deutschen Volkes.

\appendix

\section{The Dictionary: Geometry, Gauge Theory, SQM}
\label{sec:Dictionary}

This appendix concisely summarizes the theories discussed in this paper. We list the field theoretic interpretations of the geometric data of the local $G_2$-manifold, both in 4d and 7d, as well as their SQM interpretations in table \ref{table:2}.

\vspace{10pt}
{\centering
\begin{table}
 \begin{tabular}{| c | c | c | c |}
 \hline
\makecell{ ALE-fibered\\ $G_2$-manifold}  & \makecell{ 7d $\CN=1$ \\ Twisted SYM}   & \makecell{4d $\CN=1$ \\ Effective Theory} & \makecell{$\CN=2=(1,1)$\\ Colored SQM  } \\  \hline\hline
 \makecell{Redsidual Singularity\\ along $M_3$ } & \makecell{Non-abelian\\gauge symmetry  } & \makecell{Non-abelian\\gauge symmetry  } & Bulk sector \\
 \hline
\makecell{Singularity\\ enhancement} & \makecell{Approx. localized\\ zero mode on $M_3$}  & Matter & \makecell{Perturbative\\ ground state} \\ 
 \hline
Associative $S^3$ & \tn{Flow lines of }$\phi$ & Mass terms & \makecell{Instanton and\\ differential $\del_{MW}$ } 
\\
 \hline
Associative $S^3$ & \tn{Y-Flow tree of }$\phi$ & Yukawa coupling & \makecell{Generalized instanton \\ and cup-product $\cup$ }  \\
 \hline
 Associative $S^3$ & \tn{Flow-tree of }$\phi$ & \makecell{Higher-point \\ coupling }& \makecell{Generalized instanton, \\ Massey product $m_n$ } \\
 \hline
\makecell{Globally defined\\ ALE 2-cycles} & Split spectral cover & \makecell{Maximal $\#$ of \\ $U(1)$ symmetries} &\makecell{$\dim \mathfrak{g}_{\tn{ADE}}$ \\ Witten SQMs} \\ 
 \hline
\makecell{Monodromy mixed\\ ALE 2-cycles}  & 
\makecell{Non-split spectral\\ cover, Higgs field \\ with branch cuts} & \makecell{Submaximal $\#$ of \\ $U(1)$ symmetries} & \makecell{Combination of\\ Witten SQMs}   \\ 
 \hline 
\end{tabular}\caption{List of correspondences between field theory, geometry and SQM.}\label{table:2}
\end{table}}

\section{Homology Groups of Cyclically Branched, $n$-sheeted Coverings}
\label{app:Homology}

\begin{figure}
  \centering
  \includegraphics[width=15.5cm]{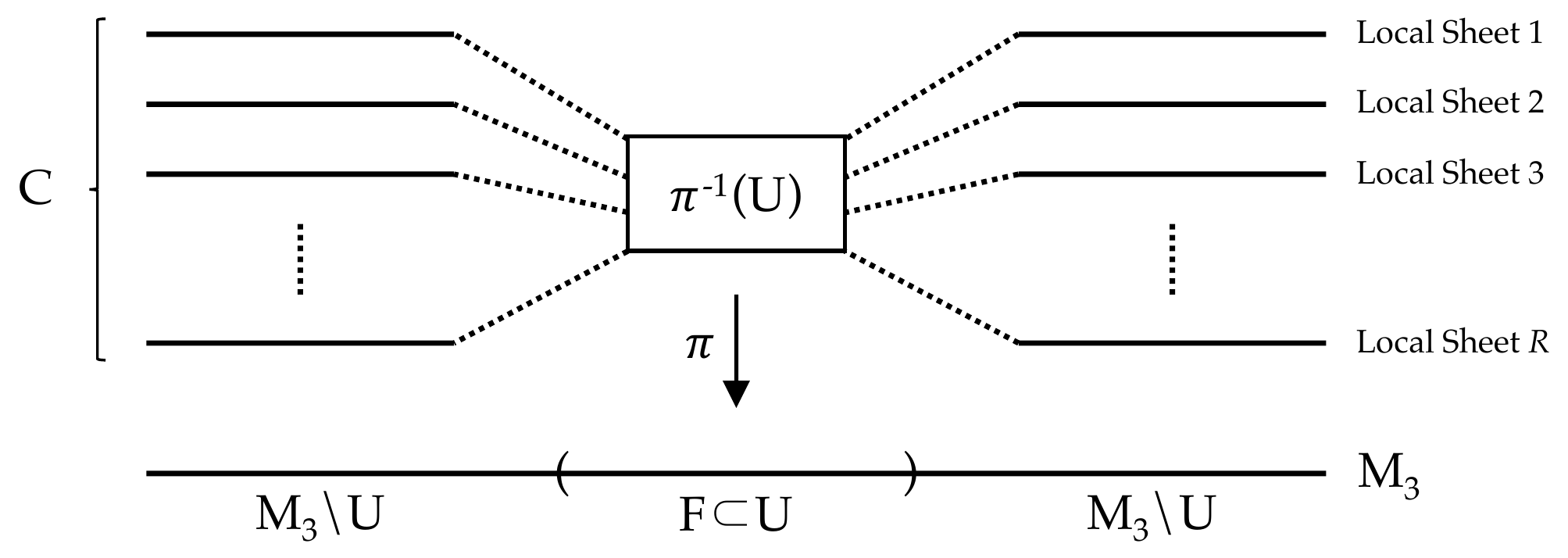}
    \caption{We sketch the covering used to compute the homology groups of non-split spectral covers using the Mayer-Vietoris sequence. The long exact sequence is applied iteratively to resolve the contributions to the homology groups originating from the branch cut structure associated to each Seifert surface $F$. The covering is given by $\tilde\CC=\pi^{-1}(U)\,\cup\,\pi^{-1}(V)$ where $U$ is a small open set containing the Seifert surface $F$ and $V$ is an open set containing $M_3\setminus U$. The intersection $U\cap V=S^2$ lifts to $n$ copies of the 2-spheres in the spectral cover $\CC$. }\label{fig:Homology}
\end{figure}

In this appendix we discuss the computation of the homology groups \eqref{eq:Homologies} using the Mayer-Vietoris sequence. Consider an $n$-sheeted covering $\pi:\tilde\CC\rightarrow M_3$ where the $n$-sheets are glued cyclically along the $l$ Seifert surfaces $F_i$ bounded by the links $L_i=\del F_i$ as given in  \eqref{eq:GluedCoverSec3}. We proceed iteratively and resolve the contributions to the homology groups originating from each link separately. Let $U$ be a small open set containing the link $L_i$ and $V$ an open set containing $M_3\setminus U$ such that their intersection is a 2-sphere $U\cap V=S^2$. We apply the Mayer-Vietoris sequence to the open sets $A=\pi^{-1}(U)$ and $B=\pi^{-1}(V)$ which intersect along $n$ copies of the 2-sphere $A\cap B=(S^2)^n$. This decomposition of the cover $\tilde \CC$ is sketched in figure \ref{fig:Homology}. The relevant long exact sequence then reads
\be\ba\label{eq:LES}
0\rightarrow
\, & H_3\lb(S^2)^{n\,},\Z\rb \rightarrow H_3(A,\Z)\oplus H_3(B,\Z) \rightarrow H_3(\tilde\CC,\Z) \rightarrow  \\
&H_2\lb (S^2)^{n\,},\Z\rb\rightarrow H_2(A,\Z)\oplus H_2(B,\Z) \rightarrow H_2(\tilde\CC,\Z) \rightarrow  \\
&H_1\lb (S^2)^{n\,},\Z\rb\rightarrow H_1(A,\Z)\oplus H_1(B,\Z) \rightarrow H_1(\tilde\CC,\Z)  \rightarrow  \\
&H_0\lb (S^2)^{n\,},\Z\rb \rightarrow H_0(A,\Z)\oplus H_0(B,\Z) \rightarrow H_0(\tilde\CC,\Z) \rightarrow 0 \,,
\ea\ee
and starting from its third line one extracts, using that 2-spheres are simply connected, the exact sequence
\be
0\rightarrow H_1(A,\Z)\oplus H_1( B,\Z) \rightarrow H_1(\tilde\CC,\Z)  \rightarrow  \Z^n \rightarrow\Z  \oplus \Z \rightarrow \Z \rightarrow 0\,,
\ee
which in turn yields
\be\ba\label{eq:FirstCohoCov}
H_1(\tilde\CC,\Z)&=\Z^{b_1(B)+n-1}\oplus \tn{Tor}(H_1(A,\Z)\oplus H_1(B,\Z))\,.
\ea\ee
The homology group $\tn{Tor}\,H_1(A,\Z)$ is a topological invariant of the link $L_i\subset A$ and discussed and tabled in \cite{knots, KnotTable}. We denote this homology group by $H_1^{(n)}(L_i)$. Next we apply the Mayer-Vietoris sequence to a decomposition of $B=A'\cup B'$ where $A'$ projects to a small neighbourhood $U'$ containing a new link $L_j$ and the open set $B'$ projects to an open set covering $\pi(B)\setminus U'$. This separates out, as in \eqref{eq:FirstCohoCov}, the contribution of the link $L_j$ to the first homology group of the cover $\CC$ and repeating this procedure for all links we find
\be\label{eq:Hom1}
H_1(\tilde\CC,\Z)=\Z^{(n-1)(l-1)}\oplus \bigoplus_{i=1}^l \tn{Tor}\, H_1^{(n)}(L_i) \,.
\ee
The remaining cohomology groups of the cover $\tilde\CC$ are fixed by Poincar\'e duality and the universal coefficient theorem. They are given by
\be
H_0(\tilde\CC,\Z)=\Z\,, \qquad H_2(\tilde\CC,\Z)=\Z^{(n-1)(l-1)}\,, \qquad H_3(\tilde\CC,\Z)=\Z\,. 
\ee
Next we remove a tubular neighbourhood $T_i\cong D^2\times S^1$ containing the link $L_i$ from the space $\tilde \CC$. We restrict to the case in which the link consists of a single knot $L_i=K_i$. We apply the Mayer-Vietoris sequence to the covering $\tilde\CC=\CC \cup T_i$, where $\CC=\tilde\CC\setminus T_i'$ and $T_i'$ is obtained by shrinking the radius of the disk $D^2$ in the solid torus $T_i$. The relevant long exact sequence reads
\be\ba
0 ~& \rightarrow ~\Z~ \rightarrow ~\Z ~\rightarrow ~ H_2(\CC,\Z) ~\rightarrow ~ H_2(\tilde \CC,\Z) ~\rightarrow ~ \Z^2 \\ & \xrightarrow[]{\iota} ~ \Z\,\oplus\, H_1(\CC,\Z)~\rightarrow ~ H_1(\tilde \CC,\Z)~\rightarrow ~ \Z ~\rightarrow ~ \Z^2 ~\rightarrow ~ \Z ~\rightarrow ~ 0\,,
\ea\ee
where $\iota$ embedds the $A,B$ cycle of the torus $T^2\cong T_i\cap T_i'$ into the spaces $ T_i$ and $\CC$ and has vanishing kernel. With this we extract the exact sequences
\be\ba
0 ~& \rightarrow ~\Z~ \rightarrow ~\Z ~\rightarrow ~ H_2(\CC,\Z) ~\rightarrow ~  H_2(\tilde \CC,\Z) ~\rightarrow ~ 0\,, \\
0 ~&\rightarrow ~ \Z^2 ~ \xrightarrow[]{\iota} ~ \Z\,\oplus\, H_1(\CC,\Z)~\rightarrow ~ H_1(\tilde \CC,\Z)~\rightarrow ~ \Z ~\rightarrow ~ \Z^2 ~\rightarrow ~ \Z ~\rightarrow ~ 0\,,
\ea\ee
and the cohomologies 
\be
H_2(\CC,\Z)\cong H_2(\tilde \CC,\Z)\,, \qquad H_1(\CC,\Z)=H_1(\tilde \CC,\Z)\oplus \Z\,.
\ee
Each excised circle contributes a 1-cycle and we find \eqref{eq:Homologies}.
\section{Comments on Colored SQMs}

In this appendix we show the supersymmetric invariance of the SQM Lagrangian \eqref{eq:SQMLagrangian} and derive the supercharges \eqref{eq:SQMSuperscharges}. We further discuss the quantization of the SQM, its Hamiltonian and Euclidean Lagrangian. The presented discussion is standard and in many parts parallels that of \cite{Hori:2000kt, Rietdijk:1992jp}.

\subsection{Supersymmetry Variations}
\label{app:SUSY}

We now show that the Lagrangian \eqref{eq:SQMLagrangian} is invariant under the supersymmetry transformations \eqref{eq:SQMVariations} with the associated supercharges \eqref{eq:SQMSuperscharges}. To make the computations more tractable we work in flat space $g_{ij}=\delta_{ij}$ and making use of the BPS equations \eqref{eq:BPS} the Lagrangian \eqref{eq:SQMLagrangian} can then be rewritten as
\be\ba\label{eq:LagrangianCompact}
\CL&= \frac{1}{2}\dot{x}^i\dot{x}_i+i\bar{\psi}^i\dot \psi_i+i\bar\lambda^\alpha\dot\lambda_\alpha-\dot x_i A_\lambda^i-\frac{1}{2}\phi_\lambda^i \phi_{i,\lambda}-\lb \CQ_{i}\phi_{j}\rb_\lambda \bar{\psi}^i\psi^j\,.
\ea\ee
For notation see section \ref{sec:SetUp}. The operator $\CQ_i=\del_i+[\varphi_i,\cdot\,]$ involves the complexified Higgs field $\varphi=\phi+iA$.

The variations \eqref{eq:SQMVariations} of the fermions $\lambda,\bar\lambda$ are such that the generator $\tilde T_\alpha=-ic_{\alpha\beta\gamma}\bar\lambda^\beta\lambda^\gamma$ introduced in \eqref{eq:FermionGenerator} varies as 
\be\ba\label{eq:GeneratorsTransform}
\delta \tilde T=\epsilon\bar\psi^i\lbb \varphi_i,\tilde T\rbb_\lambda+\bar\epsilon\psi^i\lbb\bar\varphi_i,\tilde T\rbb_\lambda\,,
\ea\ee
which implies
\be
\delta X_\lambda=\epsilon \bar\psi^i\lb Q_iX\rb_\lambda-\bar\epsilon \psi^i \lb \bar Q_iX\rb_\lambda\,,
\ee
for any Lie algebra valued quantity $X_\lambda\equiv X^\alpha\tilde T_\alpha$. With this the individual parts of the Lagrangian \eqref{eq:LagrangianCompact} are checked to vary as
\be\ba
\delta \lb -\frac{1}{2}\phi_\lambda^2 -\lb Q_{i}\phi_{j}\rb_\lambda \bar{\psi}^i\psi^j\rb&=i\epsilon\bar\psi^i\dot x^j \lb Q_{i}\phi_{j}\rb_\lambda+i\bar\epsilon \psi^i\dot x^j \lb \bar Q_{i}\phi_{j}\rb_\lambda \\
\delta\lb \frac{1}{2}\dot x^i\dot x_i +i\bar\psi^i\dot\psi_i\rb&=\dot\epsilon \dot x^i \bar\psi_i-\dot{\bar\epsilon} \dot x^i \psi_i+ i\bar\epsilon  \dot\psi_i \phi_\lambda^i+i\epsilon \dot{\bar\psi}_i\phi_\lambda^i \\
\delta\lb i\bar\lambda\dot\lambda-\dot x A_\lambda \rb&=i \dot{\tilde T}_\alpha\lb \epsilon \bar\psi^i\varphi_i^\alpha+\bar\epsilon \psi^i\bar\varphi_i^\alpha \rb \\ &~~~~-\delta \dot x_i A_\lambda^i-\dot x^j\lb \epsilon \bar\psi^i \lb Q_iA_j\rb_\lambda - \bar\epsilon \psi^i \lb\bar Q_iA_j\rb_\lambda \rb
\ea\ee
where we have used the BPS equation and the Jacobi-identity repeatedly. Integrating by parts and making further use of the BPS equations we derive
\be
\delta \CL=\dot\epsilon \dot x \bar\psi-\dot{\bar\epsilon} \dot x \psi-i\dot{\bar\epsilon} \psi \phi_\lambda-i\dot\epsilon \bar\psi\phi_\lambda=-i\dot\epsilon \lbb \bar\psi\lb i\dot x+\phi_\lambda \rb\rbb -i\dot{\bar\epsilon}\lbb \psi\lb -i\dot x+\phi_\lambda \rb\rbb\,,
\ee
which verifies the form of the supercharge given in \eqref{eq:SQMSuperscharges}.

\subsection{Canonical Quantization} 
\label{sec:canonicalquant}

Here we discuss the canonical quantization of the Lagrangian \eqref{eq:SQMLagrangian}. Taking Grassmann derivatives to act on the right the conjugate momenta to the fields $x,\psi,\lambda$ are found to be
\be\label{eq:ConjugateMomenta}
p^i=\pi_x^i=\dot{x}^i-A_\lambda^i+i\Gamma^i_{~jk}\bar\psi^j\psi^k\,, \qquad \pi_{\psi}=i\bar{\psi}\,, \qquad \pi_\lambda=i\bar{\lambda}\,,
\ee
which promoted to operators on a Hilbert space $\CH$ lead to the (anti-)commutation relations
\be\ba\label{eq:AntiCommutators}
\lbb x^i, p_j  \rbb&= i\delta^i_j\,, \qquad ~~~\:\: \lbb p_i, p_j  \rbb =R_{ijkl}\psi^k\bar\psi^l\,, \qquad\qquad\quad\:\, \qquad \lbb x^i, x^j \rbb=0\,, \\
\lbbb \psi^i, \bar\psi^j \rbbb &= g^{ij} \,, \qquad~\, \lbbb \psi^i, \psi^j \rbbb = 0 \,, \:\,\,\qquad\qquad\qquad\qquad\qquad \lbbb \bar\psi^i, \bar\psi^j \rbbb = 0 \,,  \\
\lbbb \lambda^\alpha, \bar\lambda^\beta \rbbb &= \kappa^{\alpha\beta} \,, \qquad \lbbb \lambda^\alpha , \lambda^\beta \rbbb = 0 \,, \:\,\,\qquad\qquad\qquad\qquad\quad~\, \lbbb \bar\lambda^\alpha,\bar\lambda^\beta \rbbb = 0 \,,  \\
\ea \ee
with all other (anti-)commutators vanishing. The brackets in \eqref{eq:AntiCommutators} are Dirac brackets. As unphysical Hilbert space $\CH$ we choose
\be
\CH=\Omega(M_3) \otimes \tn{Cliff}(d)\,,
\ee
where $\tn{Cliff}(d)$ is the standard representation of a Clifford algebra of dimension $2^d$ where $d=\dim\mathfrak{g}_{\tn{ADE}}$. The Hermitian inner product $\braket{\,\cdot\,,\cdot\,}$ on $\CH$ is given by
\be\label{eq:InnerProd}
\braket{\omega_1\otimes v_1 ,\omega_2\otimes v_2}=(v_1,v_2) \times \int_{M_3} *\,\overline{\omega}_1\wedge  \omega_2 \,,
\ee
where the inner product $(\,\cdot\,,\cdot\,)$ on $\tn{Cliff}(d)$ will be shortly described as the canonical inner product on a standard Fock space. As later explained it restricts to the Killing form on the Lie algebra $\mathfrak{g}_{\tn{ADE}}$ identified with the 1-particle subspace of the Fock space. The operators realising the (anti-)commutation relations \eqref{eq:AntiCommutators} are
\be\ba\label{eq:Quantization}
x^i&=x^i \times\,, \qquad \qquad\, p_i=-i\nabla_i \,, \\
\psi^i&=g^{ij} \iota_{\del/\del x^j} \,, \qquad  \bar{\psi}^i=dx^i \wedge \,, \\
\lambda^\alpha &=a^\alpha \,, \qquad\qquad~\, \bar{\lambda}_\alpha= a_\alpha^\dagger \,. \\
\ea\ee
Here $a^\alpha,a_\alpha^\dagger $ are standard anti-commuting lowering and raising operators. The Clifford algebra $\tn{Cliff}(d)$ is then constructed from a vacuum state annihilated by all lowering operators via the action of the raising operators. Setting the norm of this vacuum state to 1 fixes the inner product $(\,\cdot\,,\cdot\,)$ on $\tn{Cliff}(d)$.

The Lagrange multiplier $\zeta$ in the Lagrangian \eqref{eq:SQMLagrangian} gives rise to the constraint
\be\label{eq:PhysicalStates}
 \lb a_\alpha^\dagger a^\alpha-n\rb\ket{\tn{Physical State}}=0\,,
\ee
with respect to standard normal ordering conventions. This condition leads to the definition of the physical Hilbert space $\CH_{\tn{phys.}}\subset \CH$ spanned by physical states. Setting $n=1$ picks out a $d$-dimensional physical subspace from $\tn{Cliff}(d)$ which we identify with the Lie algebra $\mathfrak{g}_{\tn{ADE}}$. These are the states containing a single $\lambda,\bar\lambda$ excitation. The physical Hilbert space thus becomes the space of adjoint valued complex forms on $M_3$
\be\label{eq:PhysicalHilberSpace}
\CH_{\tn{phys.}}=\Lambda\lb M_3,\tn{ad}\,P_{\text{ADE}}\rb \,.
\ee
To determine how the supercharges $Q,Q^\dagger$ act on $\CH$ we note that the anti-commutation relations of \eqref{eq:AntiCommutators} imply
\be
[\tilde T_\alpha,\tilde T_\beta]=ic_{\alpha\beta\gamma}\tilde T^\gamma\,,
\ee
for the contraction $\tilde T$ defined in \eqref{eq:FermionGenerator}. This allows for the identification
\be
\tilde T_\alpha = \tn{ad}_{T_\alpha}= [T_\alpha,\cdot\,]\,,
\ee
where $T_\alpha\in\mathfrak{g}_{\tn{ADE}}$ are hermitian generators. Combining this with the form of the conjugate momenta \eqref{eq:ConjugateMomenta} the supercharges \eqref{eq:SQMSuperscharges} are thus realized as the operators
\be
\CQ=d+ \lbb\varphi\wedge\,,\cdot\, \rbb\,, \qquad \CQ^\dagger=d^\dagger- \lbb\iota_{\bar{\varphi}}\,,\cdot\, \rbb\,.
\ee
Further the 1-particle states $\bar{\lambda}^\sigma\ket{0}=T^\sigma\in\mathfrak{g}_{\tn{ADE}}$ of the physical Hilbert space are identified with Lie algebra generators $T^\sigma$ due to the relation
\be
\tilde T^\alpha \bar{\lambda}^\beta\ket{0}=ic^{\alpha\beta}_{~~\,\gamma}\bar\lambda^\gamma\ket{0}\,,
\ee
in a local trivialzation of $\tn{ad}\,P_{\text{ADE}}$\,.

A complete basis of the physical Hilbert space \eqref{eq:PhysicalHilberSpace} is given by
\be\label{eq:HilberSpaceBasis}
\CB_{\tn{phys.}}=\lbbb \bar\lambda^\alpha\ket{x^l},\, \bar\lambda^\alpha\bar\psi^i\ket{x},\, \bar\lambda^\alpha\bar\psi^i\bar\psi^j\ket{x},\, \bar\lambda^\alpha\bar\psi^i\bar\psi^j\bar\psi^k\ket{x}\rbbb
\ee
where $\alpha=1,\dots,d$ and $i,j,k=1,2,3$ and $x\in M_3$. We separated these delta functions by their degree as differential forms. Note that all states contain a $\bar\lambda$ excitation as a consequence of the constraint \eqref{eq:PhysicalStates}.

\subsection{The Hamiltonian and Euclidean Lagrangian}
\label{sec:HamEu}

The Hamiltonian generated by Legendre transformation of the Lagrangian \eqref{eq:SQMLagrangian} reads
\be\ba\label{eq:SQMHamiltonian}
H &= \frac{1}{2}\lb p^i+A_\lambda^i -i\Gamma^{i}_{~jk}\bar\psi^j\psi^k\rb\lb p_{i}+A_{\lambda,i}-i\Gamma_{ijk}\bar\psi^j\psi^k\rb\\
&~~~~+\frac{1}{2}\phi_{i,\lambda}\phi_{\lambda}^i+\lb D_{i}\phi_{j}\rb_\lambda \bar{\psi}^i\psi^j-\zeta \lb \bar{\lambda}^\alpha\lambda_\alpha-n\rb\,,
\ea\ee
which is the Laplacian associated to the covariant derivative $D$ of \eqref{eq:ConnectionsSQM} deformed by the Higgs field $\phi$. Here we have used the BPS equations to simplify the expression. 

Euclidean versions of the Lagrangian \eqref{eq:SQMLagrangian}, Variations \eqref{eq:SQMVariations} and supercharges \eqref{eq:SQMSuperscharges} follow by making the replacement $\tau\rightarrow -i\tau$ in the action. We have
\be\ba\label{eq:EuclideanSQMLagrangian}
\CL^E&= \frac{1}{2}\dot{x}^i\dot{x}_i+\bar{\psi}^i\nabla_\tau \psi_i+\bar{\lambda}^\alpha D_\tau\lambda_\alpha -i\lb F_{ij}\rb_{\lambda} \bar{\psi}^i\psi^j+\frac{1}{2} R_{ijkl} \psi^{i} \bar{\psi}^{j} \psi^{k} \bar{\psi}^{l} \\
&~~~\,+\lb D_{i}\phi_{j}\rb_\lambda \bar{\psi}^i\psi^j+\frac{1}{2} \phi_\lambda^i \phi_{\lambda,i}+\zeta \lb \bar{\lambda}^\alpha\lambda_\alpha-n\rb\,,
\ea\ee
and
\be\ba\label{eq:EuclideanSQMVariations}
\delta^E x^i&=\epsilon\bar{\psi}^i-\bar{\epsilon}\psi^i\,, \\
\delta^E\psi^i&=\epsilon\lb -\dot{x}^i+ \phi_\lambda^i\rb -\epsilon\Gamma_{j k}^{i} \bar{\psi}^{j} \psi^{k}\,,\\
\delta^E\bar{\psi}^i&=\bar{\epsilon}\lb \dot{x}^i+ \phi_\lambda^i\rb-\bar{\epsilon}\Gamma_{jk}^{i} \bar{\psi}^{j} \psi^{k}\,,\\
\delta^E\lambda^\alpha&=-i\epsilon c^{\alpha}_{\,~\beta\gamma} \bar{\psi}^i \varphi_i^\beta\lambda^\gamma-i\bar{\epsilon} c^{\alpha}_{\,~\beta\gamma} \psi^i   \bar\varphi_i^\beta \lambda^\gamma \,, \\
\delta^E\bar{\lambda}^\alpha&= -i\epsilon c^{\alpha}_{\,~\beta\gamma}\bar{\psi}^i  \varphi_i^\beta \bar{\lambda}^\gamma -i\bar{\epsilon } c^{\alpha}_{\,~\beta\gamma}\psi^i \bar\varphi_i^\beta\bar{\lambda}^\gamma\,,
\ea\ee 
of which we highlight the $\psi,\bar\psi$ variations which are key to deriving the generlized instantons of this SQM. This follows from defining the positive definite combination
\be
V=\psi^i\Big( \overline{\CQ \psi}\Big)_i= \psi_i\lb\dot x^i+ \phi_\lambda^i -\Gamma_{jk}^{i} \bar{\psi}^{j} \psi^{k}\rb
\ee
which varies as
\be
\delta^E V=-\dot x^i\dot x_i-\bar\psi^i\nabla_\tau\psi_i+\lb D_i\phi_j\rb_\lambda\bar\psi^i\psi^j-i\lb F_{ij}\rb_\lambda+\phi_{i,\lambda}\phi^i_\lambda+\frac{1}{2} R_{ijkl} \psi^{i} \bar{\psi}^{j} \psi^{k} \bar{\psi}^{l}=\CQ V\,,
\ee
and can be used to deform the euclidean action $S^E\rightarrow S^E-t\CQ V$\,. This deformation of the action leaves the Euclidean partition function or more generally the Euclidean path integral with $\CQ$-closed insertions invariant. In the $t\rightarrow \infty$ limit these path integrals localize on the BPS locus $\delta\psi=0$\,.

\section{Comments on Split Higgs Bundles}
\label{sec:R1}

Here we give details on the colored SQM in the setting of split Higgs bundles. We discuss how Witten's SQM \cite{Witten:1982im, Hori:2000kt} arises when considering the one-particle dynamics of individual color sectors. In each color sector the standard relation between the low energy physics of the SQM and Morse theory holds. This correspondence is generalized by phenomena between sectors of different color, which come in form of generalized instantons in the SQM. The relevant mathematical setting is now Morse theory with multiple morse functions as discussed in \cite{Fukaya_morsehomotopy}.

\subsection{One-partical Dynamics and Witten's SQM}
   
Split Higgs bundles are characterized by Higgs fields $\phi=\phi_IH^I\in \Omega^1(M_3,\mathfrak{g}_{\tn{ADE}})$ valued in the Cartan subalgebra of the gauge algebra $\mathfrak{g}_{\tn{ADE}}$ where the closed 1-forms $\phi_I$ are globally defined on $M_3$ and possibly singular due to source terms in the BPS equations \eqref{eq:MagSource}. Here $H^I$ denote Cartan generators with $I=1,\dots,R$ with $R=\tn{rank}\,\mathfrak{g}_{\tn{ADE}}$. Generically, these backgrounds Higgs the gauge symmetry $G_{\tn{ADE}}$ to $U(1)^R$ and the adjoint representation of the gauge group decomposes as
\be\label{eq:SplittingAp2}
\ba
G_{\tn{ADE}} ~&\rightarrow~  U(1)^R\,, \cr 
{\rm Ad}\,G_{\tn{ADE}}  ~&\rightarrow~  {\rm Ad}\Big( U(1)^R \Big) \oplus \bigoplus_{\alpha\in\Phi} {\bf R}_{\alpha} \,,
\ea
\ee
where $\Phi$ denotes the root system of $\mathfrak{g}_{\tn{ADE}}$ and ${\bf R}_\alpha$ a one-dimensional representation of $U(1)^R$ whose $R$-component charge vector given by the root $\alpha$. The color contracted Higgs field $\phi_\lambda$ for these backgrounds is given by
\be\label{eq:ColorContractedDiagHiggs}
\phi_\lambda=\sum_{\alpha\in \Phi}\alpha_I\phi^I \bar\lambda^\alpha\lambda_\alpha\,,
\ee
where $n_\alpha=\bar\lambda^\alpha\lambda_\alpha=\sum_\beta \kappa_{\alpha\beta}\bar\lambda^\alpha\lambda^\beta$ is a number operator counting the $\bar\lambda$-excitations of a state. The Lagrangian of the colored SQM for this class of Higgs backgrounds takes the form
\be\ba\label{eq:LagrangianUndecomp}
\CL&= \frac{1}{2}\dot{x}^i\dot{x}_i+i\bar{\psi}^i\nabla_\tau \psi_i+i\bar{\lambda}^\sigma \dot\lambda_\sigma-\lb \nabla_{(i}\phi_{j)}\rb_\lambda \bar{\psi}^i\psi^j-\frac{1}{2} \phi_\lambda^i \phi_{\lambda,i} \\
&~~~\,-\frac{1}{2} R_{ijkl} \psi^{i} \bar{\psi}^{j} \psi^{k} \bar{\psi}^{l}
+\zeta \lb \bar{\lambda}^\sigma\lambda_\sigma-n\rb.
\ea\ee
Here $\sigma$ runs over all Lie algebra generators. As a consequence of the diagonal form \eqref{eq:ColorContractedDiagHiggs} of the Higgs field the Lagrangian can be decomposed into color specific components. We collect the color independent terms in
\be
\CL_{\tn{Kin.}}=\frac{1}{2}\dot{x}^i\dot{x}_i+i\bar{\psi}^iD_\tau{\psi}_i-\frac{1}{2} R_{ijkl} \psi^{i} \bar{\psi}^{j} \psi^{k} \bar{\psi}^{l}-n\zeta\,. 
\ee
The colors associated with Cartan generators $H^I$, labelled by $I=1,\dots, R$, feature in the free fermionic Lagrangian
\be
\CL_{\tn{Bulk}}=\sum_{I=1}^R\lb i\bar\lambda^I\dot\lambda_I+\zeta\bar\lambda^I\lambda_I\rb\,.
\ee
We collect the terms involving the number operator $n_\alpha=\bar\lambda^\alpha\lambda_\alpha$ in
\be
\CL_{(\alpha)}= i\bar{\lambda}^\alpha \dot\lambda_\alpha+\zeta  \bar{\lambda}^\alpha\lambda_\alpha- \lb\alpha_I \nabla_{i}\phi_{j}^I \bar{\psi}^i\psi^j\rb\bar\lambda^{\alpha}\lambda_\alpha- \frac{1}{2}(\alpha_I\phi^I)^2 \lb\bar\lambda^{\alpha}\lambda_\alpha\rb^2\,,
\ee
where there is no sum running over $\alpha$, and all the terms involving a mix of distinct number operators in
\be
\CL_{(\alpha\beta)}=-(\alpha^I\phi_I)(\beta^I\phi_I)\lb\bar\lambda^{\alpha}\lambda_\alpha\rb\lb\bar\lambda^{\beta}\lambda_\beta\rb\,, 
\ee
where $\alpha\neq\beta$. The initial Lagrangian \eqref{eq:LagrangianUndecomp} is then simply the sum of these pieces
\be
\CL=\CL_{\tn{Kin.}}+\CL_{\tn{Bulk}}+\sum_{\alpha\in\Phi}\CL_{(\alpha)}+\sum_{\substack{\alpha,\beta\in\Phi \\ \alpha\neq \beta}}\CL_{(\alpha\beta)}\,.
\ee
The Hamiltonian $H$ is the Legendre transform of the Lagrangian $\CL$ and decomposes similarly
\be\label{eq:SplitHam}
H=H_{\tn{Kin.}}+H_{\tn{Bulk}}+\sum_{\alpha\in\Phi}H_{(\alpha)}+\sum_{\substack{\alpha,\beta\in\Phi \\ \alpha\neq \beta}}H_{(\alpha\beta)}\,.
\ee
Setting $n=1$ in the Lagrangian \eqref{eq:LagrangianUndecomp} and Hamiltonian \eqref{eq:SplitHam} restricts the physical Hilbert space to states containing a single $\bar\lambda$-excitation. We denote physical states of a fixed color by
\be
\ket{\chi,\alpha}=\chi\bar\lambda^\alpha\ket{0}\,, \qquad \chi=\chi_{I}\bar\psi^I\,.
\ee
Here $I$ is a multi index and $\chi\ket{0}$ quantizes to a differential form on $M_3$. Consider two physical states $\ket{\eta,\alpha},\ket{\chi,\beta}$ of distinct color $\alpha,\beta$, then trivially $\braket{\chi,\alpha|H|\eta,\beta}=0$ whereby time evolution preserves color. Further the matrix elements between states of the same color $\ket{\chi,\alpha},\ket{\eta,\alpha}$ only features two of the Hamiltonian pieces in \eqref{eq:SplitHam}, we have
\be\label{eq:SimpleHam}
\braket{\chi,\alpha|H|\eta,\alpha}=\braket{\chi,\alpha|\lbb H_{\tn{Kin.}}+H_{(\alpha)}\rbb|\eta,\alpha}\,.
\ee
Commuting the $\bar\lambda,\lambda$ past each other in \eqref{eq:SimpleHam} we find a matrix element between two differential $p$-forms $\chi,\eta$ and the operator
\be
H|_{{\bf R}_\alpha}=H_{\tn{Kin.}}+H_{(\alpha)}/n_\alpha\,.
\ee
When viewed as an operator on the physical Hilbert subspaces of fixed color $\alpha$ the color restricted Hamiltonian $H|_{{\bf R}_\alpha}$ quantizes to
\be\label{eq:SuperchargeRestricted}
H|_{{\bf R}_\alpha}=\lbbb \CQ|_{{\bf R}_\alpha} ,\bar \CQ|_{{\bf R}_\alpha} \rbbb\,, \qquad \CQ|_{{\bf R}_\alpha}=d+\alpha^I\phi_I\wedge\,,
\ee
which acts on differential $p$-forms. In similar fashion one derives the Hamiltonian acting on colors associated with the Cartan subalgebra of the gauge algebra to be the Laplace-Beltrami operator. Whenever the Cartan components of the Higgs field are exact $\phi_I=df_I$ the supercharge $\CQ|_{{\bf R}_\alpha}$ in \eqref{eq:SuperchargeRestricted} reduces to that of Witten's SQM with superpotential $\alpha^If_I$. For split Higgs bundles we can therefore associate an SQM to every Lie algebra generator $E^\alpha$ with $\alpha\in\Phi$ as well as $R$ copies of a free SQM corresponding to generators of the Cartan subalgebra. 

The group $G_\tn{ADE}$ may also be partially Higgsed to $G_{\tn{GUT}}\times U(1)^k$. In this case one again obtains an SQM for every Lie algebra generator, where now generators spanning $\tn{Ad}\,G_{\tn{GUT}}\times U(1)^k$ are associated with free SQMs and generators spanning representations of $G_{\tn{GUT}}\times U(1)^k$ are associated with the same SQM whose supercharge is determined by the Higgs background and its vector of $U(1)^k$ charges.

As a simple example of this degenerate setting consider a Higgs fields $\phi=\phi_{\mathfrak{t}}\mathfrak{t}$ which is turned on along a Cartan generator $\mathfrak{t}$ in such a way that the gauge group and its adjoint representation breaks
\be\label{eq:SplittingAp}
\ba
G_{\tn{ADE}} ~&\rightarrow~  G_{\tn{GUT}}\times U(1)\,, \cr 
{\rm Ad}\,G_{\tn{ADE}}  ~&\rightarrow~  {\rm Ad}\,G_{\tn{GUT}} \oplus {\rm Ad}\,U(1) \oplus {\bf R}_{q} \oplus \overline{\bf R}_{-q} \,.
\ea
\ee
In this case we associate with the Lie algebra generators spanning ${\rm Ad}\, G_{\tn{GUT}}^{\,}\oplus^{\,}{\rm Ad}\,U(1)$ the SQM with Lagrangian
\be
\CL|_{{\rm Ad}\, G_{\tn{GUT}}^{\,}\oplus^{\,}{\rm Ad}\,U(1)}=\frac{1}{2}\dot{x}^i\dot{x}_i+i\bar{\psi}^iD_\tau{\psi}_i-\frac{1}{2} R_{ijkl} \psi^{i} \bar{\psi}^{j} \psi^{k} \bar{\psi}^{l}\,.
\ee
To the representations ${\bf R}_q,\overline{\bf R}_{-q}$ we associate to each an SQM given by
\be\ba\label{eq:ReducedLagrangian}
\CL|_{{\bf R}_q}&=\frac{1}{2}\dot{x}^i\dot{x}_i+i\bar{\psi}^i\nabla_\tau{\psi}_i-\frac{1}{2} R_{ijkl} \psi^{i} \bar{\psi}^{j} \psi^{k} \bar{\psi}^{l}- q\lb \nabla_{i}\phi_j\rb_\mathfrak{t}\bar{\psi}^i\psi^j-\frac{1}{2}q^2\phi_\mathfrak{t}^2\,,\\
\CL|_{\overline{\bf R}_{-q}}&=\frac{1}{2}\dot{x}^i\dot{x}_i+i\bar{\psi}^i\nabla_\tau{\psi}_i-\frac{1}{2} R_{ijkl} \psi^{i} \bar{\psi}^{j} \psi^{k} \bar{\psi}^{l}+ q\lb \nabla_{i}\phi_j\rb_\mathfrak{t}\bar{\psi}^i\psi^j-\frac{1}{2}q^2\phi_\mathfrak{t}^2\,,
\ea\ee
respectively. This example naturally generalizes to higher rank Higgsing and is the point of view partially taken in \cite{Braun:2018vhk}, where the relevant SQMs where correctly noticed, however, without fitting these together in the frame work of colored SQMs.

\subsection{Perturbative Ground States and their Morse-Witten Complex}
\label{sec:RecapMW}
The Morse-Witten complex of a colored SQM probing a split Higgs bundle is the direct sum of the Morse-Witten complexes of the Witten SQMs embedded within it. Consider the Witten SQM associated to the root $\alpha$ as in \eqref{eq:SuperchargeRestricted} with supercharge
\be\label{eq:RestrictedSupercharge}
\CQ|_{{\bf R}_\alpha}=d+\alpha^I\phi_I\wedge\,,
\ee
where we require $\alpha^I\phi_I$ to be a Morse 1-form, i.e. the zeros of $\alpha^I\phi_I$ are isolated and when expressing $\alpha^I\phi_I=\alpha^Idf_I$ through locally defined function $f_I$ the Hessian of $\alpha^If_I$ is non-degenerate. In other words, $\alpha^I\phi_I$ is locally derived from a Morse function potential. The Morse index $\mu_\alpha(p)$ of an isolated zero $p\in M_3$ of $\alpha^I\phi_I$ is defined to be the number of negative eigenvalues of the non-degenerate Hessian of $\alpha^If_I$ at $p$. 

In a neighbourhood of a vanishing point $p\in M_3$, parametrized by normal coordinates centred at $p$ diagonalizing the Hessian, we have the approximation 
\be\label{eq:NormalCoords}
\alpha^If_I(x)=c_1^{(\alpha)}(x^1)^2+c_2^{(\alpha)}(x^2)^2+c_3^{(\alpha)}(x^3)^2+\CO(|x|^3)\,.
\ee
An unnormalized perturbative ground state is then given to leading order in $|x|$ by
\be
\ket{p,\mu_\alpha}=\exp\lbb-\lb |c_1^{(\alpha)}|(x^1)^2+|c_2^{(\alpha)}|(x^2)^2+|c_3^{(\alpha)}|(x^3)^2 \rb\rbb dx^{I}
\ee
where the set $I\subset \lbbb 1,2,3\rbbb$ lists the subset of indices $i=1,2,3$ for which $c_i^{(\alpha)}<0$, it contains $|I|=\mu_\alpha(p)$ elements. This perturbative ground state is embedded into the colored SQM by tensoring it with the Lie algebra generator $E^\alpha$ associated to the root $\alpha$. We write
\be
\ket{ p,\lambda^\alpha,\mu}=\exp\lbb-\lb |c_1^{(\alpha)}|(x^1)^2+|c_2^{(\alpha)}|(x^2)^2+|c_3^{(\alpha)}|(x^3)^2 \rb\rbb dx^{I}\otimes E^\alpha\,,
\ee
for a perturbative ground state of the colored SQM, dropping the index $\alpha$ on $\mu$ as the color $\alpha$ is now explicitly featured. These states generate the chain complex of the colored Morse-Witten complex
\be\label{eq:MWComplexes}
C^\mu(M_3,\phi)=\bigoplus_{\alpha\in\Phi} C^\mu_{(\alpha)}(M_3)\,,\qquad C^\mu_{(\alpha)}(M_3)=\bigoplus_{A} \R\cdot  \ket{p_A,\lambda^\alpha,\mu}\,,
\ee
where the index $A$ runs over all vanishing points of $\alpha_I\phi^I$ with Morse index $\mu$. The chain complexes $C^\mu_{(\alpha)}(M_3)$ are the Chain complexes of the uncolored Witten SQMs \eqref{eq:SuperchargeRestricted} associated to the individual colors. The supercharge $\CQ$ of the colored SQM induces the boundary map of this complex
\be\label{eq:MWcomplex1}
\begin{tikzcd}[row sep=0.25in]
C^3(M_3,\phi)\arrow[r,swap,yshift=-2,"\bar\CQ"]   & C^2(M_3,\phi) \arrow[l,swap,yshift=2,"\CQ"] \arrow[r,swap,yshift=-2,"\bar\CQ"] & C^1(M_3,\phi) \arrow[l,swap,yshift=2,"\CQ"] \arrow[r,swap,yshift=-2,"\bar\CQ"] & C^0(M_3,\phi) \arrow[l,swap,yshift=2,"\CQ"]\,.
\end{tikzcd}
\ee 
However, the Higgs field \eqref{eq:ColorContractedDiagHiggs} is diagonal and as a consequence the supercharge preserves color and the complex contains the subcomplexes
\be\label{eq:MWcomplex2}
\begin{tikzcd}[row sep=0.25in]
C^3_{(\alpha)}(M_3,\phi)\arrow[r,swap,yshift=-2,"\bar\CQ|_{{\bf R}_\alpha} "]   & C^2_{(\alpha)}(M_3,\phi) \arrow[l,swap,yshift=2,"\CQ|_{{\bf R}_\alpha}"] \arrow[r,swap,yshift=-2,"\bar\CQ|_{{\bf R}_\alpha}"] & C^1_{(\alpha)}(M_3,\phi) \arrow[l,swap,yshift=2,"\CQ|_{{\bf R}_\alpha}"] \arrow[r,swap,yshift=-2,"\bar\CQ|_{{\bf R}_\alpha}"] & C^0_{(\alpha)}(M_3,\phi) \arrow[l,swap,yshift=2,"\CQ|_{{\bf R}_\alpha}"]\,,
\end{tikzcd}
\ee 
where the boundary map follows from restricting the supercharge to each color sectors as given in \eqref{eq:RestrictedSupercharge}. As a consequence the Morse-Witten complex of the colored SQM is simply a direct sum of the Morse-Witten complexes it contains. This is made completely explicit through the action of the supercharge on perturbative ground states
\be\label{eq:SuperChargeAction}
\CQ\ket{p,\lambda^\alpha,\mu} =\sum_{q\in M_3}\sum_{\gamma_{pq}}(\pm)_{\gamma_{pq}}\exp\lbb -t\int_{\,p}^{\,q} \alpha^I\phi_I\big|_{\gamma_{pq}}\rbb\ket{q,\lambda^\alpha,\mu+1}+\CO(1/t)\,,
\ee
where $q$ runs over the vanishing points of $\alpha^I\phi_I$ of Morse index $\mu+1$ and $t$ is parameter which was introduced by rescaling the Higgs field $\phi\rightarrow t\phi$. The second sum runs over all ascending gradient flow lines $\gamma_{pq}$ of $\alpha^I\phi_I$ connecting the points $p,q$ and $(\pm)_{\gamma_{pq}}$ is a sign related to the orientation of the flow line in the moduli space of flow lines.

When the Higgsing is partial, as e.g. in \eqref{eq:SplittingAp}, the Morse-Witten complexes associated to the Lie algebra generators $E^\alpha$ spanning the same representations of the remnant gauge group are identical. For example with \eqref{eq:SplittingAp} we would have two complexes with boundary operators $\CQ|_{{\bf R}_q}=d+q\phi\wedge$ and $\CQ|_{\overline{\bf R}_{-q}}=d-q\phi\wedge$ which are related by the Hodge star.

The cohomology groups of the complex \eqref{eq:MWcomplex1} follow from those of the complex \eqref{eq:MWcomplex2}. They are Novikov cohomology groups with resepect to the closed 1-form $\alpha^I\phi_I$. The ranks of these cohomologies are constrained by certain symmetries \cite{LNov1970,dur4050}. Whenever $M_3$ is connected, compact, orientable and without boundary and the Higgs field has no singularities the Euler character and thereby the chiral index in 4d vanishes. Singular backgrounds are necessary to generate a non-vanishing Euler character and therefore a chiral spectrum in 4d.

\section{Torsion in KK Reductions}
\label{sec:Torsion}

In this appendix we comment on the role played by torsion factors in KK reductions. The set-up under consideration is a 3-manifold $\CC$ whose first homology contains the torsion factors
\be\label{eq:TorHomRed}
\tn{Tor}\,H_1(\CC,\Z)=\Z_{m_1}\oplus \dots \oplus \Z_{m_p}\,.
\ee
The torsion homology groups contribute additional differential forms for the compactification to 4d. These are not detected by the de Rham cohomology groups $H^i_{\tn{dR}}(\CC)$ and are not described by Hodge theory. Indeed, consider a torsion cycle $\gamma\in \tn{Tor}\,H_1(\CC_k,\Z)$ of order $m$, i.e. $m\gamma=\del \Sigma$ where $\Sigma$ is a 2-cycle and briefly assume $\tn{Tor}\,H_1(\CC_k,\Z)\cong \Z_m$. Any 1-form $\alpha\in \Omega^1(\CC_k)$ integrated against the torsion cycle vanishes by Stoke's theorem
\be
\int_\gamma \alpha=\frac{1}{m}\int_{m\gamma}\alpha=\int_\Sigma d\alpha=0\,.
\ee
However to any such torsion cycle one can associate a differential form via the universal coefficient theorem
\be\label{eq:UniversalCoefficient}
\tn{Tor}\,H^{i+1}(\CC_k,\Z)\cong \tn{Hom}\lb \tn{Tor}\,H_{i}(\CC_k,\Z),\mathbb{Q}/\Z\rb\,,
\ee
which here yields a 2-form $\beta\in \Omega^2(\CC_k)$ associated with the torsional 1-cycle $\gamma$. The form $\beta$ acts on torsional cycles $\gamma$ as
\be\label{eq:Simple}
\beta(\gamma)=\frac{1}{m}\int_\Sigma \beta~~~\tn{mod}\,1\,, \qquad \beta(\gamma)\in \Z/m=\lbbb \frac{0}{m},\frac{1}{m}\,, \dots\,, \frac{m-1}{m}\rbbb\,.
\ee
This form is exact, but not co-closed, and an eigenvector of the Laplacian $\Delta$ on $\CC_k$, i.e. for a positive mass $\mu$ we have
\be\label{eq:Example}
m\beta=d\alpha\,, \qquad \Delta \beta=-\mu \beta \,, \qquad  \Delta \alpha=-\mu \alpha\,.
\ee
Torsional forms and their role in compactifications are discussed in \cite{Wen:1985qj, freed1986,  Marchesano:2006ns, Camara:2011jg}. 

We now discuss the generalizations of \eqref{eq:Example} to the more general homology groups \eqref{eq:TorHomRed}. For 3-manifolds Seifert introduced in \cite{zbMATH03011631} the linking form $L$, generalizing \eqref{eq:Simple}, which is defined by
\be
L~:~ \tn{Tor}\,H_1(\CC,\Z) \times \tn{Tor}\,H_1(\CC,\Z) \rightarrow \mathbb{Q}/\Z\,, \qquad L(\gamma_i,\gamma_j)= \frac{\gamma_i\cdot \Gamma_j}{m_j}~~~\tn{mod}\, 1\,,
\ee
where $\Gamma_j$ is a 2-chain such that $\del \Gamma_j=m_j\gamma_j$ and the operation denoted by $``\,\cdot\,"$ abbreviates intersections. The universal coefficient theorem \eqref{eq:UniversalCoefficient} now yields $p$ 2-forms $\beta_i$ to which there exist $p$ 1-forms $\alpha_i$ such that the relations \eqref{eq:Example} generalize to
\be\label{eq:Relation}
d\alpha_i=(L^{-1})_i^{~j}\beta_j\,, \qquad (L^{-1})_i^{~j}\in \Z\,,\qquad \int_{M_3}\alpha_i\wedge \beta_j=\delta_{ij}\,, \qquad i,j=1,\dots,p\,.
\ee
These $2p$ forms again span eigenspaces of the Laplace operator and are characterized by a positive definite mass matrix $M$ as
\be
\Delta \alpha_i=-M_{i}^{~j}\alpha_j\,,\qquad\Delta \beta_i=-\widetilde M_{i}^{~j}\beta_j\equiv-(LML^{-1})_{i}^{~j}\beta_j\,. 
\ee

\bibliographystyle{JHEP}
\bibliography{G2}

\end{document}